\documentclass[aps,prx,reprint,superscriptaddress]{revtex4-2}

\usepackage[T1]{fontenc}
\usepackage[utf8]{inputenc}
\usepackage{graphicx}
\usepackage{amsmath,amssymb}
\usepackage{mathtools}
\usepackage{braket}
\usepackage{slashed}
\usepackage{cancel}
\usepackage{mathrsfs}
\usepackage{latexsym}

\usepackage{booktabs}
\usepackage{multirow, makecell}
\usepackage{array}
\newcolumntype{R}[1]{>{\raggedleft\arraybackslash}p{#1}}
\newcolumntype{L}[1]{>{\raggedright\arraybackslash}p{#1}}
\newcolumntype{C}[1]{>{\centering\arraybackslash}p{#1}}

\usepackage{xcolor}
\usepackage{soul}
\usepackage{empheq}
\usepackage{simpler-wick}
\usepackage{placeins}
\usepackage{afterpage}

\usepackage[colorlinks=true, allcolors=blue]{hyperref}

\allowdisplaybreaks
\begin{document}
\title{\textbf{Cavity elimination in cavity-QED: a self-consistent input-output approach}}

\author{E. Rambeau }
\email{eliott.rambeau@universite-paris-saclay.fr}
\affiliation{Universit\'e Paris Cit\'e, Centre de Nanosciences et de Nanotechnologies, F-91120 Palaiseau, France} 
\affiliation{Université Paris-Saclay, CNRS, Centre de Nanosciences et de Nanotechnologies, 91120, Palaiseau, France}

\author{L. Lanco}
\email{loic.lanco@u-paris.fr}
\affiliation{Universit\'e Paris Cit\'e, Centre de Nanosciences et de Nanotechnologies, F-91120 Palaiseau, France} 
\affiliation{Université Paris-Saclay, CNRS, Centre de Nanosciences et de Nanotechnologies, 91120, Palaiseau, France}

\begin{abstract}
    Simplifying composite open quantum systems through model reduction is central to enable their analytical and numerical understanding. In this work, we introduce a self-consistent approach to eliminate the cavity degrees of freedom of cavity quantum electrodynamics (CQED) devices in the non-adiabatic regime, where the cavity memory time is comparable with the timescales of the atom dynamics. To do so, we consider a CQED system consisting of a two-level atom coupled to a single-mode cavity, both subsystems interacting with the environment through an arbitrary number of ports, within the input–output formalism. A self-consistency equation is derived for the reduced atom dynamics. This allows retrieving an exact expression for the effective Purcell-enhanced emission rate and, under reasonable approximations, a set of self-consistent dynamical equations and input-output relations for the effective two level atom. The resulting reduced model captures non-Markovian features, characterized through an effective Lindblad equation exhibiting two decoherence rates, a positive and a negative one. In the continuous-wave excitation regime, we benchmark our approach by computing effective steady states and output flux expressions beyond the low-power excitation regime, for which a semi-classical treatment is usually applied. We also compute two-time correlations and spectral densities, showing an excellent agreement with full cavity quantum electrodynamics simulations, except in the strong-coupling, high-excitation regime. Our results provide a practical framework for reducing the size of CQED models, which could be generalized to more complex atom and cavity configurations.
\end{abstract}

\maketitle

\section{Introduction}
Quantum light–matter interactions in cavity-confined electromagnetic modes have led to both fundamental and technological advances, in platforms ranging from cavity quantum electrodynamics (CQED)~\cite{Haroche2006,Walther2006} to its superconducting version, circuit-QED~\cite{Blais2021}. Cavity QED studies the interaction between stationary quantum systems like atoms~\cite{Reiserer2015,Kimble1998}, ions~\cite{Blatt2012}, or solid state emitters~\cite{Janitz2020,Bradac2020,Gonzalez-Tudela2024,Senellart2017} and the quantized modes of a resonator. This has enabled the observation of key phenomena of quantum physics, such as vacuum Rabi splitting~\cite{Thompson1992,Khitrova2006,Yoshie2004}, quantum non-demolition (QND) measurements~\cite{Gleyzes2007,Grangier1998}, Dicke superradiance~\cite{mivehvar2021}, one-atom lasing~\cite{McKeever2003}, or the study of controlled decoherence~\cite{Brune1996}. In circuit QED, artificial atoms based on superconducting circuits are coupled to microwave resonators, providing a high degree of engineering control and scalability, and enabling new regimes of quantum measurement and control~\cite{Campagne-Ibarcq2013,Sarlette2012}. Beyond their role in fundamental science, both cavity- and circuit-QED platforms are regarded as promising building blocks for future quantum technologies~\cite{Reiserer2015,Kimble2008,ladd2010,blais2020}, whether the cavities are used as emitters~\cite{Somaschi2016,Couteau2023,Thomas2022} or receivers~\cite{Duan2004,LeJeannic2022,Monroe2002,Pichler2017} of quantum light.

Cavity QED provides textbook examples of composite open quantum systems, yielding intensive theoretical studies~\cite{Carmichael2007,Haroche2006,Larson2024}. Its dynamics has been solved exactly in some specific cases, e.g when a two level atom in a single-mode cavity is used as a single-photon source, emitting both through and outside the cavity mode after an initial excitation event~\cite{Carmichael1989,Andreani1999,Auffeves2008}. To describe CQED devices both as emitters and receivers, a powerful approach consists in using the input--output formalism~\cite{Gardiner1985,Gardiner2004}. This formalism allows expressing the output fields of a system in terms of its inputs, creating a very practical method to compute fluxes and correlations~\cite{Gardiner1985,Caneva2015,Flayac2013,Xu2015}. It also enables deriving Heisenberg–Langevin equations~\cite{Carmichael1993b,Walls2008} describing the system's dynamics while properly taking into account vacuum fluctuations. This approach, also often applied to waveguide-QED~\cite{Fan2010,Lalumiere2013}, can be generalized to describe cascaded quantum systems~\cite{Gardiner1993,Carmichael1993} and quantum networks in general~\cite{Reiserer2015,Gough2009}, particularly within the SLH formalism~\cite{Combes2017}. In the low-excitation regime, that is when the excited state population is negligible, the input-output description of cavity-QED devices can be combined with a semi-classical approximation to derive analytical expressions for various scattering amplitudes, such as reflectivity and transmission coefficients~\cite{Waks2006,Majumdar2012}. In presence of increasing excitation power, the semi-classical approximation can be used to understand physically interesting effects such as giant optical non-linearities~\cite{Auffeves-Garnier2007}. However, it could also lead to unverified predictions artificially emerging in the theory when the quantum fluctuations of the atomic state are neglected~\cite{Lugiato1983,Savage1988}. To our knowledge, no exact solutions have been found to describe the dynamics and the optical response of cavity-QED devices in the general case, even when considering a single 2-level atom in a single-mode cavity, with a single input field.

Adiabatic elimination approaches early emerged to overcome this issue and eliminate the cavity modes in cavity-QED~\cite{Carmichael2007,Cirac1992,Warszawski2000}, thereby reducing the system's Hilbert space and leading to an analytically solvable dynamics. The development of adiabatic elimination techniques in quantum physics has constituted a whole research field that was initiated decades ago and used to understand lasers and masers~\cite{Lax1967,Haken1975}, atom cooling~\cite{Cohen-Tannoudji1992} as well as three-level atomic systems~\cite{Dalibard1989}. This field is built on the assumption that there exists two separate time-scales in the studied physical system, allowing to define an effective system where all the fast processes have been neglected and whose slow dynamics can be more easily studied. A general adiabatic elimination procedure for open quantum systems was early proposed in~\cite{Mirrahimi2009} using a geometric approach, subsequently developed in~\cite{Azouit2017} and applied to various systems in~\cite{Piccione2022,LeRegent2023}. This framework relies on a bijective mapping between the exact density matrix and a reduced manifold: using singular perturbation theory~\cite{Tikhonov1952,Fenichel1979} this allows deriving the effective slow dynamics to a desired order in the perturbation parameter. There now exist a variety of different elimination techniques which can be applied to a broad range of systems~\cite{Finkelstein-Shapiro2020,Saideh2020,Tokieda2025,Jager2022,Burgarth2021,Macieszczak2016,Brion2007,Douglas2015,Reiter2012,Cirac1992,Wiseman1993,Reiter2012,Macieszczak2016,Damanet2019,Grigoletto2023,Grigoletto2024,Dridi2010}, beyond the usual adiabatic approximation. Consequently, signatures of non-Markovianity could be identified across several of these frameworks, including the geometric~\cite{Azouit2017}, projection-based~\cite{Finkelstein-Shapiro2020}, and exact model reduction~\cite{Grigoletto2023} approaches. The generality of these reduction schemes makes them very powerful, yet complicates their consistent integration within the input–output theory: this poses a challenge for deriving the effective dynamics of the output fields that constitute the primary observables in CQED experiments. 

In this work, we introduce a \emph{self-consistent cavity elimination} method for the description of CQED devices. This framework remains valid when the cavity memory time is comparable to the atom dynamical timescales, thereby extending the validity of effective descriptions beyond the standard adiabatic regime. We consider a two-level system in a single-mode, multi-port cavity, which we describe using Heisenberg–Langevin equations in the input–output formalism. Under controlled approximations, we derive an effective atomic evolution characterized by generalized Purcell factor and Rabi frequency. Furthermore, we obtain output-operator expressions that depend solely on the input fields and atomic operators, thereby capturing the complete system dynamics within a reduced Hilbert space. We demonstrate that, under continuous-wave coherent excitation, the self-consistent approach provides both analytical and numerical predictions which are found to match the numerical predictions of the full CQED model at low excitation power, even in the strong coupling-regime. They are also shown to remain valid at high power in the weak-coupling regime. This is made possible by the fact that the self-consistent approach partially takes into account the system's non-Markovianity, which manifests itself through the presence of a negative decoherence rate in the effective dynamics. From a practical standpoint, this approach offers key advantages: (i) it yields new physical insights thanks to the possibility of deriving analytical expressions describing various quantities and phenomena; (ii) it allows performing efficient numerical simulations within a drastically-reduced Hilbert space.

The paper is organized as follows. In Sec.~\ref{Chapter_2}, we introduce the model of a two-level atom coupled to a single-mode, multi-port cavity, and we establish the corresponding Heisenberg–Langevin equations and input–output relations. In Sec.~\ref{Chapter_3}, we develop the self-consistent cavity elimination approach, establishing an exact self-consistency equation and a useful approximation which allows deriving the general analytical results used throughout this work. Section~\ref{Chapter_4} analyzes the dynamics of the effective, reduced system under coherent excitation, proving its non-Markovian evolution. Secs.~\ref{Chapter_5}–\ref{Chapter_7} are devoted to compute analytically and test numerically respectively the output intensities, the spectral densities - showing an analytical explanation of the Mollow triplet asymmetry - and second-order correlation functions. These calculations show an excellent agreement with the full CQED simulations in the weak-coupling regime, up to relatively high excitation powers. In Sec.~\ref{Chapter_8}, we discuss the current limitations of the self-consistent cavity-elimination approach. These limitations are illustrated with analytical and numerical predictions in the strong-coupling regime, this time showing an agreement only at low power with the full CQED model. Conclusions are given in Sec.~\ref{Chapter_9}.

\section{A two level system in a cavity: input-output description}\label{Chapter_2}
In this section, we introduce the cavity quantum electrodynamics (CQED) model employed throughout the paper. We then present the input–output formalism, which enables the derivation of Heisenberg–Langevin equations and input–output relations. Finally, we describe a preliminary cavity–elimination approach based on the adiabatic approximation. Throughout this work, we adopt units such that $\hbar=1$.
\subsection{Model}
In this paper, we consider a cavity quantum electrodynamics (CQED) system consisting of a two-level system (TLS) coupled to a single-mode optical cavity. The system interacts with its environment through multiple ports that allow for both driving and dissipation. The dynamics of this composite system is governed by the following Hamiltonian~\cite{Gardiner1985}:
\begin{equation}
    \hat{\mathcal{H}} = \hat{\mathcal{H}}_{\textrm{sys}}+\hat{\mathcal{H}}_{\textrm{B}}+\hat{\mathcal{H}}_{\textrm{int}}
    \label{Full-Hamiltonian}
\end{equation}
In this expression, $\hat{\mathcal{H}}_{\textrm{sys}}$ is the Jaynes–Cummings Hamiltonian~\cite{Larson2024,Carmichael2007} under the rotating wave approximation (RWA)~\cite{Carmichael2007} with respect to a reference frequency $\omega_{\textrm{ref}}$. This frequency is arbitrary but commonly taken as the atom frequency~\cite{Carmichael2007} ($\omega_{\rm a}$), the cavity frequency ($\omega_{\rm c}$)~\cite{Auffeves2008,Lodahl2015} or input frequency~\cite{Majumdar2012} ($\omega_{\rm in}$) in the presence of an excitation. This Hamiltonian is:
\begin{equation}
\begin{split}
        &\hat{\mathcal{H}}_{\textrm{sys}} = \hat{\mathcal{H}}_{\textrm{a}}+\hat{\mathcal{H}}_{\textrm{c}}+\hat{\mathcal{H}}_{\textrm{a-c}}\\
        &\hat{\mathcal{H}}_{\textrm{a}} = (\omega_{\textrm{a}}-\omega_{\textrm{ref}})\hat{\sigma}^\dagger\hat{\sigma}\\
        &\hat{\mathcal{H}}_{\textrm{c}} = (\omega_{\textrm{c}}-\omega_{\textrm{ref}})\hat{a}^\dagger\hat{a}\\
        &\hat{\mathcal{H}}_{\rm a-c} = ig(\hat{\sigma}^\dagger\hat{a}-\hat{a}^\dagger\hat{\sigma})
    \end{split}
    \label{Jaynes-Cumming-Hamiltonian}
\end{equation}
The cavity annihilation operator $\hat{a}$ satisfies $[\hat{a},\hat{a}^\dagger] = \hat{\mathbb{I}}$, while the atomic lowering operator $\hat{\sigma}$ obeys $\left\{\hat{\sigma},\hat{\sigma}^\dagger\right\} = \hat{\mathbb{I}}$. The parameter $g\geq 0$ denotes the atom–cavity coupling strength. To this system's Hamiltonian, we add the bath as a bosonic field with a Hamiltonian $\hat{\mathcal{H}}_{\textrm{B}}$ and its interaction with the atom-cavity system through $ \hat{\mathcal{H}}_{\textrm{int}}$. To do so, we use the input-output theory~\cite{Gardiner1985} and define two types of system's ports:
\begin{itemize}
    \item The cavity mode interacts with its environment by exchanging photons through a set of ports, hereafter denoted as cavity-coupled (CC) ports. Each CC port $j$ is described by an infinite set of bosonic annihilation operators $\hat{b}_j(\omega)$, that satisfy the canonical commutation relations $[\hat{b}_{j}(\omega),\hat{b}_{j'}^\dagger(\omega)]=\delta(\omega-\omega')\delta_{jj'}$. 
    \item The two level system (TLS) directly interacts with its environment by exchanging photons through a set of ports, hereafter denoted as atom-coupled (AC) ports. Each AC port $l$ is described by an infinite set of bosonic annihilation operators $\hat{c}_l(\omega)$, that satisfy the canonical commutation relations $[\hat{c}_{l}(\omega),\hat{c}_{l'}^\dagger(\omega)]=\delta(\omega-\omega')\delta_{ll'}$
\end{itemize}
Such description is rather general as it allows considering any geometry, with an arbitrary number of atom-coupled and cavity-coupled ports. In the rotating wave approximation at frequency $\omega_{\rm ref}$, we write:
\begin{equation}\label{H-int-def}
    \begin{split}
        \hat{\mathcal{H}}_{\textrm{B}} & = \sum_j \int_{-\infty}^\infty d\omega (\omega-\omega_{\rm ref})\hat{b}^\dagger_{\rm j}(\omega)\hat{b}_{\rm j}(\omega)\\ &\quad+\sum_l \int_{-\infty}^\infty d\omega (\omega-\omega_{\rm ref})\hat{c}^\dagger_{\rm j}(\omega)\hat{c}_{\rm j}(\omega)\\
        \hat{\mathcal{H}}_{\textrm{int}} & = i\sum_j\int_{-\infty}^\infty d\omega\sqrt{\frac{\kappa_j}{2\pi}}\left[\hat{b}_j^\dagger(\omega)\hat{a}-\hat{a}^\dagger\hat{b}_j(\omega)\right]\\ &\quad+i\sum_l\int_{-\infty}^\infty d\omega\sqrt{\frac{\gamma_l}{2\pi}}\left[\hat{c}_l^\dagger(\omega)\hat{\sigma}-\hat{\sigma}^\dagger\hat{c}_l(\omega)\right]
    \end{split}
\end{equation}
$\hat{\mathcal{H}}_{\textrm{int}}$ is assumed to be linear with respect to all the $\hat{b}_{j},\hat{b}_{j}^\dagger,\hat{c}_{l},\hat{c}_{l}^\dagger$. Following~\cite{Gardiner1985}, we consider Markovian damping meaning that both the bath-cavity coupling constants $\kappa_j\geq 0$ and the bath-atom coupling constants $\gamma_l\geq 0$ are frequency-independent. The frequency integrals are extended from $(0,+\infty)$ to $(-\infty,+\infty)$, an approximation valid when the rotating frame frequency $\omega_{\rm ref}$ is much larger than the relevant bandwidths of the system (input spectral width, damping...), a condition typically satisfied in quantum optics~\cite{Gardiner1985}.\newline

Following Ref.~\cite{Gardiner1985}, we now define the input operators for each atom-coupled and cavity-coupled port:
\begin{equation}
    \begin{split}
        \hat{b}_{\textrm{in}}^{(j)} (t) &= \frac{1}{\sqrt{2\pi}}\int_{-\infty}^\infty d\omega e^{-i\omega(t-t_0)}\hat{b}_{j,0} (\omega)\\
        \hat{c}_{\textrm{in}}^{(l)} (t) &= \frac{1}{\sqrt{2\pi}}\int_{-\infty}^\infty d\omega e^{-i\omega(t-t_0)}\hat{c}_{l,0} (\omega)
    \end{split}
\end{equation}
where $\hat{b}_{j,0} (\omega),\hat{c}_{l,0} (\omega)$ are the operators $\hat{b}_j(\omega),\hat{c}_l(\omega)$ evaluated at an initial time $t_0<t$. We also define the output operators for each AC and CC port:
\begin{equation}
    \begin{split}
        \hat{b}_{\textrm{out}}^{\rm (j)} (t) &= \frac{1}{\sqrt{2\pi}}\int_{-\infty}^\infty d\omega e^{-i\omega(t-t_1)}\hat{b}_{j,1} (\omega)\\
        \hat{c}_{\textrm{out}}^{\rm (l)} (t) &= \frac{1}{\sqrt{2\pi}}\int_{-\infty}^\infty d\omega e^{-i\omega(t-t_1)}\hat{c}_{l,1} (\omega)
    \end{split}
\end{equation}
where $\hat{b}_{j,1} (\omega),\hat{c}_{l,1} (\omega)$ are the operators $\hat{b}_j(\omega),\hat{c}_l(\omega)$ evaluated at a final time $t_1>t$ after the interaction. All the operators $\hat{b}_{\textrm{in}}^{(j)},\hat{c}_{\textrm{in}}^{(l)}$ and $ \hat{b}_{\textrm{out}}^{(j)}, \hat{c}_{\textrm{out}}^{(l)}$ satisfy the canonical commutation relations $[\hat{O}(t),\hat{O}^\dagger(t')]=\delta(t-t')$ (with $\hat{O}=\{\hat{b}_{\textrm{in}}^{(j)},\hat{c}_{\textrm{in}}^{(l)},\hat{b}_{\textrm{out}}^{(j)}, \hat{c}_{\textrm{out}}^{(l)}\}$). They are normalized such that the incoming and outgoing photon fluxes through the atom-coupled and cavity-coupled ports, expressed in photons per unit of time, are~\cite{Walls2008}: 
\begin{equation}
    \begin{split}
        \Phi_{\rm CC,in}^{(j)}(t) & = \langle\hat{b}_{\textrm{in}}^{(j)\dagger}(t)\hat{b}_{\textrm{in}}^{(j)}(t)\rangle\\
        \Phi_{\rm CC,out}^{(j)}(t) &= \langle\hat{b}_{\textrm{out}}^{(j)\dagger}(t)\hat{b}_{\textrm{out}}^{(j)}(t)\rangle \\
        \\
        \Phi_{\rm AC,in}^{(l)}(t) & = \langle\hat{c}_{\textrm{in}}^{(l)\dagger}(t)\hat{c}_{\textrm{in}}^{(l)}(t)\rangle\\
         \Phi_{\rm AC,out}^{(l)}(t) &= \langle\hat{c}_{\textrm{out}}^{(l)\dagger}(t)\hat{c}_{\textrm{out}}^{(l)}(t)\rangle
    \end{split}
\end{equation} 
The output operators are also very practical to compute first-order coherences $\langle\hat{b}_{\textrm{out}}^{(j)\dagger}(t)\hat{b}_{\textrm{out}}^{(j)}(t+\tau)\rangle$ and $\langle\hat{c}_{\textrm{out}}^{(l)\dagger}(t)\hat{c}_{\textrm{out}}^{(l)}(t+\tau)\rangle $, as well as second-order coherences $\langle\hat{b}^{(j)\dagger}_{\textrm{out}}(t)\hat{b}^{(j)\dagger}_{\textrm{out}}(t+\tau) \hat{b}^{(j)}_{\textrm{out}}(t+\tau)\hat{b}^{(j)}_{\textrm{out}}(t)\rangle$ and $\langle\hat{c}^{(l)\dagger}_{\textrm{out}}(t)\hat{c}^{(l)\dagger}_{\textrm{out}}(t+\tau) \hat{c}^{(l)}_{\textrm{out}}(t+\tau)\hat{c}^{(l)}_{\textrm{out}}(t)\rangle$. 

\begin{figure}[h]
    \centering
    \includegraphics[width=1\columnwidth]{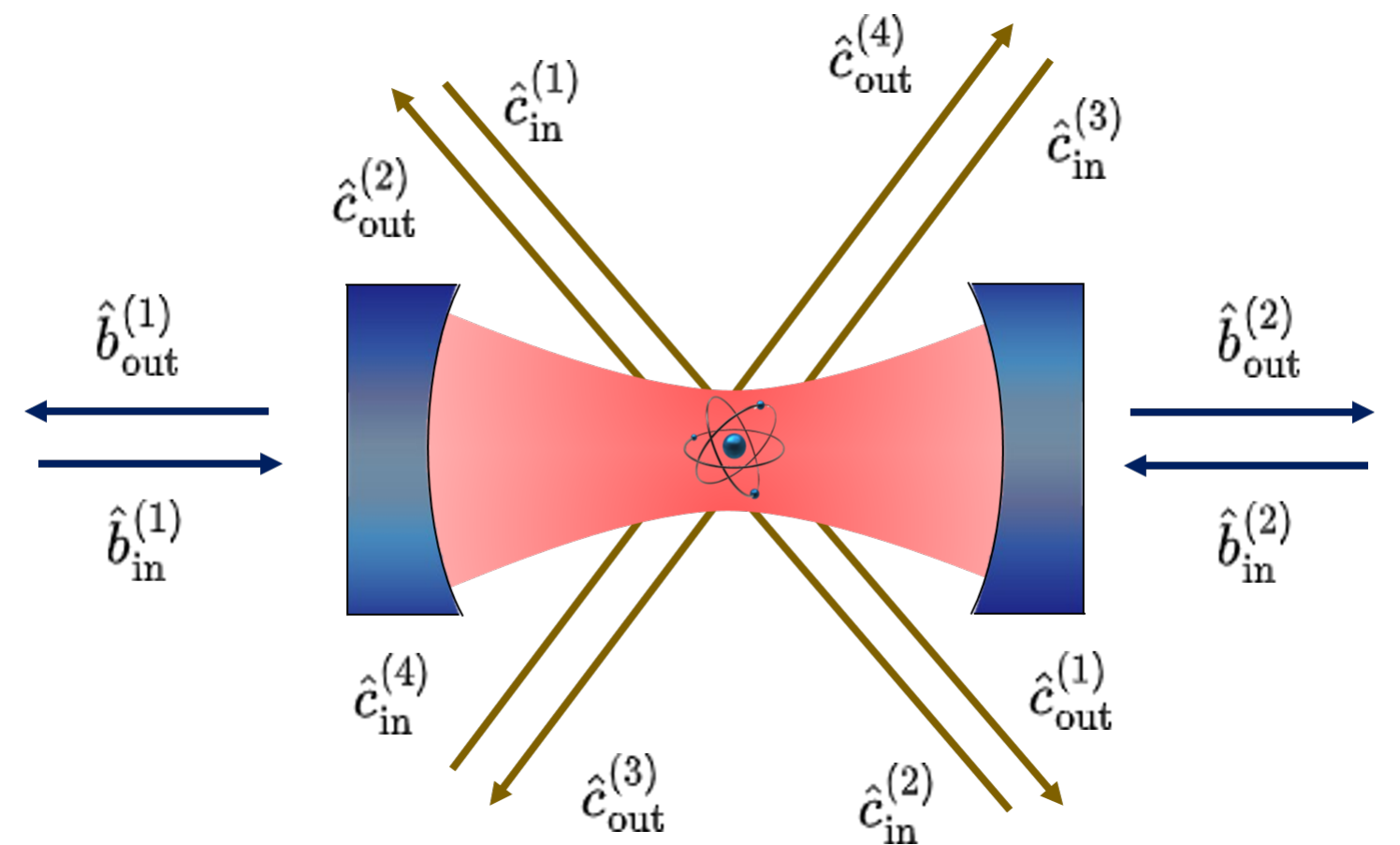}
    \caption{ Scheme of an atom coupled to a single mode cavity, both the atom and the cavity being coupled through several ports to the external environment. The atom is here coupled to the environment through four ports, each one with damping rates $\gamma_l$ with $l\in\{1,2,3,4\}$ and input-output operators $\hat{c}^{(l)}_{\textrm{in}},\hat{c}^{(l)}_{\textrm{out}}$. Similarly, the cavity is here coupled to the environment through two ports, with damping rates $\kappa_j$ with $j\in\{1,2\}$ and input-output operators $\hat{b}^{(j)}_{\rm in},\hat{b}^{(j)}_{\rm out}$. The number of ports can be arbitrarily increased to model more realistic geometries.}\label{Fig-model-description}
\end{figure}

Fig.~\ref{Fig-model-description} shows an example of such an open atom-cavity system with two CC ports and four AC ports. In general each set of ports can be arbitrarily large, allowing to take into account losses of the cavity and spontaneous emission of the atom into the 3D environment, in various geometries. Taking into account all the CC ports with individual cavity-bath coupling constants $\kappa_j$, we define the cavity damping rate as $\kappa=\sum_j\kappa_j$. Similarly, taking into account all the AC coupled ports with atom-bath coupling constants $\gamma_{l}$, we define the rate of emission of the atom outside the fundamental mode of the cavity as $\gamma_{\rm a}=\sum_l\gamma_l$. 

The geometry shown in Fig.~\ref{Fig-model-description}, which is standard in free space CQED, illustrates a typical configuration for the input and output fields. In absence of coupling for the CC ports ($\kappa_{j}=0$), the output fields $\hat{b}_{\rm out}^{(j)}$ in Fig.~\ref{Fig-model-description} correspond to light that has been directly reflected by the cavity mirrors. Contrarily, in absence of coupling for the AC ports ($\gamma_{l}=0$), the output field $\hat{c}_{\rm out}^{(l)}$ in Fig.~\ref{Fig-model-description} correspond to light that has been directly transmitted without interacting with the atom. Since the formalism introduced here is general, more realistic geometries can be taken into account with any number of cavity-coupled or atom-coupled ports.

\subsection{Langevin equations and input-output relations}

To describe the dynamics of the studied system, we follow the procedure detailed in Refs.~\cite{Gardiner1985,Gardiner2004} to derive, with details in the Appendix~\ref{Subsec-appendix-derivation-Heisenberg-Langevin}, the Heisenberg-Langevin equations for our system:

\begin{subequations}\label{Heisenberg-Langevin system}
\begin{empheq}[left=\empheqlbrace]{align}
\Dot{\hat{a}} &= -\kappa' \hat{a}-g \hat{\sigma} - \sum_j\sqrt{\kappa_{j}} \hat{b}_{\textrm{in}}^{(j)}\label{eq:a_dot} \\
\Dot{\hat{\sigma}} &= -\gamma' \hat{\sigma} - g \hat{\sigma}_z \hat{a} + \hat{\sigma}_z\sum_l\sqrt{\gamma_l}\hat{c}_{\rm in}^{(l)}\label{eq:sigma_dot} \\
\Dot{\hat{\sigma}}_z &= 2g(\hat{\sigma}^\dagger\hat{a}+\hat{a}^\dagger\hat{\sigma})-\gamma_{\rm a}(\hat{\sigma}_z+\hat{\mathbb{I}})\label{eq:sigma_z_dot}\\
&\quad -2(\hat{\sigma}^\dagger\sum_l\sqrt{\gamma_{l}}\hat{c}_{\textrm{in}}^{(l)}+\sum_l \sqrt{\gamma_{l}}\hat{c}_{\textrm{in}}^{(l)\dagger}\hat{\sigma})\nonumber
\end{empheq}
\end{subequations}
where the explicit time dependence $(t)$ has been dropped for convenience. In these relations, we use the Pauli operator $\hat{\sigma}_{z} = \hat{\sigma}^\dagger\hat{\sigma}-\hat{\sigma}\hat{\sigma}^\dagger$ and the complex damping rates $\kappa'$ and $\gamma'$ such that:
\begin{equation}
    \begin{split}
        \kappa' &= \frac{\kappa}{2}+i(\omega_{\textrm{c}}-\omega_{\textrm{ref}})\\
        \gamma' &= \frac{\gamma_{\rm a}}{2}+i(\omega_{\textrm{a}}-\omega_{\textrm{ref}})\\
    \end{split}
\end{equation}
Each Heisenberg–Langevin equation contains three main contributions: (i) a free evolution of the atom (proportional to $\gamma'$) or the cavity (proportional to $\kappa'$), including damping, at rates $\gamma_{\rm a}/2$ and $\kappa/2$, and coherent oscillations, at detunings $\omega_{\mathrm{a}}-\omega_{\mathrm{ref}}$ and $\omega_{\mathrm{c}}-\omega_{\mathrm{ref}}$; (ii) an interaction term proportional to $g$; and (iii) Langevin forces from the AC and CC input ports, that play a role in the driving and damping of the system. We note that, when taking the expectation values, one obtains a system of optical Bloch equations that can be strongly simplified, depending on the inputs fields. For instance, any CC port $j$ with vacuum input leads to $\langle\hat{b}_{\rm in}^{(j)}\rangle =0$ and similarly, any AC port $l$ with vacuum input leads to $\langle\hat{c}_{\rm in}^{(l)}\rangle=0$. 
 
We note that the Heisenberg-Langevin equation for $\hat{\sigma}_z$ can be obtained either via the same procedure than the one used in Refs.~\cite{Gardiner1985,Gardiner2004}, detailed in the Appendix~\ref{Subsec-appendix-derivation-Heisenberg-Langevin}, or directly from Eq.~\eqref{eq:sigma_dot} and the definition of $\hat{\sigma}_z$. These two derivations agree thanks to an important feature of the input-output formalism, namely the fact that the product derivation rule holds for any system operator, i.e: 
\begin{equation}
    \begin{split}
       \frac{d}{dt}\hat{O}_1\hat{O}_2 = \Dot{\hat{O}}_1\hat{O}_2+\hat{O}_1\Dot{\hat{O}}_2
        \label{calculus_rule}
    \end{split}
\end{equation}
holds for any pair of system operators $\hat{O}_1,\hat{O}_2$. In the present case, we show in the Appendix~\ref{Subsec-appendix-commutation-relation} that this is ensured by the commutation relations (for any system operator $\hat{O}$ and any AC or CC input operator),
\begin{equation}    
\begin{split}
    [\frac{\gamma_{l}}{2}\hat{\sigma}+ \sqrt{\gamma_{l}}\hat{c}_{\textrm{in}}^{(l)},\hat{O}] = 0 \hspace{1cm} [\frac{\kappa_{j}}{2} \hat{a} 
    + \sqrt{\kappa_{j}} \hat{b}_{\textrm{in}}^{(j)},\hat{O}] = 0
\end{split}
\end{equation}
This also allows the fundamental (anti-)commutation rules to be preserved:
\begin{equation}
    \frac{d}{dt}[\hat{a},\hat{a}^\dagger]=0~~\textrm{ and }~~\frac{d}{dt}\{\hat{\sigma},\hat{\sigma}^\dagger\}=0
    \label{eq:derive-identity}
\end{equation}
which ensures that $\hat{\sigma}_z=2\hat{\sigma}^\dagger\hat{\sigma}-\hat{\mathbb{I}}$ and thus $\Dot{\hat{\sigma}}_z=2(\Dot{\hat{\sigma}}^\dagger\hat{\sigma}+\hat{\sigma}^\dagger\Dot{\hat{\sigma}})$ at all times. 

Using the input-output formalism, we follow~\cite{Gardiner1985} and derive the input-output relations in appendix~\ref{Subsec-input-output-relations}, 
\begin{equation}
       \begin{split}
            \hat{b}_{\textrm{out}}^{(j)} &= \hat{b}_{\textrm{in}}^{(j)}+\sqrt{\kappa_{j}}\hat{a}\\
            \hat{c}_{\textrm{out}}^{(l)} &= \hat{c}_{\textrm{in}}^{(l)}+\sqrt{\gamma_{l}}\hat{\sigma}
       \end{split}
       \label{input-output-equation}
\end{equation}
These relations determine the output fields $\hat{b}_{\textrm{out}}^{(j)},\hat{c}_{\textrm{out}}^{(l)}$, based on the input fields $\hat{b}_{\textrm{in}}^{(j)},\hat{c}_{\textrm{in}}^{(l)}$ and the system operators $\hat{a}$ and $\hat{\sigma}$. While the input fields are typically specified, determining the system operators at any time requires solving the Heisenberg-Langevin equations~\eqref{Heisenberg-Langevin system}. To this date, the dynamics of a driven cavity-QED system has not been solved exactly, though it can be efficiently simulated through a truncation of the cavity Hilbert space, truncation that needs to be adapted depending on the number of incoming photons. Another fundamental issue for the understanding of the CQED dynamics is that Eq.~\eqref{input-output-equation} provides expressions of the cavity-coupled output operators $\hat{b}_{\rm out}^{(j)}$ that depend only on $\hat{a}$. These expressions give no direct information about internal interactions between the cavity and the atom, thus limiting the interpretation of how the atom affects the output field.

In the specific case where at most one excitation (of atomic or photonic nature) is present in the system at any time, one can overcome these issues by considering the Hilbert space of a three-level system (no excitation $\ket{g,0}$, one photon and a ground-state atom $\ket{g,1}$, zero photon and an excited-state atom $\ket{e,0}$). Such a representation was introduced to exactly solve the CQED dynamics in the case of a device used as an emitter in Refs.~\cite{Lodahl2015,Carmichael1989,Andreani1999}. It was then generalized to devices with weak input fields~\cite{Auffeves2008}. 
Another approach to simplify the CQED dynamics is based on the adiabatic elimination of the cavity, which has been extensively studied~\cite{Carmichael2007,Tokieda2024,Azouit2017}. This approach relies on the assumption that the cavity and the atom dynamics have two different timescales. We present one form of this approach in the next subsection, adapting the treatment of Ref.~\cite{Carmichael2007} to our input-output model.

\subsection{Adiabatic cavity elimination}

We present an adiabatic elimination procedure, strongly inspired by~\cite{Carmichael2007}. Adiabatic cavity elimination aims to describe the atom dynamics entirely within its own Hilbert space, while still retaining the influence of the cavity. To do so, we are looking for an expression of $\hat{a}$ in the atom's Hilbert space. We start by integrating Eq.~\eqref{eq:a_dot}, which yields the exact relation:

\begin{equation}
    \begin{split}
        \hat{a} &=\hat{a}(0)e^{-\kappa' t}-g\int_0^t\hat{\sigma}(t-t')e^{-\kappa' t'}dt'\\
        &\quad-\sum_j\sqrt{\kappa_{j}}\int_0^t \hat{b}_{\textrm{in}}^{(j)}(t-t')e^{-\kappa' t'}dt'
    \end{split}
    \label{Integrated_langevin_eq}
\end{equation}
with contributions from the cavity’s initial state (first term), the atom–cavity interaction (second term), and the cavity-coupled input fields (last term). We note that the adiabatic elimination approach will only be valid once the cavity operator $\hat{a}$ has kept no memory of its initial value $\hat{a}(0)$ and is governed entirely by its interaction with the atom and the input fields. This regime is reached for times $t \gg 1/\kappa$, where $\frac{1}{\kappa}$ is the effective photon lifetime in the bare cavity. This approximation leads to:
 \begin{equation}
    \begin{split}
        \hat{a} &=-g\int_0^\infty\hat{\sigma}(t-t')e^{-\kappa' t'}dt'\\
        &-\sum_j\sqrt{\kappa_{j}}\int_0^\infty \hat{b}_{\textrm{in}}^{(j)}(t-t')e^{-\kappa' t'}dt'
    \end{split}
    \label{eq-a-with-time-approx}
\end{equation}
The input-field-related term is a convolution, describing the well-known empty-cavity filtering~\cite{Gardiner2004,Raymer2013}. Its exact expression depends on the input form: in order to stay general, we define the filtered input field for each CC port, $\hat{b}_{\textrm{in}}^{(j\textrm{,filt})}$, as:
\begin{equation}
    \begin{split}
        \sqrt{\kappa_{j}}\int_0^\infty \hat{b}_{\textrm{in}}^{(j)}(t-t')e^{-\kappa' t'}dt' = \frac{\sqrt{\kappa_{j}}}{\kappa'}\hat{b}_{\textrm{in}}^{(j\textrm{,filt})}
    \end{split}
    \label{def-b-in-filt}
\end{equation}
Using these definitions we express the cavity operator in the absence of atom-cavity coupling, i.e $g=0$, leading to $\hat{a}=\hat{a}_{\rm c}$ with, by definition: 
\begin{equation}
     \hat{a}_{\rm c} =  -\sum_{j}\frac{\sqrt{\kappa_{j}}}{\kappa'}\hat{b}_{\textrm{in}}^{(j\textrm{,filt})}
     \label{eq:a-g=0}
\end{equation}
As such, $\hat{a}_{\rm c}$ corresponds to the cavity annihilation operator that would be obtained in the absence of the atom: it takes into account all the CC inputs, after cavity filtering.

Reciprocally, in the specific case where $g=0$, the atom evolves independently from the cavity through:
\begin{equation}
    \begin{split}
        \Dot{\hat{\sigma}} &= -\gamma'\hat{\sigma}+\hat{\sigma}_z\frac{\hat{\Omega}_{\rm a}}{2}\\
        \Dot{\hat{\sigma}}_z &= -\gamma_{\rm a}(\hat{\sigma}_z+\hat{\mathbb{I}})-\hat{\sigma}^\dagger \hat{\Omega}_{\rm a}-\hat{\Omega}_{\rm a}^\dagger\hat{\sigma}
    \end{split}
\end{equation}
where we define: $\hat{\Omega}_{\textrm{a}} = 2\sum_l\sqrt{\gamma_{l}}\hat{c}_{\textrm{in}}^{(l)}$, an operator taking into account all the atom-coupled inputs. Such simplified Heisenberg-Langevin equations include both coherent driving through the operator $\hat{\Omega}_{\rm a}$, and damping at the rate $\gamma_{\rm a}$. Taking the expectation values yields the effective system of Bloch equations:
\begin{equation}
    \begin{cases}
    \langle \Dot{\hat{\sigma}}\rangle  &= -\gamma'\langle \hat{\sigma}\rangle +\frac{1}{2}\langle \hat{\sigma}_z\hat{\Omega}_{\rm a}\rangle \\
         \langle \Dot{\hat{\sigma}}_z\rangle  &= -\gamma_{\rm a}(\langle \hat{\sigma}_z\rangle +1)-\langle \hat{\sigma}^\dagger \hat{\Omega}_{\rm a}\rangle -\langle \hat{\Omega}_{\rm a}^\dagger\hat{\sigma}\rangle 
    \end{cases}
\end{equation}
Furthermore, if all the input fields $\hat{c}_{\textrm{in}}^{(l)}$ are driven by coherent light or vacuum, $\langle \hat{\sigma}_z\hat{\Omega}_{\rm a}\rangle =\langle \hat{\sigma}_z\rangle \langle \hat{\Omega}_{\rm a}\rangle =\langle \hat{\sigma}_z\rangle {\Omega}_{\rm a}$ and $\langle\hat{\sigma}\hat{\Omega}_{\rm a}\rangle = \langle\hat{\sigma}\rangle\Omega_{\rm a}$ with $\Omega_{\rm a} = \langle \hat{\Omega}_{\rm a}\rangle$, thus:
\begin{equation}
    \begin{cases}
    \langle \Dot{\hat{\sigma}}\rangle  &= -\gamma'\langle \hat{\sigma}\rangle +\frac{1}{2}\langle \hat{\sigma}_z\rangle {\Omega}_{\rm a}\\
         \langle \Dot{\hat{\sigma}}_z\rangle  &= -\gamma_{\rm a}(\langle \hat{\sigma}_z\rangle +1)-\langle \hat{\sigma}^\dagger\rangle  {\Omega}_{\rm a}-{\Omega}_{\rm a}^*\langle \hat{\sigma}\rangle 
    \end{cases}
    \label{System-adiabtic-elim}
\end{equation}
In the case where $g=0$ we thus retrieve the standard Bloch equations for a TLS~\cite{Carmichael2007}, where $\Omega_{\rm a}$ denotes the Rabi frequency induced by direct excitation of the atom through the AC ports.

In the rest of this subsection, we follow Ref.~\cite{Carmichael2007} to study the situation where $g\ne0$, restraining ourselves to the cases where the cavity can be adiabatically eliminated, that is where the atom dynamics is far slower than the cavity memory loss. This situation corresponds to making the approximation $\hat{\sigma}(t-t')\approx\hat{\sigma}(t)$ on the integral $\int_{0}^{\infty}\hat{\sigma}(t-t')e^{-\kappa't'}dt'$, meaning that the atom does not evolve during one photon lifetime in the cavity. As can be seen from Eq.~\eqref{Heisenberg-Langevin system}, this approximation requires that $\gamma_{\rm a},g\ll\kappa$. We also follow~\cite{Carmichael2007} by considering the case $\omega_{\rm ref} = \omega_{\rm a}$, so that Im($\gamma'$)=0, to avoid taking into account fast oscillations in the phase of $\hat{\sigma}(t)$. This adiabatic approximation also requires that the Rabi oscillations experienced by the atom remain very slow compared to $\kappa^{-1}$. We then compute the integral as $\int_{0}^{\infty}\hat{\sigma}(t-t')e^{-\kappa't'}dt'=\hat{\sigma}(t)\int_{0}^{\infty}e^{-\kappa't'}dt'=\frac{\hat{\sigma}}{\kappa'}$. In such a case, Eq.~\eqref{eq-a-with-time-approx} takes a very simple form allowing to express the annihilation operator $\hat{a}$, at all times, as a function of the bare cavity annihilation operator $\hat{a}_{\rm c}$ and the atomic decay operator $\hat{\sigma}$, through:
\begin{equation}
         \hat{a} = \hat{a}_{\rm c}-\frac{g}{\kappa'}\hat{\sigma}
    \label{adiabatic-elim-a-expr}
\end{equation}
showing that the system in this configuration converges to a state where the cavity and the atom evolve together. In addition, substituting Eq.~\eqref{adiabatic-elim-a-expr} into Eq.~\eqref{eq:sigma_dot} gives:
\begin{equation}
    \begin{split}
        \Dot{\hat{\sigma}} = -\Gamma^{'\rm (2)}\hat{\sigma}+\hat{\sigma}_z\frac{\hat{\Omega}_{\rm a}-2g\hat{a}_{\rm c}}{2}
    \end{split}
    \label{Adibatic-elim-langevin-eq}
\end{equation}
with
\begin{equation}
    \Gamma^{'\rm (2)} = \frac{\gamma_{\rm a}}{2}+\frac{2g^2}{\kappa+2i(\omega_{\rm c}-\omega_{\rm a})} = \frac{\Gamma^{\rm (2)}}{2}+i(\omega_{\rm a'}^{\rm (2)}-\omega_{\rm a})
    \label{Gamma-prime-order-2}
\end{equation}
a complex constant characterizing the effective behavior of the atom in its Hilbert space only. In Eq.~\eqref{Gamma-prime-order-2}, the real part of $\Gamma'^{\rm (2)}$ is defined as $\frac{\Gamma^{\rm (2)}}{2}$, with $\Gamma^{\rm (2)}$ a modified damping rate from which we identify the two possible emission channels for the atom: 
\begin{equation}
    \Gamma^{\rm (2)} = \gamma_{\rm a}+\frac{4g^2}{\kappa^2+4(\omega_{\rm c}-\omega_{\rm a})^2}\kappa
\end{equation}
In this expression, the first term corresponds to the emission outside the cavity mode, with rate $\gamma_{\rm a}$, and the second one to emission into the cavity mode, enhanced by the Purcell effect~\cite{Purcell1946,Haroche1989}, with rate $\frac{4g^2\kappa}{\kappa^2+4(\omega_{\rm c}-\omega_{\rm a})^2}$. This result is the same as in Ref.~\cite{Carmichael2007} when taking $\gamma_{\rm a}=0$.
The imaginary part of $\Gamma^{'\rm (2)}$ describes coherent oscillations at a rate that takes into account a cavity-induced frequency shift:
\begin{equation}
    \omega_{\rm a'}^{\rm (2)}-\omega_{\rm a}=-\frac{4g^2}{\kappa^2+4(\omega_{\rm c}-\omega_{\rm a})^2}\left(\omega_{\rm c}-\omega_{\rm a}\right)
\end{equation} 
We note that in the resonant case $\omega_{\rm a}=\omega_{\rm c}$, we retrieve the well-known formula for the Purcell-enhanced emission rate $\Gamma^{\rm (2)} = \gamma_{\rm a}+ \frac{4g^2}{\kappa}$ and no frequency shift $\omega_{\rm a'}=\omega_{\rm a}$~\cite{Carmichael2007}. We note that, because of the adiabatic approximation, these constants are not exact and are said to be at second order in $\frac{g}{\kappa}$, hence the notation $\Gamma^{'\rm (2)},\Gamma^{\rm (2)},\omega_{\rm a'}^{\rm (2)}$. To our knowledge, the highest order achieved in the adiabatic elimination of such CQED system is the fourth order in $\frac{g}{\kappa}$~\cite{Tokieda2024}. 

Using these results, we can derive the equation of evolution of $\Dot{\hat{\sigma}}_z$. As before, the two methods -- (i) inserting Eq.~\eqref{adiabatic-elim-a-expr} in Eq.~\eqref{eq:sigma_z_dot} or (ii) using Eq.~\eqref{Adibatic-elim-langevin-eq} and the definition $\hat{\sigma}_z=\hat{\sigma}^\dagger\hat{\sigma}-\hat{\sigma}\hat{\sigma}^\dagger$ -- give the same results, ensuring the consistency of the theory. Taking the expectation values, one obtains the effective system of Bloch equations:
\begin{equation}
    \begin{cases}
        \langle \Dot{\hat{\sigma}}\rangle  &= -\Gamma^{'\rm (2)}\langle \hat{\sigma}\rangle +\frac{1}{2}\langle \hat{\sigma}_z( \hat{\Omega}_{\rm a}-2g\hat{a}_{\rm c})\rangle \\
         \langle \Dot{\hat{\sigma}}_z\rangle  &=-2\Gamma^{\rm (2)}(\langle \hat{\sigma}_z\rangle +1)-\langle ( \hat{\Omega}_{\rm a}-2g\hat{a}_{\rm c})^\dagger\hat{\sigma}\rangle \\
         &-\langle \hat{\sigma}^\dagger( \hat{\Omega}_{\rm a}-2g\hat{a}_{\rm c})\rangle 
    \end{cases}
    \label{Effective-system-Bloch-eq}
\end{equation}
Through the operator $\hat{\Omega}_{\rm a}-2g\hat{a}_{\rm c}$, this effective system of Bloch equations describes driving from both the AC and CC ports, and gives an effective evolution of the atom in a reduced Hilbert space. 

Using Eq.~\eqref{adiabatic-elim-a-expr} we also obtain a new version of the input-output equations~\eqref{input-output-equation},
\begin{equation}
    \hat{b}_{\textrm{out}}^{(j)} =\hat{b}_{\textrm{out,c}}^{(j)} -\frac{2g\sqrt{\kappa_{j}}}{\kappa+2i(\omega_{\rm c}-\omega_{\rm a})}\hat{\sigma}
    \label{eq-inout-output-adiabatic}
\end{equation}
where, by definition, $\hat{b}_{\textrm{out,c}}^{(j)}$ represents the output field obtained for the bare cavity (i.e when $g=0$):
\begin{equation}
  \hat{b}_{\textrm{out,c}}^{(j)} = \hat{b}_{\textrm{in}}^{(j)}+\sqrt{\kappa_{j}}\hat{a}_{\rm c}
  \label{def-bout-c}
\end{equation}
Eq.~\eqref{eq-inout-output-adiabatic} provides an effective input-output relation, that is similar to the AC ports input-output relation of Eq.~\eqref{input-output-equation} and to, more generally, waveguide-QED-like input-output relations~\cite{Lalumiere2013,Fan2010}. Together with Eq.~\eqref{Effective-system-Bloch-eq}, this allows deriving a complete formalism, in the general case of an arbitrary number of AC and CC ports, that gives direct access to useful quantities such as output fluxes and correlations. Computing such quantities can be done in two ways: (i) solving analytically the dynamics of Eq.~\eqref{Effective-system-Bloch-eq}, if the input field expressions allow it, or (ii) solving it numerically within a Hilbert space of dimension 2, drastically reduced compared to the full CQED Hilbert space.

While this technique provides satisfying results, it is limited to the bad cavity limit with a weak drive. In particular, Eq.~\eqref{Effective-system-Bloch-eq} together with the adiabatic elimination condition impose that the atom should have a small Purcell enhancement factor $|\Gamma^{'\rm (2)}|\ll \kappa$, which is generally not desired in cavity-QED experiments~\cite{Somaschi2016,Lodahl2015,Liu2018}. While higher-order, perturbative adiabatic elimination techniques~\cite{Tokieda2024,Azouit2017} allow describing regimes closer to the strong atom-cavity coupling, such approach generally only remains valid for limited values of the parameter $\frac{g}{\kappa}$. To overcome this, we present in the next sections a non-perturbative self-consistent technique that allows computing quantities also in the weak-coupling, strong-driving regime, and in the strong-coupling, weak-driving regime.

\section{Self consistent cavity elimination}\label{Chapter_3}
In this section, we introduce a self-consistent cavity elimination approach. We begin by introducing the ansatz made on the evolution of $\hat{\sigma}$, including effective driving and damping terms, enabling us to derive a self-consistency equation. We retrieve an analytical expression for the effective Purcell factor and cavity-induced frequency shift. Furthermore, under the approximation that the effective driving evolves slowly compared to $\kappa$, we compute effective input-output relations and effective Heisenberg–Langevin equations, describing the atom dynamics. Finally, we test our model in the semi-classical, low-power limit, where our model is shown to exactly retrieve known results~\cite{Waks2006,Majumdar2012}.
\subsection{Ansatz: effective damping and driving}

We introduce here the ansatz at the heart of the self-consistent cavity elimination approach. We aim to solve Eq.~\eqref{eq:a_dot}, which reduces in the long time limit ($t\gg\frac{1}{\kappa}$) to:
\begin{equation}
    \hat{a} = \hat{a}_{\rm c}-g\int_0^\infty\hat{\sigma}(t-t')e^{-\kappa' t'}dt'
    \label{eq-a-to-solve-final}
\end{equation}
Solving this integral requires an explicit form for $\hat{\sigma}(t-t')$. In order to extend the analysis, we use the following ansatz to describe an effective atomic evolution:
\begin{equation}\label{Ansatz}
        \Dot{\hat{\sigma}}= -\Gamma'\hat{\sigma}+\hat{F}    
\end{equation}
In this ansatz, $\Gamma'$ is an effective, complex damping coefficient whose real part represents the effective damping rate and whose imaginary part governs the effective frequency of the reduced system. Of course, it is expected that $\Gamma'\rightarrow\Gamma'^{\rm (2)}$ (Eq.~\eqref{Gamma-prime-order-2}), at the second order in $\frac{g}{\kappa}$, in the adiabatic limit where $g,\gamma_{\rm a} \ll \kappa$. In any case, we will limit ourselves to a regime where $\gamma_{\rm a}<\kappa$, i.e the good-emitter regime where the cavity is the fast subsystem to be eliminated.

In addition, in the ansatz of Eq.~\eqref{Ansatz}, the operator $\hat{F}$ captures all the contributions of the driving, required for the effective atom dynamics to match the one of the full atom–cavity system. It should include, for instance, terms inducing Rabi oscillations. It is to be noted that such a separation between driving and damping is natural for a cavity-QED device in the weak coupling regime, where monoexponential damping is expected~\cite{Carmichael1989}. This is not true in the strong coupling regime, where, for instance, vacuum Rabi oscillations are obtained if the atom is initially put in the excited state $\ket{e}$ with no input field and the cavity initially in the vacuum state~\cite{Thompson1992,Sanchez-Mondragon1983,Brune1996b,Wallraff2004}. In such a case, one can still write the ansatz of Eq.~\eqref{Ansatz}, yet it is expected that the operator $\hat{F}$ will take a more complex form describing the non-mono-exponential damping that cannot be covered by the term $-\Gamma'\hat{\sigma}$. 

In the following, we apply the ansatz of Eq.~\eqref{Ansatz}, which can also be considered as a definition of the operator $\hat{F}$, whether or not the device is considered in the weak- or strong-coupling regime. The analytical expressions of $\Gamma'$ and $\hat{F}$ will follow from a self-consistency condition described in the next subsection.

\subsection{Self-consistency equation and approximate solution}\label{Sec:approximation}

We now present the analytical results of cavity elimination following the self-consistency approach. Starting from the ansatz~\eqref{Ansatz}, we evaluate in the Appendix~\ref{Subsec-appendix-integral-calc} the integral:
\begin{equation}\label{int-result}
     \begin{split}
          \int_0^\infty\hat{\sigma}(t-t')e^{-\kappa' t'}dt' &= \frac{\hat{\sigma} }{\kappa'-\Gamma'}-\frac{1}{\kappa'(\kappa'-\Gamma')}\Big(\hat{F}\\ 
          &\quad-\frac{\dot{\hat{F}}}{\kappa'}+\frac{\ddot{\hat{F}}}{\kappa'^2}+..÷\Big)
     \end{split}
 \end{equation}
This result is fundamental since it allows solving Eq.~\eqref{eq:a_dot}, by inserting it in Eq.~\eqref{eq-a-to-solve-final}, leading to a new expression of $\hat{a}$ that depends only on the TLS operators and the ansatz parameters: 
\begin{equation}\label{express-a-sum-F}
            \hat{a} = \hat{a}_\textrm{c}-g\frac{\hat{\sigma} }{\kappa'-\Gamma'}+\frac{g}{\kappa'(\kappa'-\Gamma')}(\hat{F}-\frac{\Dot{\hat{F}}}{\kappa'}+\frac{\Ddot{\hat{F}}}{\kappa'^2}+...)
\end{equation}
We have checked that, by deriving this expression and using the identity $\frac{d}{dt}\hat{b}_{\textrm{in}}^{(j\textrm{,filt})} = \kappa'[\hat{b}_{\textrm{in}}^{(j)} - \hat{b}_{\textrm{in}}^{(j\textrm{,filt})}]$, Eq.~\eqref{express-a-sum-F} allows recovering the original Eq.~\eqref{eq:a_dot}. This confirms that the solution in Eq.~\eqref{express-a-sum-F} can be considered exact, provided that the infinite sum $\hat{F}-\frac{\Dot{\hat{F}}}{\kappa'}+...$ converges, in the limit $t\gg\kappa^{-1}$ where the memory of the initial operator $\hat{a}(0)$ has been lost.
We now insert this expression of $\hat{a}$ in Eq.~\eqref{eq:sigma_dot} and identify the result with the ansatz \eqref{Ansatz}, obtaining the self-consistency equation:
\begin{equation}\label{Self-consistency-eq}
    \begin{split}
    \Dot{\hat{\sigma}}& = -\Gamma'\hat{\sigma}+\hat{F} \\
    &=-(\gamma'+\frac{g^2}{\kappa'-\Gamma'})\hat{\sigma}-x\hat{\sigma}_z(\hat{F}-\frac{\Dot{\hat{F}}}{\kappa'}+\frac{\Ddot{\hat{F}}}{\kappa'^2}+...)\\
    &\quad+\hat{\sigma}_z\left(\frac{ \hat{\Omega}_\textrm{a}}{2}-g\hat{a}_\textrm{c} \right)
    \end{split}
\end{equation}
In this equation, we introduced the dimensionless parameter:
\begin{equation}\label{Def:x}
    x = \frac{g^2}{\kappa'(\kappa'-\Gamma')}
\end{equation}
which serves as a measure of the light–matter coupling strength. Indeed, in the limit $\left|x\right|\ll 1$, our results will usually reduce to the familiar two-level atom theory. We note that this limit will be reached if $g\ll\kappa$ but also for a strongly-coupled system whose atom-cavity detuning is very large compared to $\kappa$.

Provided that the infinite sum converges, Eq.~\eqref{Self-consistency-eq} ensures that both Eq.~\eqref{eq:a_dot} and Eq.~\eqref{eq:sigma_dot} are satisfied. This technique shifts the difficulty from solving the system of equations~\eqref{Heisenberg-Langevin system} to identifying $\Gamma'$ and $\hat{F}$ from the self-consistency equation. It will then allow calculating $\hat{a}$ using Eq.~\eqref{express-a-sum-F}, from which all the cavity coupled output operators $\hat{b}_{\rm out}^{(j)}$ can be deduced.

We now introduce an approximation, which is central to the present work, as it simplifies drastically the problem: $\hat{F}-\frac{\Dot{\hat{F}}}{\kappa'}+\frac{\ddot{\hat{F}}}{\kappa'^2}-...\approx\hat{F}$. This approximation considers that the effective drive operator $\hat{F}$ slowly evolves at the timescale of the bare cavity photon lifetime $\kappa^{-1}$. The validity range of such an approximation will be discussed later on, based on analytical calculations in the semi-classical approximation (Sec.~\ref{Subsec:Semiclassical}) as well as on numerical simulations in a few practical cases (Secs.~\ref{Sec:practical-case-intensity},\ref{Sec:practical-case-SDF},\ref{Sec:practical-case-g2},\ref{Sec:plot-strong})

Under this approximation, the self-consistency equation Eq.~\eqref{Self-consistency-eq} reduces to:
\begin{equation}\label{identification-eq}
         \begin{split}
           -\Gamma'\hat{\sigma}+(\hat{\mathbb{I}}+x\hat{\sigma}_z)\hat{F}&=-(\gamma'+\frac{g^2}{\kappa'-\Gamma'})\hat{\sigma}\\
             &\quad+\hat{\sigma}_z\left(\frac{ \hat{\Omega}_\textrm{a}}{2}-g\hat{a}_\textrm{c} \right)
    \end{split}
\end{equation}

\subsection{Effective damping rate and cavity-induced frequency shift}\label{Subsec:Effective-damping}

We here discuss the effective damping rate expression found by solving Eq.~\eqref{identification-eq}, identifying the proportionality coefficients in front of the operator $\hat{\sigma}$. This leads to the following equation:
\begin{equation}\label{Gamma-prime-identification-eq}
    \Gamma' = \gamma'+\frac{g^2}{\kappa'-\Gamma'}
\end{equation}
This equation reduces to a second-order polynomial equation for $\Gamma'$ which has two solutions denoted $\Gamma'_\pm$, with:
\begin{equation}\label{Gamma_prime_expressions}
        \Gamma'_\pm =\frac{\kappa'+\gamma'\pm\sqrt{(\kappa'-\gamma')^2-4g^2}}{2}
\end{equation}
Interestingly, these two values are exactly the two eigenvalues found e.g in Refs.~\cite{Auffeves2008,Lien2016}. In these works, the dynamics of a cavity-QED system is solved in a truncated Hilbert space with only three occupied states $\ket{g,0}$ (absence of excitation), $\ket{e,0}$ (atomic excitation) and $\ket{g,1}$ (photonic excitation). Such an approximation becomes exact in the absence of driving field and if the system contains at most one initial excitation. In such a case, as shown in Ref.~\cite{Auffeves2008}, the dynamics can be solved through an analogy with the dynamics of two coupled classical cavities. In particular, a set of linear coupled equations for $\langle\Dot{\hat{\sigma}}\rangle$ and $\langle\Dot{\hat{a}}\rangle$ is obtained whose dynamics is governed by the two eigenvalues $\Gamma'_\pm$ (here presented in a more compact form than in Ref.~\cite{Auffeves2008}). 
In the simple case of no atom-cavity detuning ($\omega_{\rm a}=\omega_{\rm c}$), these two eigenvalues allow defining rigorously the weak and strong coupling regimes in the resonant atom-cavity case. Indeed, the case where $|\kappa-\gamma|>4g$ characterizes the weak coupling regime since it leads to two different real values $\Gamma'_+$ and $\Gamma'_-$, leading only to damping of the system. Correspondingly, the case where $|\kappa-\gamma|<4g$ characterizes the strong coupling regime since $\Gamma'_-$ and $\Gamma'_+$ are complex conjugates, leading not only to damping but also to vacuum Rabi oscillations~\cite{Lodahl2015,Eberly1980,Brune1996b}. In Sec.~\ref{Sec:plot-strong} we will show simulations illustrating two dips in the reflectivity response of a strongly-coupled system with $\omega_{\rm a}\ne\omega_{\rm c}$ and $\kappa\gg\gamma_{\rm a}$. In such a case, one can directly visualize that $\Gamma'_+$ describes the "cavity-like" reflectivity dip centered on the frequency $\omega_{\rm c'}$ pushed away from $\omega_{\rm c}$ due to the presence of a vacuum Rabi splitting, with $\omega_{\rm c'}$ given by $\omega_{\rm c'}-\omega_{\rm ref} = \rm Im(\Gamma'_+)$. On the contrary $\Gamma'_-$ describes the "atom-like" reflectivity dip centered on the frequency $\omega_{\rm a'}$ pushed away from $\omega_{\rm a}$, with $\omega_{\rm a'}$ given by $\omega_{\rm a'}-\omega_{\rm ref}=\rm Im(\Gamma'_-)$. The "cavity-like" and "atom-like" behaviors are also materialized by the widths of the respective reflectivity dips, determined by $\rm Re(\Gamma'_+)>\rm Re(\Gamma'_-)$. As such, the "cavity-like" dip has a larger width, closer to $\kappa$, than the "atom-like" one. The same classification also applies in the weak-coupling, detuned case where the frequency-shifts induced by the atom-cavity coupling are smaller, leading to $\omega_{\rm a'}\approx\omega_{\rm a}$ and $\omega_{\rm c'}\approx\omega_{\rm c}$. Finally, when $g=0$ we retrieve that $\Gamma'_+=\kappa'$ and $\Gamma'_-=\gamma'$, here again confirming the interpretation of these effective rates as associated to "cavity-like" and "atom-like" behaviors.

Generally speaking, the self-consistency equation~\eqref{Gamma-prime-identification-eq} can be verified using both choices, $\Gamma'=\Gamma'_-$ or $\Gamma'=\Gamma'_+$, though each choice implies a different equation for the dimensionless parameter $x=\frac{g^2}{\kappa'(\kappa'-\Gamma')}$, and for the self-consistency equation to be fullfilled by the effective driving operator $\hat{F}$. It is only natural to choose, however, the atom-like solution $\Gamma'=\Gamma'_-$ which corresponds to the slowest damping (as shown in the Appendix~\ref{Subsec-Study-Gamma}, $\rm Re(\Gamma'_-)\leq \rm Re(\Gamma'_+)$, the equality being obtained only in the resonant strong coupling case where $\rm Re(\Gamma'_-) = \rm Re(\Gamma'_+) = \frac{\kappa+\gamma_{\rm a}}{4}$). Choosing $\Gamma'=\Gamma'_-$ is particularly relevant in the weak-coupling regime or with strong detunings $|\omega_{\rm c}-\omega_{\rm a}|\gg\kappa$, where $\rm Re(\Gamma'_+)$ is of the order of $\frac{\kappa}{2}$, leading to a dynamics that is entirely damped at the relevant timescales considered in this work, i.e $t\gg\kappa^{-1}$. 

In the following, we will choose $\Gamma'=\Gamma'_-$ and define the effective atom damping $\Gamma$ and the effective atom frequency $\omega_{\rm a'}$ through:

\begin{equation}
    \Gamma' = \frac{\Gamma}{2}+i(\omega_{\rm a'}-\omega_{\rm ref})
\end{equation}

We note that in the limit where $(g,\gamma_{\rm a})\ll\kappa$ this solution reduces, at second order, to $\Gamma'=\Gamma^{\rm '(2)}$, defined in Eq.~\eqref{Gamma-prime-order-2}, as expected in such conditions where the adiabatic approximation can be made. We also note that expanding $\Gamma'=\Gamma'_-$ to the fourth order in $\frac{g}{\kappa}$ allows reproducing recent results based on high-order adiabatic elimination techniques~\cite{Tokieda2024}. This suggests the existence of a link between self-consistent cavity elimination and high-order adiabatic elimination approaches.

\subsection{Effective input-output relation}\label{Subsec:effective-input-output}

Once the value of $\Gamma'$ has been fixed, through the choice of the solution $\Gamma'=\Gamma'_-$, the dimensionless parameter $x=\frac{g^2}{\kappa'(\kappa'-\Gamma')}$ is also fixed. We can then identify the effective driving term $\hat{F}$ from Eq.~\eqref{identification-eq} by solving the self-consistency equation. Under the approximation of a slowly-evolving effective driving, $\hat{F}-\frac{\Dot{\hat{F}}}{\kappa'}+\frac{\Ddot{\hat{F}}}{\kappa^{'2}}-...\approx\hat{F}$, this equation simplifies to:
\begin{equation}
    (\hat{\mathbb{I}}+x\hat{\sigma}_z)\hat{F}= \hat{\sigma}_z\left(\frac{ \hat{\Omega}_\textrm{a}}{2}-g\hat{a}_\textrm{c} \right)\\
\end{equation}
Then, inverting $(\hat{\mathbb{I}}+x\hat{\sigma}_z)$ yields:
\begin{equation}\label{F-expression}
       \begin{split}
            \hat{F}&= \frac{\hat{\sigma}_z-x\hat{\mathbb{I}}}{1-x^2}\left(\frac{ \hat{\Omega}_\textrm{a}}{2}-g\hat{a}_\textrm{c} \right)\\
            \iff \hat{F}&= (\hat{\sigma}_z-x\hat{\mathbb{I}})\frac{\hat{\Omega}}{2}
       \end{split}
\end{equation}
with
\begin{equation}\label{Def:Omega}
    \begin{split}
        \hat{\Omega} & = 2\frac{g\sum_{j}\frac{\sqrt{\kappa_{\textrm{j}}}}{\kappa'}\hat{b}_{\textrm{in}}^{\textrm{(j,filt)}}+\sum_l\sqrt{\gamma_{\textrm{l}}}\hat{c}_{\textrm{in}}^{\textrm{(l)}}}{1-x^2}\\
        &= \frac{\hat{\Omega}_\textrm{a}-2g\hat{a}_\textrm{c}}{1-x^2}
    \end{split}
\end{equation}
The operator $\hat{\Omega}$ defined here collects all possible input contributions: the filtered CC inputs $\hat{b}_{\rm in}^{(j,{\rm filt})}$ and the AC inputs $\hat{c}_{\rm in}^{(l)}$. It is, as will be shown later, responsible for Rabi oscillations, whose effective Rabi frequency is governed by $\Omega=\langle\hat{\Omega}\rangle$. Substituting Eq.~\eqref{F-expression} into the cavity operator expression Eq.~\eqref{express-a-sum-F} (within the approximation $\hat{F}-\Dot{\hat{F}}/\kappa'+...\approx \hat{F}$) yields:
\begin{equation}\label{a-expression-elim}
       \hat{a}  
     =\hat{a}_\textrm{c} -\frac{g}{\kappa'-\Gamma'}\left[\hat{\sigma}-
     (\hat{\sigma}_z-x\hat{\mathbb{I}})\frac{\hat{\Omega}}{2\kappa'}\right]
\end{equation}
which, when inserted in Eq.~\eqref{input-output-equation}, gives an effective input–output relation for the CC ports:
\begin{equation}\label{effective-input-output}
     \begin{split}
         \hat{b}_{\textrm{out}}^{(j)}& =  \hat{b}_{\textrm{out,c}}^{(j)}-\sqrt{\Gamma_j}\left[\hat{\sigma}-(\hat{\sigma}_z-x\hat{\mathbb{I}})\frac{\hat{\Omega}}{2\kappa'}\right]\\
   \iff \hat{b}_{\textrm{out}}^{(j)} & =  \hat{b}_{\textrm{out,c}}^{(j)}-\sqrt{\Gamma_j} \hat{\sigma}'
     \end{split}
\end{equation}
with
\begin{equation}\label{sigma-prime-Gamma-j-def}
    \begin{split}
        &\hat{\sigma}' =\hat{\sigma}-(\hat{\sigma}_z-x\hat{\mathbb{I}})\frac{\hat{\Omega}}{2\kappa'}\\
        &\Gamma_j = \frac{g^2}{(\kappa'-\Gamma')^2}\kappa_j=\frac{\kappa_j}{\kappa'-\Gamma'}(\Gamma'-\gamma')
    \end{split}
\end{equation}
Here, $\hat{b}_{\textrm{out,c}}^{(j)}$ describes the output field that would be obtained in the absence of atom-cavity coupling, as already introduced in Eq.~\eqref{def-bout-c}. Interestingly, Eq.~\eqref{effective-input-output} qualitatively differs from the input-output relations obtained through adiabatic elimination~\eqref{eq-inout-output-adiabatic}. Indeed, the operator $\hat{\sigma}'$ defined in Eq.~\eqref{sigma-prime-Gamma-j-def} differs from the lowering operator $\hat{\sigma}$ and takes into account contributions from all the inputs through the operator $\hat{\Omega}$. The prefactor $\sqrt{\Gamma_j}$, in front of $\hat{\sigma}'$, also differs from the prefactor $\frac{2g}{\kappa+2i(\omega_{\rm c}-\omega_{\rm a})}\sqrt{\kappa_j}$ of Eq.~\eqref{eq-inout-output-adiabatic}, unless $\gamma_{\rm a},g\ll\kappa$.
 Still, Eq.~\eqref{effective-input-output} retains an essential feature of the adiabatic, waveguide-QED-like expression of Eq.~\eqref{eq-inout-output-adiabatic}: the fact that it gives an analytical expression of the atom-induced contribution to the output field, $\sqrt{\Gamma_j}\hat{\sigma}'$, which interferes with the bare cavity output field $\hat{b}_{\rm out,c}^{(j)}$. Also, the advantage of cavity elimination remains preserved since $\hat{\sigma}'$ only depends on the global input operator $\hat{\Omega}$ and on the TLS ones ($\hat{\sigma},\hat{\sigma}_z,\hat{\mathbb{I}}$).

To compute steady-state fluxes and time-dependent correlations, we must now derive the corresponding effective Langevin equations governing the effective dynamics after cavity elimination.

\subsection{Effective two-level dynamics}\label{Subsec:Effective-two-level-dynamics}
We now derive the effective dynamics of the reduced system. In the complete CQED model we remind that the evolution of $\hat{\sigma}_z$ (Eq.~\eqref{eq:sigma_z_dot}) can be consistently obtained either from Eq.~\eqref{eq:sigma_dot}, using the fact that $\hat{\sigma}_z =2\hat{\sigma}^\dagger\hat{\sigma}-\hat{\mathbb{I}}$ together with Eqs.~\eqref{calculus_rule}, \eqref{eq:derive-identity}, or directly from the Hamiltonian of Eq.~\eqref{Full-Hamiltonian}. In our effective cavity-eliminated model, the evolution of $\hat{\sigma}_z$ can be obtained by either inserting Eq.~\eqref{a-expression-elim} into Eq.~\eqref{eq:sigma_z_dot} or by using $\hat{\sigma}_z =2\hat{\sigma}^\dagger\hat{\sigma}-\hat{\mathbb{I}}$ together with Eqs.~\eqref{Ansatz},\eqref{calculus_rule},\eqref{eq:derive-identity}, preserving the consistency. Both approaches yield the same results, leading to the following effective Heisenberg-Langevin equations:
\begin{equation}\label{Effective-Langevin-system}
    \begin{cases}
        \Dot{\hat{\sigma}} &= -\Gamma'\hat{\sigma}+ \frac{1}{2}(\hat{\sigma}_z-x\hat{\mathbb{I}})\hat{\Omega}\\
         \Dot{\hat{\sigma}}_z &=-\Gamma(\hat{\sigma}_z+\hat{\mathbb{I}})-\hat{\Omega}^\dagger\hat{\sigma}(1+x)^*
         -\hat{\sigma}^\dagger\hat{\Omega}(1+x)
    \end{cases}
\end{equation}
These equations are qualitatively different from the standard Heisenberg-Langevin equations describing a TLS in absence of a cavity (Eq.~\eqref{Adibatic-elim-langevin-eq}) due to the presence of the operator $\hat{\sigma}_z-x\hat{\mathbb{I}}$ instead of $\hat{\sigma}_z$ in the expression of $\dot{\hat{\sigma}}$. 
This is only in the limit $\left|x\right|\ll 1$ that the system reduces to the one obtained for a bare TLS.
The effective Bloch equations are obtained by taking the expectation values:

\begin{equation}\label{Effective-Bloch-system}
    \begin{cases}
        \langle \Dot{\hat{\sigma}}\rangle  &= -\Gamma'\langle \hat{\sigma}\rangle + \frac{1}{2}(\langle \hat{\sigma}_z\hat{\Omega}\rangle -x\langle \hat{\Omega}\rangle )\\
         \langle \Dot{\hat{\sigma}}_z\rangle  &=-\Gamma(\langle \hat{\sigma}_z\rangle +1)-\langle \hat{\Omega}^\dagger\hat{\sigma}\rangle (1+x)^*\\
        & \quad  -\langle \hat{\sigma}^\dagger\hat{\Omega}\rangle (1+x)
    \end{cases}
\end{equation}

In the case where all input ports are driven by either coherent light or vacuum, $\langle \hat{\sigma}_z \hat{\Omega}\rangle = \langle \hat{\sigma}_z\rangle \Omega$ and $\langle \hat{\sigma}^\dagger \hat{\Omega}\rangle = \langle \hat{\sigma}\rangle^* \Omega$, this reduces to two coupled equations that fully describes the dynamics in a reduced Hilbert space. This will be discussed in detail in section~\ref{Chapter_4}. 

Furthermore, Eqs.~\eqref{effective-input-output} and~\eqref{Effective-Bloch-system} hold without assumptions on the input field, making them useful for a broad range of scenarios. One could use them to theoretically describe cascaded systems, by taking the output of a cavity to be the input of an other one, as described in Refs.~\cite{Gardiner1993,Carmichael1993}. Cascaded systems have direct applications: they can be used to model any type of input to a cavity-QED device~\cite{Kiilerich2020}, and to describe the generation and evolution of entanglement in quantum network architectures~\cite{DiFidio2010,Brion2012,Gonta2013}. In general, the cascading of an arbitrary number of systems is efficiently described in the framework of the SLH theory~\cite{Combes2017}, with applications to distributed quantum networks~\cite{Reiserer2015,DiFidio2024}. Generalizing our input-output-based cavity-elimination approach to a full SLH description could potentially help the analysis of such networks with a drastic dimension reduction, thereby enabling analytical calculations and numerical simulations. This application, however, goes beyond the scope of the present paper.

\subsection{Semiclassical low-power limit}\label{Subsec:Semiclassical}

A regime that is particularly interesting to study, independently from the coherent or non-coherent nature of the driving fields, is the low-excitation limit where the excited state is negligibly populated. This typically arises from low-power CW excitation, or, as more relevant to practical applications in quantum information processing, when low-power or highly-detuned excitation pulses~\cite{Koch2007,Blais2004} are used to drive the system. This corresponds to a limit where $\langle \hat{\sigma}_z\rangle $ is considered to be -1 at all times, and where the semi-classical equation  $\langle \hat{\sigma}_z\hat{\Omega}\rangle =\langle \hat{\sigma}_z\rangle \langle \hat{\Omega}\rangle =-\langle \hat{\Omega}\rangle $ applies~\cite{Carmichael2007}, even if the excitation pulses are not coherent. The system of Bloch equations~\eqref{Effective-Bloch-system} then reduces to:
\begin{equation}
    \langle \Dot{\hat{\sigma}}\rangle  = -\Gamma'\langle \hat{\sigma}\rangle -\frac{\Omega}{2}(1+x)
    \label{Eq:sigma-dot-semi-classical}
\end{equation}
whose solution is:
\begin{equation}
    \langle \hat{\sigma} \rangle  = \langle \hat{\sigma}(0)\rangle e^{-\Gamma' t}-\frac{1+x}{2}\int_0^t\Omega(t')e^{-\Gamma'(t-t')}dt'
    \label{evolution-semi-classical}
\end{equation}
This solution describes the evolution of the atomic dipole, in response to an arbitrary excitation, in the low-excitation limit. The calculated average value $\langle\hat{\sigma}\rangle$ can then be used together with the input–output relation of Eq.~\eqref{effective-input-output} to compute the average output amplitude for each CC port:
\begin{equation}
    \begin{split}
     \langle \hat{b}_{\textrm{out}}^{(j)}\rangle  =  \langle \hat{b}_{\textrm{out,c}}^{(j)}\rangle -\sqrt{\Gamma_j}\left[\langle \hat{\sigma}\rangle +(1+x)\frac{\Omega}{2\kappa'}\right]
    \label{b-out-semi-classical}
    \end{split}
\end{equation}
and similarly for each AC port:
\begin{equation}
    \langle \hat{c}_{\rm out}^{(l)}\rangle = \langle\hat{c}_{\rm in}^{(l)}\rangle+\sqrt{\gamma_{l}}\langle\hat{\sigma}\rangle
\end{equation}
These equations imply that the device's optical response does not instantaneously follow the value of $\Omega$ at a given time. Indeed, the output field amplitudes $\langle \hat{b}_{\textrm{out}}^{(j)}\rangle$ and $\langle \hat{c}_{\textrm{out}}^{(l)}\rangle$ also depend on $\langle \hat{\sigma}\rangle $ which, as evident from Eq.~\eqref{evolution-semi-classical}, follows $\Omega$ with a delay governed by the effective damping rate $\Gamma$. This delay is directly related to the Wigner time delay~\cite{Strauss2019} and can be interpreted as the effective time required to build a dipole in response to an optical excitation, as in Eq.~\eqref{evolution-semi-classical}.\newline

We now turn to the case of a low-power input field slowly varying compared to $\Gamma^{-1}$. In such regime, $\langle \hat{b}_{\rm in}^{(j,{\rm filt})}\rangle = \langle \hat{b}_{\rm in}^{(j)}\rangle$ and the average value $\langle\hat{\sigma}\rangle$ can be considered to be always in the stationary regime of Eq.~\eqref{Eq:sigma-dot-semi-classical}, i.e. $\langle\hat{\sigma}\rangle=-\frac{\Omega(1+x)}{2\Gamma'}$ at all times. Together with Eq.~\eqref{Def:Omega} and with the identity $\kappa'\Gamma'(1-x)=\kappa'\gamma'+g^2$  (see Eqs.~\eqref{Def:x},\eqref{Gamma-prime-identification-eq}), this allows expressing the average value $\langle\hat{\sigma}\rangle$ in the following form:
\begin{equation}
    \langle \hat{\sigma}\rangle  = \frac{-g\sum_j\sqrt{\kappa_j}\langle \hat{b}^{(j)}_{\textrm{in}}\rangle +\kappa'\langle \sum_j \sqrt{\gamma_{l}}\hat{c}_{\textrm{in}}^{(l)}\rangle }{\kappa'\gamma'+g^2}
    \label{sigma-semi-classical}
\end{equation}
Inserting this result in Eq.~\eqref{a-expression-elim} gives the corresponding average value of $\langle \hat{a}\rangle $ for a low-power, slowly-varying input field: 
\begin{equation}
    \langle \hat{a}\rangle   = \frac{-\gamma'\sum_j\sqrt{\kappa_j}\langle \hat{b}^{(j)}_{\textrm{in}}\rangle +g\langle \sum_l \sqrt{\gamma_{l}}\hat{c}_{\textrm{in}}^{(l)}\rangle }{\kappa'\gamma'+g^2}
     \label{a-semi-classical}
\end{equation}
Importantly, these average values could also be derived from the exact Heisenberg-Langevin equations of Eq.~\eqref{eq:sigma_dot} and~\eqref{eq:a_dot}, by taking their expectations values considering the low-excitation limit, where the semi-classical approximation $\langle \hat{\sigma}_z\hat{a}\rangle =\langle \hat{\sigma}_z\rangle \langle \hat{a}\rangle =-\langle \hat{a}\rangle $ holds. In this case, the system can be used as a closed set of equations which, for a slowly-varying input field allowing to consider $\langle \Dot{\hat{\sigma}}\rangle=0$ and $\langle \Dot{\hat{a}}\rangle=0$, leads to the same solutions for $\langle \hat{\sigma}\rangle$ (Eq.~\eqref{sigma-semi-classical}) and $\langle \hat{a}\rangle$ (Eq.~\eqref{a-semi-classical})~\cite{Auffeves-Garnier2007}. It means that, when long excitation pulses are used in the low-excitation limit, the self-consistent cavity elimination approach allows retrieving exactly the predictions of the complete CQED model, in both the strong and weak coupling regimes. This result is important since this low excitation limit is used in many applications to derive reflection and/or transmission amplitudes~\cite{Duan2004,Auffeves-Garnier2007}. In our multi-port input-output approach, such amplitudes can equivalently be deduced by inserting~\eqref{a-semi-classical} in the input-output equation Eq.~\eqref{input-output-equation} or by inserting~\eqref{sigma-semi-classical} in the eliminated input-output equation~\eqref{effective-input-output}. For instance, assuming that only the cavity-coupled ports are driven yields:
\begin{equation}
            \langle \hat{b}_{\textrm{out}}^{(j)}\rangle = \langle \hat{b}_{\textrm{in}}^{(j)}\rangle -\frac{\gamma'}{\kappa'\gamma'+g^2}
           \sum_{i}\sqrt{\kappa_i\kappa_j}\langle \hat{b}_{\textrm{in}}^{(i)}\rangle 
\end{equation}
a very practical expression that strongly simplifies when only one CC input is driven. We also note that our framework not only recovers known results in this limit, but also generalizes them: Eqs.~\eqref{evolution-semi-classical}–\eqref{b-out-semi-classical} provide access to the full time-dependence of $\langle \hat{\sigma}\rangle$ and $\langle \hat{b}_{\rm out}^{(j)}\rangle$ for an arbitrary optical drive in the low-excitation limit.

While our theory is general, further progress requires specification of the input fields. In the remainder of this work we will mainly focus on coherent driving.

\section{General study under coherent excitation}\label{Chapter_4}
In this section we assume coherent or vacuum excitation drives from all the AC and CC ports. This will allow us to derive a simplified system of Bloch equations, which we will show to be non-Markovian. Finally, we will discuss the requirement to numerically simulate such a system.

\subsection{Effective Bloch equations}

Under coherent excitation, the atomic operators and input fields are decoupled:
\begin{equation}
    \begin{split}
        \langle \hat{\Omega}\hat{\sigma}^\dagger \rangle  = \Omega\langle \hat{\sigma}^\dagger \rangle ~~~&\textrm{ and }~~~\langle \hat{\Omega}\hat{\sigma} \rangle  = \Omega\langle \hat{\sigma}\rangle  \\
        &\textrm{ and }~~~\langle \hat{\Omega}\hat{\sigma}_z \rangle  = \Omega\langle \hat{\sigma}_z\rangle  \\
    \end{split}
\end{equation}
This leads to the following system of Bloch equations:
\begin{equation}
        \begin{cases}
        \langle \Dot{\hat{\sigma}}\rangle  &=-\Gamma'\langle \hat{\sigma}\rangle + \frac{\Omega}{2}(\langle \hat{\sigma}_z\rangle -x)\\
        \langle \Dot{\hat{\sigma}}_z\rangle  &= -\Gamma(\langle \hat{\sigma}_z\rangle +1) -(1+x)\Omega\langle \hat{\sigma}^\dagger\rangle 
        -\langle \hat{\sigma}\rangle (1+x)^*\Omega^*\\    
        \end{cases}
        \label{Bloch-equation-final-system}
\end{equation}
These equations describe the evolution of the effective atom alone, interacting with its environment including the effect of the cavity, now considered part of that environment.\newline

A crucial point is that this effective atom dynamics is no longer Markovian. The Markov approximation assumes that the environment has no memory~\cite{Gardiner2004,Carmichael1999}, yet this assumption fails once the cavity is included in the environment and when the cavity lifetime $\kappa^{-1}$ is no longer negligible compared to the effective atom dynamics timescales $\Gamma^{-1}$ and $\Omega^{-1}$. We discuss the system's non-Markovianity in the next subsection.

\subsection{Non-Markovian Master equation}\label{sub-sec-non-markov}

It has been shown, e.g in Refs.~\cite{Hall2014,Rivas2014}, that, in the non-Markovian case, the time-local master equation can be put into a canonical Lindblad-like form:
\begin{equation}\label{Lindblad-def}
    \Dot{\rho} = -i[\mathcal{H},\rho]+\sum_{k=1}^{d^2-1}\gamma_k (t) \left(\hat{L}_k(t) \rho\hat{L}_k(t)^\dagger-\frac{1}{2}\{\hat{L}_k(t)^\dagger \hat{L}_k(t),\rho\}\right)
\end{equation}
where both the decoherence operators $\hat{L}_k(t)$ and the rates $\gamma_k(t)$ may depend on time, and the decoherence operators are defined such that they form an orthonormal basis of traceless operators, 
\begin{eqnarray}
    \textrm{Tr}(\hat{L}_k(t)) = 0~~\textrm{ and }~~\textrm{Tr}(\hat{L}_i^\dagger(t)\hat{L}_j(t)) = \delta_{ij}
\end{eqnarray}
Importantly, the rates are not constrained to be positive: the negativity of at least one $\gamma_k(t)$ is a hallmark of non-Markovianity~\cite{Hall2014,Rivas2014}, and is commonly interpreted as a signature of re-coherence in the system~\cite{Piilo2008,Tokieda2024,Semin2017}.\newline 

This picture is consistent with our cavity-eliminated description. Since the reduced system now only consists of the atom, coherence can be regained through its interaction with the environment — more precisely, with the cavity mode. To find a signature of such non-Markovianity, we identify an effective Hamiltonian and a set of effective decoherence operators that reproduce the Bloch equations~\eqref{Bloch-equation-final-system}, one of the decoherence operators being associated to a negative decoherence rate.\newline

The effective Hamiltonian reads
\begin{equation}
    \hat{\mathcal{H}}_{\textrm{eff}}=(\omega_{\textrm{a}'}-\omega_{\textrm{ref}})\hat{\sigma}^\dagger\hat{\sigma}-\frac{i}{2}\left(\Omega(1+\frac{x}{2})\hat{\sigma}^\dagger-\Omega^*(1+\frac{x}{2})^*\hat{\sigma}\right)
    \label{effective-h-def}
\end{equation}
This Hamiltonian has the same structure than a regular Hamiltonian for a driven two level system, with an effective Rabi frequency $\Omega(1+\frac{x}{2})$ and an effective atom frequency $\omega_{\rm a'}$. To reproduce Eq.~\eqref{Bloch-equation-final-system}, two effective decoherence operators associated to two effective damping rates are also required. We denote $\gamma_{\rm M}\geq 0$ (associated to a decoherence operator $\hat{L}_{\rm M}$) the rate participating to a purely Markovian evolution, and $\gamma_{\rm NM}\leq 0$ (associated to a decoherence operator $\hat{L}_{\rm NM}$) the one implying non-Markovian evolution. We find:
\begin{align}\label{effective_decoherence_rates}
        \gamma_{\rm M} &=\frac{\Gamma}{2}\left(1+\sqrt{1+2\frac{\left|\Omega x\right|^2}{\Gamma^2}}\right)\\
        \gamma_{\rm NM} &=\frac{\Gamma}{2}\left(1-\sqrt{1+2\frac{\left|\Omega x\right|^2}{\Gamma^2}}\right)\nonumber
\end{align}
associated to their corresponding effective decoherence operators:
\begin{align}\label{effective-L-def}
        \hat{L}_{\rm M} &= -\frac{1}{\sqrt{\gamma_{\rm M}-\gamma_{\rm NM}}}\left[\sqrt{\gamma_{\rm M}}~e^{i\phi}\hat{\sigma} -\sqrt{\left|\gamma_{\rm NM}\right|}\frac{\hat{\sigma}_z}{\sqrt{2}} \right]\\
        \hat{L}_{\rm NM} &= \frac{1}{\sqrt{\gamma_{\rm M}-\gamma_{\rm NM}}}\left[\sqrt{\left|\gamma_{\rm NM}\right|}~e^{i\phi} \hat{\sigma} + \sqrt{\gamma_{\rm M}}\frac{\hat{\sigma}_z}{\sqrt{2}} \right]\nonumber
\end{align}
with
\begin{align}
        \phi &=-\textrm{arg}(x\Omega)
\end{align}

As suggested by its notation, $\gamma_{\rm NM}$, the second rate is negative for all values of $\frac{\left|\Omega x\right|^2}{\Gamma^2}\ne 0$, hence being a signature of non-Markovianity, if $\Omega\ne 0$ and $x\ne 0$. Increasing either the atom–cavity coupling strength or the input power enhances $\frac{\left|\Omega x\right|^2}{\Gamma^2}$, pushing $\gamma_{\rm NM}$ toward more negative values and thereby amplifying non-Markovianity. At very low power ($\Omega\ll\Gamma$), $\frac{\left|\Omega x\right|^2}{\Gamma^2}\ll1$ and thus $\gamma_{\rm M}\approx\frac{\Gamma}{2}$ while $|\gamma_{\rm NM}|\ll\gamma_{\rm M}$. In such regime, the system's dynamics can actually be approximated considering a Markovian dynamics governed by $\hat{\mathcal{H}}_{\textrm{eff}}$ and $\hat{L}_{\rm M}$ alone. We note that the non-Markovianity depends at the same time on the coupling and on the input power, meaning that at very low coupling $|x|\ll 1$, one can saturate the cavity without reaching a highly non-Markovian behavior. 

Such non-Markovian behavior have already been found across several elimination frameworks, including high-order adiabatic elimination~\cite{Tokieda2024}, projection-based technique~\cite{Finkelstein-Shapiro2020}, and exact model reduction~\cite{Grigoletto2023}. Crucially, it has been shown to be an intrinsic feature of adiabatic elimination rather than an artifact of the expansion truncation~\cite{Tokieda2024}. While in our case it takes the form of a negative damping rate, the non-Markovianity can also be detected through a breakdown of complete positivity~\cite{Tokieda2024}, or non-trace preserving dynamics~\cite{Finkelstein-Shapiro2020}. We also note that these non-Markovian effects have notably been linked to entanglement between the retained subspace and the eliminated degrees of freedom~\cite{Tokieda2024,Jordan2004,McCracken2013}. Such non-Markovianity is expected for elimination techniques, as they effectively truncate a part of the system, including it into the environment. If the composite system possesses internal non-Markovian dynamics—as is typically the case—the reduced system will interact similarly with its augmented environment. In the cavity-QED context, stronger atom-cavity coupling enhances the cavity's memory of past interactions with the atom, leading to an increased non-Markovianity.
    
These effective Lindblad-like operators not only evidence the non-Markovianity of our system: they also happen to be particularly useful when it comes to simulating CQED experiments. Classical cavity quantum electrodynamics simulations cannot treat the infinite-dimensional cavity Hilbert space directly; a necessary truncation is performed, assuming that the cavity contains no more than $N$ photons. Doing so, one can use the complete CQED Hamiltonian~\eqref{Full-Hamiltonian}, together with Master equation solvers, to simulate various physical quantities. However, even within this assumption the Hilbert space to simulate has a size of $2\times(N+1)$, which can be problematic depending on the complexity of the simulations and the value of $N$. Our effective Lindblad-like operators provide a way to overcome these difficulties as they allow defining a Lindblad-like equation within a Hilbert space of dimension 2.

Starting from numerical cavity-QED simulations, one can implement numerical simulations in the reduced model using a modified master equation in the form of Eq.~\eqref{Lindblad-def} with the effective Hamiltonian of Eq.~\eqref{effective-h-def} and the effective decoherence operators and rates defined by Eqs.~\eqref{effective-L-def} to~\eqref{effective_decoherence_rates}. Finally, one has to replace the annihilation operator $\hat{a}$, initially defined in the Fowk space of the cavity mode, by the effective annihilation operator expressed, in Eq.~\eqref{a-expression-elim}, as a function of the atom operators $\hat{\sigma}$ and $\hat{\sigma}_z$. These modifications happen to be, by themselves, sufficient to fully convert a cavity-QED simulation toolbox into one where the cavity has been self-consistently eliminated.

\section{Computing output intensities under coherent CW excitation}\label{Chapter_5}
In this section we derive analytical expressions for the output intensities under a single coherent continuous wave (CW) excitation at a laser frequency $\omega_{\rm las}$. We work in the rotating frame at frequency $\omega_{\rm ref}=\omega_{\rm las}$ and start by computing the steady-state solutions, then present explicit formulae for the output fluxes, including both coherent and incoherent parts. Finally, we illustrate the theory with the practical case of a coherent CW excitation from a single cavity port. We compare the predictions of our cavity-eliminated model with the predictions of the full CQED model. This will allow testing the validity range of the self-consistent cavity elimination approach under coherent CW excitation. We note that, for brevity, we will use the acronym $\cancel{\rm C}$QED as a short notation to describe our model where the cavity is taken into account but self-consistently eliminated.

\subsection{Steady-state density matrix}
The optical Bloch equations of Eq.~\eqref{Bloch-equation-final-system} allow deriving the steady state average values, under coherent CW excitation, for the following operators:
\begin{equation}\label{steady-states}
    \begin{cases}
        \langle \hat{\sigma}^\dagger\hat{\sigma}\rangle _{\textrm{st}} &= \left|\frac{\Omega(1+x)}{2\Gamma'}\right|^2\frac{1}{1+\frac{|\Omega|^2}{\Gamma|\Gamma'|^2}\textrm{Re}(\Gamma'(1+x))}\\
         \langle \hat{\sigma}\rangle _{\textrm{st}}  &= \frac{\Omega}{2\Gamma'}(\langle \hat{\sigma}_z\rangle _{\textrm{st}}-x)\\
    \end{cases}
\end{equation}

The expectation values $\langle \hat{\sigma}^\dagger\hat{\sigma}\rangle _{\textrm{st}}$ and $\langle \hat{\sigma}\rangle _{\textrm{st}}$ fully specify the atom state in the stationary regime. We note that when $|x|\ll 1$, and thus $1+x\approx 1$ and $\langle \hat{\sigma}_z\rangle _{\textrm{st}}-x\approx\langle \hat{\sigma}_z\rangle _{\textrm{st}}$, these expressions become equivalent to the standard TLS steady-state expressions~\cite{Carmichael1999}. The regime where $|x|\ll 1$ is obtained when the device is far from the strong coupling threshold $g\ll \kappa$, but also if the atom is highly detuned from the cavity (dispersive regime). 

In the general case, the expressions of $\langle\hat{\sigma}^\dagger\hat{\sigma}\rangle_{\rm st}$ and  $\langle \hat{\sigma}\rangle_{\rm st}$ need to take into account the coupling parameter $x$. We note, however, that the expressions of Eq.~\eqref{steady-states} should not be considered exact, as the slow effective driving approximation has been made, which is expected to fail for large values of $\Omega$, when the Rabi oscillations become comparable to the damping rate $\kappa$. Still, Eq.~\eqref{steady-states} provides a useful set of equations for low and intermediate excitation regimes, allowing a description of the effective atom dynamics beyond the semiclassical approximation.
In order to simplify later calculations, we will also work with the operator $\hat{\sigma}'=\hat{\sigma}-(\hat{\sigma}_z-x)\frac{\Omega}{2\kappa'}$ and thus with the following average values in the steady state regime:
\begin{equation}\label{Sigma-prime-steady}
    \begin{cases}
        \langle \hat{\sigma}'\rangle _{\textrm{st}} &= \frac{\Omega}{2\Gamma'}(\langle \hat{\sigma}_z\rangle _{\textrm{st}}-x)\frac{\kappa'-\Gamma'}{\kappa'} \\
        \langle \hat{\sigma}'^{\dagger}\hat{\sigma}'\rangle _{\textrm{st}} &= \langle \hat{\sigma}^\dagger\hat{\sigma}\rangle _{\textrm{st}}\left[1-\left|\frac{\Omega}{\kappa'}\right|^2\textrm{Re}(x)\right]+\left|\frac{\Omega(1+x)}{2\kappa'}\right|^2\\
        &\quad+\textrm{Re}\left(\langle\hat{\sigma}^\dagger\rangle_{\rm st}\frac{\Omega(1+x)}{\kappa'}\right)
    \end{cases}
\end{equation}
In the case where $|\kappa'|\gg|\Gamma'|$, the steady-state value $\langle \hat{\sigma}'\rangle _{\textrm{st}}$ approaches the one obtained for the atomic dipole operator $\langle \hat{\sigma}\rangle _{\textrm{st}}$.

\subsection{Coherent and incoherent output intensities}

For each output port, the output flux can be computed through the expectation value $\langle \hat{b}_{\rm out}^{ (j)\dagger}\hat{b}_{\rm out}^{ (j)}\rangle $ or $\langle \hat{c}_{\rm out}^{ (l)\dagger}\hat{c}_{\rm out}^{ (l)}\rangle $, directly deduced from Eq.~\eqref{input-output-equation} and Eq.~\eqref{effective-input-output}. These fluxes can be expressed in the following forms:
\begin{equation}
    \begin{split}
        \langle \hat{b}_{\textrm{out}}^{\textrm{(j)}\dagger}\hat{b}_{\textrm{out}}^{(j)}\rangle_{\rm st}  & = |\langle \hat{b}_{\textrm{out}}^{(j)}\rangle |^2_{\textrm{st}}+|\Gamma_j|(\langle \hat{\sigma}'^{\dagger}\hat{\sigma}'\rangle _{\textrm{st}}-|\langle \hat{\sigma}'\rangle _{\textrm{st}}|^2) \\
        \langle \hat{c}^{\textrm{(l)}\dagger}_{\textrm{out}}\hat{c}^{(l)}_{\textrm{out}}\rangle_{\rm st}  & =|\langle \hat{c}^{(l)}_{\textrm{out}}\rangle |^2_{\textrm{st}}+|\gamma_{l}|(\langle \hat{\sigma}^{\dagger}\hat{\sigma}\rangle _{\textrm{st}}-|\langle \hat{\sigma}\rangle _{\textrm{st}}|^2)
    \label{Fluxes-expressions}
    \end{split}
\end{equation}
which explicitly shows the coherent and incoherent contributions for each AC and CC ports. The coherent flux for the CC port $j$ (resp AC port $l$) is $|\langle \hat{b}_{\textrm{out}}^{(j)}\rangle |^2_{\textrm{st}}$ (resp. $|\langle \hat{c}_{\textrm{out}}^{(l)}\rangle |^2_{\textrm{st}}$): it can directly be deduced from Eq.~\eqref{effective-input-output} (resp. Eq.~\eqref{input-output-equation}) and Eq.~\eqref{Sigma-prime-steady} (resp. Eq.~\eqref{steady-states}). For the AC ports, the incoherent part is proportional to $\langle \hat{\sigma}^{\dagger}\hat{\sigma}\rangle _{\textrm{st}}-|\langle \hat{\sigma}\rangle _{\textrm{st}}|^2$, as is the case for a regular TLS in the absence of a cavity. 

However, for the CC ports, the incoherent contribution is proportionnal to $\langle \hat{\sigma}'^{\dagger}\hat{\sigma}'\rangle _{\textrm{st}}-|\langle \hat{\sigma}'\rangle _{\textrm{st}}|^2$. This represents a qualitative difference with respect to waveguide-QED models, for which all ports are treated as atom-coupled ones, leading to an incoherent contribution proportionnal to $\langle \hat{\sigma}^{\dagger}\hat{\sigma}\rangle _{\textrm{st}}-|\langle \hat{\sigma}\rangle _{\textrm{st}}|^2$ in all ports. 

In each port, the coherent and incoherent contributions superpose to produce the total output flux. We note that this description is inherently more accurate than the semi-classical approximation which corresponds to neglecting any incoherent contribution, by neglecting fluctuations around the average values determined by the field amplitudes $\langle\hat{b}_{\rm out}^{ (j)}\rangle_{\rm st}$ and $\langle \hat{c}_{\rm out}^{ (l)}\rangle_{\rm st}$. In Sec.~\ref{Chapter_6} we will see how the spectrum of the incoherent light contribution can be effectively modeled for each output port.

\subsection{Practical case: excitation from a single cavity port}\label{Sec:practical-case-intensity}

We now test our theory in the practical case of a single coherent CW input through the port 1, meaning that 
\begin{equation}
        \begin{split}
            &\langle \hat{b}_{\textrm{in}}^{\textrm{(1)}}(t)\rangle =\langle \hat{b}_{\textrm{in}}\rangle \textrm{,}~~~~~\langle \hat{b}_{\textrm{in}}^{(j)}(t)\rangle  = 0 ~~~ \forall ~ j\ne 1\textrm{,}\\ 
            &\langle  \hat{c}_{\textrm{in}}^{(l)}\rangle  = 0 ~~~\forall~  l
        \end{split}
\end{equation}
We thus classify the outputs as follows: the output signal leaving through the first cavity-coupled port is called \emph{reflected}, the collection of all the signals leaving through the other CC ports is called \emph{transmitted}, and the collection of all the signals leaving through the AC ports is called \emph{emitted}. Consequently, we denote the output fluxes as
\begin{equation}\label{Name-of-ports}
    \begin{split}
        &\phi_{\rm refl} = \langle \hat{b}_{\rm out}^{\rm (1) \dagger}\hat{b}_{\rm out}^{\rm (1)}\rangle _{\rm st}\\
        &\phi_{\rm tr} = \sum_{j\ne 1} \langle \hat{b}_{\rm out}^{ (j) \dagger}\hat{b}_{\rm out}^{ (j)}\rangle _{\rm st}\\
        &\phi_{\rm em} = \sum_l \langle \hat{c}_{\rm out}^{ (l) \dagger}\hat{c}_{\rm out}^{ (l)}\rangle _{\rm st}\\
    \end{split}
\end{equation}
where energy conservation implies $\phi_{\rm in}=|\langle \hat{b}_{\textrm{in}}\rangle |^2 = \phi_{\rm refl}+\phi_{\rm tr}+\phi_{\rm em}$.\newline

In order to characterize different input powers, we define the number of incoming photons during a photon lifetime in the cavity
\begin{equation}
    N_{\textrm{in}} = \frac{\phi_{\rm in}}{\kappa}
\end{equation}

In most cases, in this paper, the emission of the atom outside of the fundamental mode of the cavity will be taken to be small in front of the other rates, $\gamma_{\rm a}\ll(g,\kappa)$. In this situation one can define the weak atom-cavity coupling regime as:
\begin{equation}
    \frac{4g}{\kappa}< 1
\end{equation} 

In this section, we consider two sets of parameters presented in Tab.~\ref{table-of-parameters}. Both are defined in a moderately weak coupling regime, $\frac{4g}{\kappa}=0.5$, with a damping rate of the cavity through its first port of $\frac{\kappa_1}{\kappa}=0.8$. The two sets of parameters differ by their atom-cavity detuning (the first set with $\omega_{\rm a}-\omega_{\rm c}=0$ and the second one with $\omega_{\rm a}-\omega_{\rm c}=\kappa/2$) and by their total emission rate through the AC ports ($\gamma_{\rm a}=\kappa/1280$ for the first set and $\gamma_{\rm a}=\kappa/32$ for the second one). As a figure of merit, we also use the cooperativity $C = \frac{2g^2}{\kappa\gamma_{\textrm{a}}}$: if $C\gg 1$ the atom has much more chance to emit in the cavity mode than directly through the AC ports. The first set of parameters is characterized by $C=40$, and the second one by $C=1$.

\begin{figure*}[t]
    \centering
    \makebox[\textwidth][c]{\includegraphics[width=2\columnwidth]{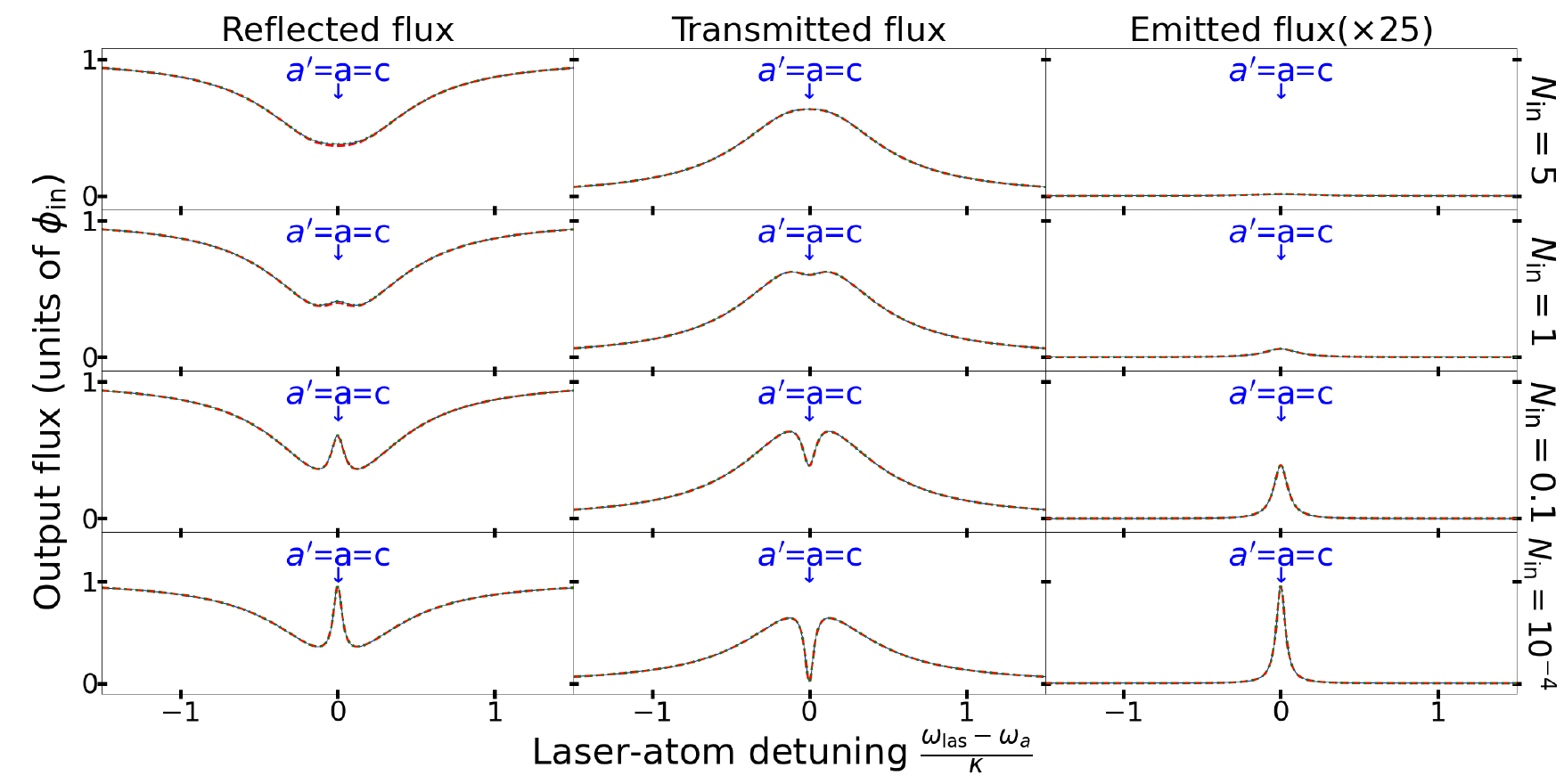}}
    \caption{Reflected (first column), transmitted (second column) and emitted (third column) fluxes in units of the input flux $\phi_{\rm in}$. The full CQED simulations (dotted red) are compared with the analytical $\textrm{\cancel{C}QED}$ calculations (blue) and the numerical $\textrm{\cancel{C}QED}$ simulations (dotted green). Each row corresponds to a different incoming photon flux $\phi_{\rm in}$ with (from the top to the bottom row): $N_{\rm in}$=5; 1; 0.1 and $10^{-4}$ respectively. The first set of parameters in Table 1 has been used, corresponding to a high cooperativity $C=40$ and an absence of atom-cavity detuning. }\label{fig-flux-weak-1}
\end{figure*}

In Fig.~\ref{fig-flux-weak-1}, we scan the response of the device characterized by the first set of parameters in Tab.~\ref{table-of-parameters}, as a function of the laser-atom detuning $\omega_{\rm las}-\omega_{\rm a}$, in units of $\kappa$. We study the reflected (first column), transmitted (second column) and emitted fluxes (third column, enlarged by a factor of 25), in units of $\phi_{\rm in}$, for various input powers (each row corresponding to $N_{\mathrm{in}}=$5; 1; 0.1; $10^{-4}$). 
Doing that, we compare: (i) the full CQED simulations (dotted red) (ii) the analytical $\textrm{\cancel{C}}$QED predictions (blue) computed using Eqs.~\eqref{Fluxes-expressions} and (iii) the numerical $\textrm{\cancel{C}}$QED simulations (dotted green) simulated using the effective Hamiltonian~\eqref{effective-h-def} and effective decoherence operators~\eqref{effective-L-def}. 
For this first set of parameters, the cavity frequency and the atom frequency are taken equal, leading, in the weak coupling regime, to an effective atom frequency $\omega_{\rm a'} = \omega_{\rm a} = \omega_{\rm c}$: these superposed resonances are highlighted, in each spectrum, by the vertical arrows associated to the three letters a, a' and c. The system also has a high cooperativity $C=40$, which is usually the ideal situation for quantum information applications~\cite{Reiserer2015,Kimble2008,Walther2006}. 
At very high power $N_{\rm in}=5$ (top row), the atom gets saturated and becomes transparent to the incoming light. In reflectivity, we thus observe the characteristic Lorentzian response of the bare cavity, whose contrast is directly determined by the output coupling efficiency $\frac{\kappa_1}{\kappa}=0.8$. At such high power, the emitted flux through the AC ports remains residual, $\phi_{\rm em}\ll\phi_{\rm in}$, and the transmitted flux is also characterized by a Lorentzian profile since $\phi_{\rm tr}\approx 1-\phi_{\rm refl}$ in this case. When the power decreases, the atom is not saturated anymore: its resonance fluorescence changes both the reflectivity and transmission spectrum by interfering with the bare cavity reflected and transmitted fluxes. The largest interference contrast is obtained at the lowest power, where the fraction of atom resonance fluorescence emitted through the AC ports, $\frac{\phi_{\rm em}}{\phi_{\rm in}}$, is also the largest, with a spectral width governed by $\Gamma$.
In all conditions, the analytical and numerical $\cancel{\mathrm{C}}$QED predictions agree perfectly, which allows verifying the consistency of the analytical formulae~\eqref{Fluxes-expressions} with respect to the effective Hamiltonian~\eqref{effective-h-def}, effective decoherence operators~\eqref{effective-L-def} and rates~\eqref{effective_decoherence_rates}. Also, these $\cancel{\rm C}$QED predictions almost perfectly agree with the full CQED model predictions. This was expected at low power where the output fluxes can all be considered coherent and where the semi-classical treatment fully applies (see discussion in Sec.~\ref{Subsec:Semiclassical}). A small discrepancy with the full CQED model appears at the largest power $N_{\rm in} =5$, which we relate to the inaccuracy in such regime of the slow effective drive approximation, $\hat{F}-\frac{\Dot{\hat{F}}}{\kappa'}+\frac{\Ddot{\hat{F}}}{\kappa^{'2}}-...\approx \hat{F}$. The impact of such inaccuracy remains very limited here, however, since the atom transition is already largely saturated at such excitation power.

\begin{figure*}[t]
    \centering
    \makebox[\textwidth][c]{\includegraphics[width=2\columnwidth]{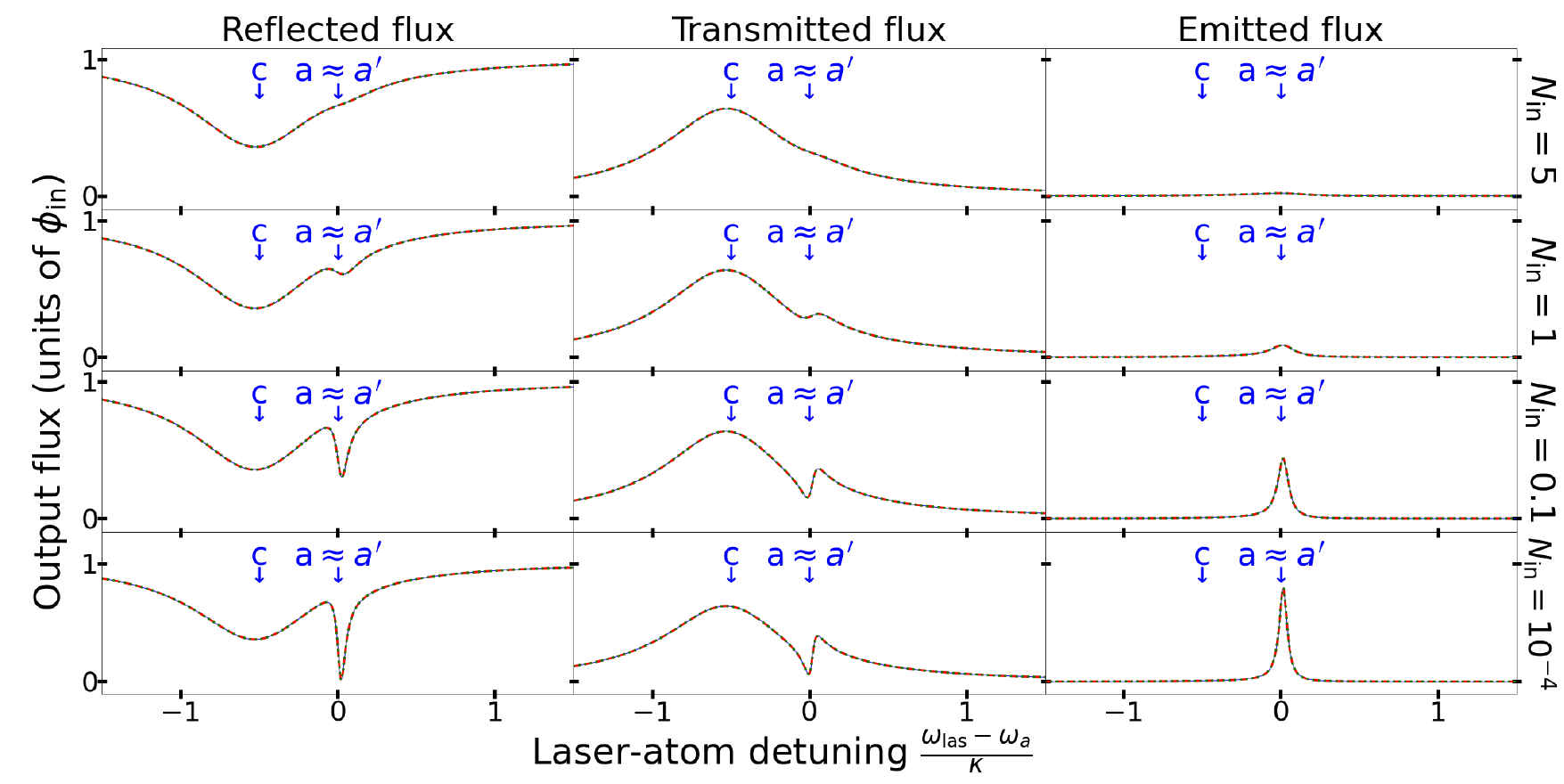}}
    \caption{Reflected (first column), transmitted (second column) and emitted (third column) fluxes in units of the input flux $\phi_{\rm in}$. The full CQED simulations (dotted red) are compared with the analytical $\textrm{\cancel{C}QED}$ calculations (blue) and the numerical $\textrm{\cancel{C}QED}$ simulations (dotted green). Each row corresponds to a different incoming photon flux $\phi_{\rm in}$ with (from the top to the bottom row): $N_{\rm in}$=5; 1; 0.1 and $10^{-4}$ respectively. The second set of parameters in Table 1 has been used, corresponding to a moderate cooperativity $C=1$ and to an atom-cavity detuning $\omega_{\rm a}-\omega_{\rm c}=0.5\kappa$. }\label{fig-flux-weak-2}
\end{figure*}

We show in Fig.~\ref{fig-flux-weak-2} the result of similar simulations using our second set of parameters in Table.~\ref{table-of-parameters}. The system is also in the weak coupling regime, with $\frac{4g}{\kappa}=0.5$ and $\frac{\kappa_1}{\kappa}=0.8$, yet with a non-zero atom-cavity detuning and a larger rate of spontaneous emission outside the cavity mode $\gamma_{\rm a}=\kappa/32$ (corresponding to a moderate cooperativity $C=1$).
The detuning between the cavity and the atom shifts the interference pattern: the constructive and destructive interferences now arise in both reflection and transmission when $\omega_{\rm laser}\approx\omega_{\rm a}\approx\omega_{\rm a'}$, which significantly differ from the cavity frequency $\omega_{\rm c}$. Due to the atom-cavity detuning the effective atom frequency is slightly shifted yet remains close to the atomic one ($\omega_{\rm a'} \approx \omega_{\rm a}+5.10^{-4}\kappa$), a feature that will change in the strong coupling regime (see Sec.~\ref{Chapter_8}). It can be seen on the right column that the lower cooperativity enhances the fraction of emitted photons, $\frac{\phi_{\rm em}}{\phi_{\rm in}}$ reaching $\sim 80\%$ at very low input power.
Contrary to Fig.~\ref{fig-flux-weak-1}, no discrepancy is observed between the full CQED and $\textrm{\cancel{C}QED}$ models, even at high power. These different discrepancies arise primarily from the change of detuning rather than the change of cooperativity. More generally, we find that for sufficiently large atom-cavity detunings, the $\cancel{\mathrm{C}}$QED model agrees with the full CQED simulations across all coupling strengths and input powers.

\begin{table}[htb]
    \centering
    \begin{tabular}{c c c c c}
        \hline
        Parameter & & First set (units of $\kappa$) & & Second set (units of $\kappa$) \\
        \hline 
        g & & 1/8 & & 1/8\\
        $\kappa_1$ & & 0.8 & & 0.8 \\
        $\gamma_{\rm a} $ & & 1/1280 & & 1/32\\ 
        $\omega_{\rm a}-\omega_{\rm c}$ & & 0 & & 1/2\\  
        \hline 
    \end{tabular}
    \caption{Sets of parameters used for the numerical simulations and for the analytical $\cancel{\rm C}$QED calculations in the weak coupling regime.}\label{table-of-parameters}
\end{table}

\section{Computing spectral densities under coherent CW excitation}\label{Chapter_6}
In this section, we aim to compute dynamical quantities, in particular the first-order coherence and the corresponding spectral density under coherent CW excitation. We solve the system of Bloch equations~\eqref{Bloch-equation-final-system} in the specific case of a resonant excitation $\omega_{\rm las}=\omega_{\rm a'}$, and use the solution to compute the first-order coherence and, subsequently, the spectral density in each port. We then benchmark our model by comparing it to simulations with the full CQED model.

\subsection{Dynamics of the system}

The analysis begins by computing the solution of the Bloch-equations system~\eqref{Bloch-equation-final-system}, which consists of two coupled differential equations for the average values $\langle\hat{\sigma}\rangle$ and $\langle\hat{\sigma}^\dagger\hat{\sigma}\rangle$. In the general case and under coherent CW excitation, $\langle \hat{\sigma} \rangle  (t)$ and $\langle \hat{\sigma}^\dagger\hat{\sigma} \rangle  (t)$ can be expressed in the following form:
\begin{widetext}
    \begin{equation}\label{Bloch-solution}
        \begin{cases}
    \langle \hat{\sigma} \rangle  (t)&=   A_1(t)(\langle \hat{\sigma}^\dagger\hat{\sigma}(0)\rangle -\langle \hat{\sigma}^\dagger\hat{\sigma}\rangle _{st})
    + A_2(t) (\langle \hat{\sigma}^\dagger(0)\rangle -\langle \hat{\sigma}^\dagger\rangle _{st})
    + A_3(t) (\langle \hat{\sigma}(0)\rangle -\langle \hat{\sigma}\rangle _{st})+\langle \hat{\sigma}\rangle _{st} \\
         \langle \hat{\sigma}^\dagger\hat{\sigma} \rangle (t) &= B_1 (t)(\langle \hat{\sigma}^\dagger\hat{\sigma}(0)\rangle -\langle \hat{\sigma}^\dagger\hat{\sigma}\rangle _{st})
          + B_2 (t)(\langle \hat{\sigma}^\dagger(0)\rangle -\langle \hat{\sigma}^\dagger\rangle _{st})
         + B_3(t) (\langle \hat{\sigma}(0)\rangle -\langle \hat{\sigma}\rangle _{st})+\langle \hat{\sigma}^\dagger\hat{\sigma}\rangle _{st}
        \end{cases}
    \end{equation}
\end{widetext}
The $A_i,B_i$ coefficients represent time-dependent complex coefficients, which tend towards 0 at long times and verify at time $t=0$: $A_1(0)=A_2(0)=B_2(0)=B_3(0)=0$ and $A_3(0)=B_1(0)=1$. In the general case, the expressions of the coefficients  $A_i,B_i$ are cumbersome to determine analytically: numerical methods are then preferable. Still, there is one particularly useful situation where the analytical treatment remains simple: the case of a resonant excitation, $\omega_{\mathrm{laser}}=\omega_{\mathrm{a'}}$. In this case, the eigenvalues of the system of Bloch equations~\eqref{Bloch-equation-final-system} simplify to:
\begin{align}\label{eigenvalues-b-in-dep}
    \lambda_0 &= -\frac{\Gamma}{2}\nonumber\\
    \lambda_\pm &= \frac{\Gamma}{4}\left(-3\pm\sqrt{1-\frac{16|\Omega|^2}{\Gamma^2}\textrm{Re}\left(1+x\right)}\right)
\end{align}
We note that when $1+x\approx 1$, i.e. when the atom is weakly coupled or strongly detuned from the cavity, one retrieves the well-known expressions obtained for a 2-level system~\cite{Carmichael1993b,Gardiner2004,Mollow1969}. We clearly see from Eq.~\eqref{eigenvalues-b-in-dep} that when the driving is small enough, $|\Omega|< \frac{\Gamma}{4\textrm{Re}(1+x)}$, the eigenvalues $\lambda_{\pm}$ are real and give rise to different damping rates, in addition to $\lambda_0=-\frac{\Gamma}{2}$ which is also real. On the contrary, when $|\Omega|> \frac{\Gamma}{4\textrm{Re}(1+x)}$, the eigenvalues $\lambda_\pm$ are complex conjugates, leading to oscillations between the ground and excited states, i.e. the well-known Rabi oscillations~\cite{Cohen-Tannoudji1998}. 
In this situation of resonant excitation at $\omega_{\rm las}=\omega_{\rm a'}$, we show in the appendix~\ref{Subsec-effective-dynamics-solution} that the coefficients $A_i,B_i$ can be expressed as:
\begin{align}\label{A-i-sol}
        A_1(t)&= \frac{ \Omega}{\lambda_--\lambda_+}\left(e^{\lambda_-t}-e^{\lambda_+ t}\right)\nonumber\\   
        A_2(t) &= \frac{(1+x)\Omega^2}{2\textrm{Re}(1+x)|\Omega|^2}\Bigg[\frac{(\lambda_+-\lambda_0)e^{\lambda_- t}-(\lambda_--\lambda_0)e^{\lambda_+ t}}{\lambda_+-\lambda_-}\nonumber\\
     & \quad -e^{\lambda_0 t}\Bigg]\nonumber\\      
       A_3(t) &= \frac{1}{2\textrm{Re}(1+x)}\Bigg[\frac{(1+x)^*}{\lambda_+-\lambda_-}((\lambda_+-\lambda_0)e^{\lambda_- t}\nonumber\\
      &\quad-(\lambda_--\lambda_0)e^{\lambda_+ t})+e^{\lambda_0 t}(1+x)\Bigg]
\end{align}
And:
\begin{align}\label{B-i-sol}
         B_1(t)  &= \frac{1}{\lambda_+-\lambda_-}\left((\lambda_+-\lambda_0)e^{\lambda_+t}-(\lambda_--\lambda_0)e^{\lambda_-t}\right)\nonumber\\
         B_2 (t) &= \frac{(1+x)\Omega}{2(\lambda_--\lambda_+)}\left(e^{\lambda_+t}-e^{\lambda_- t}\right)\nonumber \\
         B_3(t)&= B_2^*(t)
\end{align}
We note that only $(A_3,B_1)\ne 0$ when there are no inputs (i.e. $\Omega=0$), meaning that populations and coherences are decoupled in absence of driving.

In order to simplify later calculations, we also compute the evolution of the average values $\langle\hat{\sigma}'\rangle(t)$ and $\langle\hat{\sigma}'^{\dagger}\hat{\sigma}'\rangle(t)$ from the definition $\hat{\sigma}'=\hat{\sigma}-\left(\hat{\sigma}_z-x\hat{\mathbb{I}}\right)\frac{\Omega}{2\kappa'}$ (Eq.~\eqref{sigma-prime-Gamma-j-def}):

\begin{widetext}
\begin{equation}\label{sigma_prime_evolution}
    \begin{cases}
        \langle \hat{\sigma}'\rangle (t) &= C_1  (t)(\langle \hat{\sigma}^\dagger\hat{\sigma}(0)\rangle -\langle \hat{\sigma}^\dagger\hat{\sigma}\rangle _{st})
          + C_2  (t)(\langle \hat{\sigma}^\dagger(0)\rangle -\langle \hat{\sigma}^\dagger\rangle _{st})
         + C_3 (t) (\langle \hat{\sigma}(0)\rangle -\langle \hat{\sigma}\rangle _{st})+\langle \hat{\sigma}'\rangle _{st}\\
         \langle \hat{\sigma}'^{\dagger}\hat{\sigma}'\rangle (t) &= D_1  (t)(\langle \hat{\sigma}^\dagger\hat{\sigma}(0)\rangle -\langle \hat{\sigma}^\dagger\hat{\sigma}\rangle _{st})
          + D_2  (t)(\langle \hat{\sigma}^\dagger(0)\rangle -\langle \hat{\sigma}^\dagger\rangle _{st})
         + D_3 (t) (\langle \hat{\sigma}(0)\rangle -\langle \hat{\sigma}\rangle _{st})+\langle \hat{\sigma}'^{\dagger}\hat{\sigma}'\rangle_{st}
    \end{cases}
\end{equation}
\end{widetext}

With the coefficients:
\begin{equation}
    \begin{split}\label{C-i-expr}
        C_i(t) &= A_i(t)-B_i(t)\frac{{\Omega}}{\kappa'}\\
    \end{split}
\end{equation}
and:

\begin{align}\label{D-i-expr}
    D_1(t) &=\textrm{Re}(\frac{\Omega(1+x)}{\kappa'}A_1^*(t))+B_1(t)\left(1-\left|\frac{\Omega}{\kappa'}\right|^2\textrm{Re}(x)\right)\nonumber\\
        D_2(t) &=\frac{\Omega(1+x)}{2\kappa'}A_3^*(t)+\left(\frac{\Omega(1+x)}{2\kappa'}\right)^*A_2(t)\nonumber\\
        &\quad+B_2(t)\left(1-\left|\frac{\Omega}{\kappa'}\right|^2\textrm{Re}(x)\right)\nonumber\\ 
        D_3(t) &=D_2^*(t)
\end{align}
We note that Eqs.~\eqref{C-i-expr},~\eqref{D-i-expr} hold in general, yet it is only in the resonant case that the coefficients $C_i,D_i$ have a simple analytical expression, using Eqs.~\eqref{A-i-sol},\eqref{B-i-sol}.

\subsection{First order coherences and spectral densities}

We now compute the first-order coherences for both types of ports under coherent CW excitation. For each AC port $l$, the first-order coherence can be decomposed in two contributions, coherent and incoherent:
\begin{align}\label{def-g-1-port-2}
         \langle \hat{c}^{(l),\dagger}_{\textrm{out}}(\tau) \hat{c}^{(l)}_{\textrm{out}}\rangle _{\rm st}=\left|\langle \hat{c}_{\rm out}^{(l)}\rangle_{\rm st}\right|^2+|\gamma_l|(\langle \hat{\sigma}^{\dagger}(\tau)\hat{\sigma}\rangle _{\textrm{st}}-|\langle \hat{\sigma}\rangle _{\textrm{st}}|^2)
\end{align}
The coherent contribution $\left|\langle \hat{c}_{\rm out}^{(l)}\rangle_{\rm st}\right|^2$ is constant: it is governed by the coherent interference between the CW laser and the atom resonance fluorescence which is described by Eq.~\eqref{steady-states}. The incoherent part, on the contrary, depends on the time delay $\tau$. It can be computed using the quantum regression theorem~\cite{Breuer2002}, leading to:
\begin{align}\label{g-1-atom-incoh-general}
        \langle \hat{\sigma}^{\dagger}(\tau)\hat{\sigma}\rangle_{\rm st} -|\langle \hat{\sigma}_{\rm st}\rangle |^2 &= -A_1^*(\tau)\langle \hat{\sigma}^\dagger\hat{\sigma}\rangle _{\rm st}\langle \hat{\sigma}\rangle _{\rm st}-A_2^*(\tau)\langle \hat{\sigma}\rangle _{\rm st}^2\nonumber\\ 
        &\quad+A_3^*(\tau)(\langle \hat{\sigma}^\dagger\hat{\sigma}\rangle _{\rm st}-|\langle \hat{\sigma}\rangle _{\rm st}|^2)
\end{align}
As before, this expression simplifies in the resonant case, where we use Eq.~\eqref{A-i-sol} to compute:
\begin{align}\label{g-1-atom-ports}
        &\langle \hat{\sigma}^{\dagger}(\tau)\hat{\sigma}\rangle_{\rm st} -|\langle \hat{\sigma}\rangle_{\rm st} |^2 =\mathcal{N}_{0}e^{\lambda_0 \tau} \\
    &\quad+(\frac{\mathcal{N}_{1}}{\lambda_+-\lambda_0}+\mathcal{N}_{2})e^{\lambda_+ \tau}-(\frac{\mathcal{N}_{1}}{\lambda_--\lambda_0}+\mathcal{N}_{2})e^{\lambda_- \tau}\nonumber
\end{align}
where the constants $\mathcal{N}_{0,1,2}$ are fully expressed in the Appendix~\ref{Appendix}. It can be seen from this equation that the first-order coherence is governed by the three eigenvalues, each one having a different weight in the dynamics.

We now perform the same analysis for each CC port $j$, that is, we write the first-order coherence as:
\begin{align}\label{def-g-1-port-1}
          \langle \hat{b}^{(j),\dagger}_{\textrm{out}}(\tau) \hat{b}_{\textrm{out}}^{(j)}\rangle _{\textrm{st}}=\left|\langle \hat{b}_{\rm out}^{(j)}\rangle_{\rm st}\right|^2+|\Gamma_j|(\langle \hat{\sigma}'^{\dagger}(\tau)\hat{\sigma}'\rangle _{\textrm{st}}-|\langle \hat{\sigma}'\rangle _{\textrm{st}}|^2)
    \raisetag{-5pt}
\end{align}
Eq.~\eqref{def-g-1-port-1} has a form very similar to Eq.~\eqref{def-g-1-port-2}, with the important qualitative difference that the operator $\hat{\sigma}'$ is used instead of $\hat{\sigma}$. As before, this first-order coherence is composed of a constant coherent part $\left|\langle \hat{b}_{\rm out}^{(j)}\rangle_{\rm st}\right|^2$, that can directly be deduced from Eq.~\eqref{Sigma-prime-steady}, and an incoherent part computed using the quantum regression theorem:
\begin{align}\label{general-g-1-expression}
        &\langle \hat{\sigma}'^{\dagger}(\tau)\hat{\sigma}'\rangle _{\textrm{st}}-|\langle \hat{\sigma}'\rangle _{\textrm{st}}|^2 =\nonumber \\ 
        &C_3^*(\tau)\left[\langle\hat{\sigma}^\dagger\hat{\sigma}\rangle_{\rm st}+\langle\hat{\sigma}^\dagger\rangle_{\rm st}\left((1+x)\frac{\Omega}{2\kappa'}-\langle\hat{\sigma}'\rangle_{\rm st}\right)\right]\\
         &-\left(C_2^*(\tau)\langle\hat{\sigma}\rangle_{\rm st}+C_1^*(\tau)\langle\hat{\sigma}^\dagger\hat{\sigma}\rangle_{\rm st}\right)\left[(1-x)\frac{\Omega}{2\kappa'}+\langle\hat{\sigma}'\rangle_{\rm st}\right]\nonumber\raisetag{0.7cm}
\end{align}
In the resonant case $\omega_{\rm las} = \omega_{\rm a'}$, we can directly use the analytical expressions of the $A_i,B_i$ (and thus $C_i,D_i$) coefficients to derive:
\begin{widetext}
        \begin{equation}\label{incohrent-g-1}
            \langle \hat{\sigma}'^{\dagger}(\tau)\hat{\sigma}'\rangle_{\rm st} -|\langle \hat{\sigma}'\rangle_{\rm st} |^2=\mathcal{N}_{0}'e^{\lambda_0 \tau} 
            +(\frac{\mathcal{N}_{1}'}{\lambda_+-\lambda_0}+\mathcal{N}_{2}')(\lambda_+-\lambda_0-\kappa'^*)e^{\lambda_+ \tau}
        -(\frac{\mathcal{N}_{1}'}{\lambda_--\lambda_0}+\mathcal{N}_{2}')(\lambda_--\lambda_0-\kappa'^*)e^{\lambda_- \tau}
    \end{equation}
\end{widetext}
where the constants $\mathcal{N}_{0,1,2}'$ are fully expressed in the Appendix~\ref{Appendix}. We note here that the first-order coherences for all the AC and CC ports are governed by the same eigenvalues ($\lambda_0,\lambda_+$ and $\lambda_-$), though with different amplitudes for each port. \newline 

We now turn to the spectral density of the light scattered through each output port, defined as the Fourier transform (FT) of the corresponding first-order coherence. The coherent contribution to the output flux in each output port physically corresponds to monochromatic light that oscillates only at the input frequency $\omega_{\rm las}$: its Fourier transform is therefore proportional to $\delta(\omega-\omega_{\rm las})$~\cite{Mollow1969}. In the following, we consider only the incoherent part. For each CC port $(j)$ and each AC port $(l)$, we have:

\begin{align}\label{def-SDF-incoh}
        S_{\rm AC}^{(l,\textrm{incoh})}(\omega) &=|\gamma_l|\textrm{FT}(\langle \hat{\sigma}^{\dagger}(\tau)\hat{\sigma}\rangle _{\textrm{st}}-|\langle \hat{\sigma}\rangle _{\textrm{st}}|^2)\nonumber\\
        S_{\rm CC}^{(j,\textrm{incoh})}(\omega) &=|\Gamma_j|\textrm{FT}(\langle \hat{\sigma}'^{\dagger}(\tau)\hat{\sigma}'\rangle _{\textrm{st}}-|\langle \hat{\sigma}'\rangle _{\textrm{st}}|^2)
\end{align}
In the resonant case, one can combine these definitions together with Eqs.~\eqref{incohrent-g-1} and~\eqref{g-1-atom-ports} and use the following identity
\begin{equation}\label{FT-transform}
    \textrm{FT}(e^{\lambda t})= -\frac{1}{2\pi}\frac{1}{\lambda+i(\omega_{\rm las}-\omega)}+\textrm{c.c}
\end{equation}
where c.c is the complex conjugate. This allows computing the analytical spectral density of the scattered light for both types of ports:
\begin{widetext}
\begin{align}\label{SDF-cavity-ports}
        S_{\rm AC}^{(l,\textrm{incoh})}(\omega) &=-\frac{|\gamma_l|}{2\pi}\Bigg[\frac{\mathcal{N}_{0}}{\lambda_0+i(\omega_{\rm las}-\omega)}
    +\frac{\frac{\mathcal{N}_{1}}{\lambda_+-\lambda_0}+\mathcal{N}_{2}}{\lambda_++i(\omega_{\rm las}-\omega)}-\frac{\frac{\mathcal{N}_{1}}{\lambda_--\lambda_0}+\mathcal{N}_{2}}{\lambda_-+i(\omega_{\rm las}-\omega)}\Bigg]+\rm c.c \\
       S_{\rm CC}^{(j,\textrm{incoh})}(\omega) &=-\frac{|\Gamma_j|}{2\pi}\left[
       \frac{\mathcal{N}_0}{\lambda_0+i(\omega_{\textrm{las}}-\omega)}+ \frac{(\frac{\mathcal{N}_1'}{\lambda_+-\lambda_0}+\mathcal{N}_2')(\lambda_+-\lambda_0-\kappa'^*)}{\lambda_++i(\omega_{\textrm{las}}-\omega)}
         -\frac{(\frac{\mathcal{N}_1'}{\lambda_--\lambda_0}+\mathcal{N}_2')(\lambda_--\lambda_0-\kappa'^*)}{\lambda_-+i(\omega_{\textrm{las}}-\omega)}\nonumber
        \right]+\textrm{c.c}
    \raisetag{1cm}
\end{align}
\end{widetext}
 Eqs.~\eqref{SDF-cavity-ports} shows that the spectral density generally consists of the sum of three contributions, whose central frequencies are determined by Im($\lambda_0)$,Im($\lambda_+)$, Im($\lambda_-)$. At larger excitation power, these three contributions constitute the well-known Mollow triplet~\cite{Mollow1969,Lodahl2015}, which the self-consistent cavity elimination approach allows describing analytically for both AC and CC ports. As will be illustrated in the next subsection, the shape of the Mollow Triplet spectrum is altered by the cavity, both for the AC and CC ports.

\subsection{Practical case: resonant excitation from a single cavity port}\label{Sec:practical-case-SDF}

To benchmark the self-consistent cavity elimination approach, we compare its predictions with full cavity-QED simulations. We consider a single, coherent CW drive incident on the first cavity port, resonant with the effective atomic frequency: $\omega_{\rm las} = \omega_{\rm a'}$. As in Subsection~\ref{Sec:practical-case-intensity}, the field associated to the first CC port, $\hat{b}_{\rm out}^{\rm (1)}$, is denoted as reflected. The ones associated to $\hat{b}_{\rm out}^{ (j)}$ with $j>2$ are considered transmitted, and the ones associated to all $\hat{c}_{\rm out}^{(l)}$ are considered emitted.

\begin{figure*}[t]
    \centering
    \makebox[\textwidth][c]{\includegraphics[width=2\columnwidth]{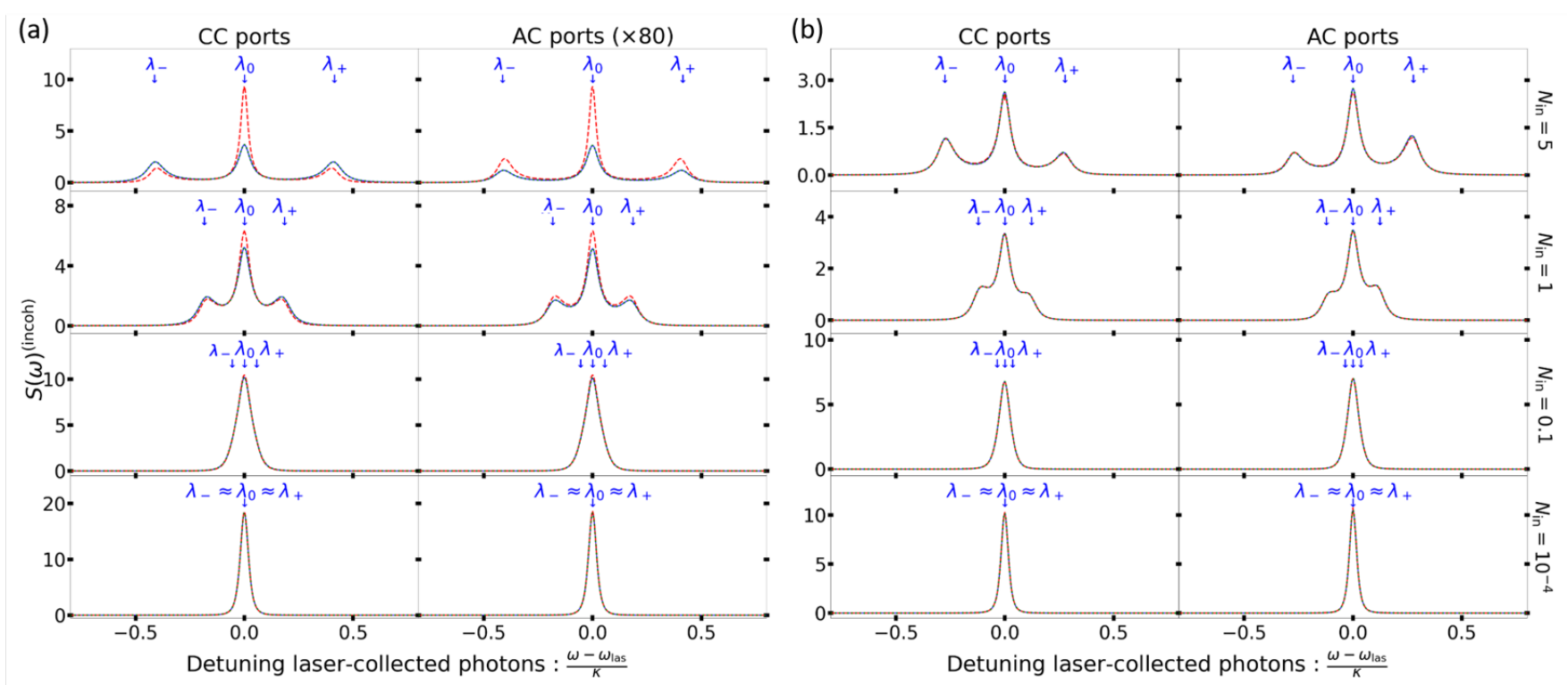}}
    \caption{Spectral density of the scattered incoherent light, summed over all CC ports (first column) and AC ports (second column). The full CQED simulations (dotted red) are compared with the analytical $\textrm{\cancel{C}QED}$ calculations (blue) and the numerical $\textrm{\cancel{C}QED}$ simulations (dotted green). Each row corresponds to a different incoming photon flux $\phi_{\rm in}$ with (from the top to the bottom row): $N_{\rm in}$=5; 1; 0.1 and $10^{-4}$ respectively. (a): First set of parameters in Table.~\ref{table-of-parameters}, with $C=40$ and $\omega_{\rm a}=\omega_{\rm c}$. (b): second set of parameters in Table.~\ref{table-of-parameters}, with $C=1$ and $\omega_{\rm a}-\omega_{\rm c}=0.5\kappa$.}\label{fig-SDF}
\end{figure*}

The results are shown in Fig.~\ref{fig-SDF}, where we plot the incoherent part of the spectral density as a function of the detuning between the detected photons and the driving laser. The two sets of parameters in Tab.~\ref{table-of-parameters} are considered, the first one for Fig.~\ref{fig-SDF}(a) and the second one for Fig.~\ref{fig-SDF}(b). In each panel, the left column displays the sum of all spectral densities, $S_{\rm CC}^{\rm (incoh)}$, including both the reflected and transmitted fields (CC ports). The right one displays the sum of all spectral densities, $S_{\rm AC}^{\rm (incoh)}$, including all the emission channels (AC ports). As with Fig.~\ref{fig-flux-weak-1} and~\ref{fig-flux-weak-2}, the rows correspond to varying input powers, from $N_{\rm in}=10^{-4}$ (bottom row) to $N_{\rm in}=5$ (top row). The spectral positions associated to the imaginary parts of the system's three eigenvalues are indicated by blue markers. 

In all conditions, these figures show a prefect agreement between the numerical $\cancel{\rm C}$QED simulations and the analytical $\cancel{\rm C}$QED expressions of Eqs.~\eqref{SDF-cavity-ports}. The comparison with the full CQED numerical simulations also shows an excellent agreement, for all ports and parameter sets, in the weak excitation regime. In such a case, there is a single emission line centered on $\omega_{\rm las}=\omega_{\rm a'}$ since $\rm Im(\lambda_+)=\rm Im(\lambda_-)=\rm Im(\lambda_0)=0$. As the excitation power increases, three different frequencies appear since $\rm Im(\lambda_+)\ne\rm Im(\lambda_-)\ne\rm Im(\lambda_0)$. This leads to Stokes (resp. anti-Stokes) emission lines at frequencies governed by $\rm Im(\lambda_+)$ (resp. $\rm Im(\lambda_-)$).
The Mollow triplet, consisting of three distinct emission lines~\cite{Mollow1969}, emerges when $N_{\rm in}$ becomes higher than 1.
This regime is associated to the appearance of fast Rabi oscillations, with an effective Rabi frequency governed by Im$(\lambda_+)\approx|\Omega|\sqrt{\textrm{Re}(1+x)}$, as deduced from Eq.~\eqref{eigenvalues-b-in-dep} in the high-excitation regime. In such regime our approximation $\hat{F}-\frac{\Dot{\hat{F}}}{\kappa'}+...\approx\hat{F}$, which considers that the effective drive operator evolves at timescale slower than $\kappa^{-1}$, is expected to fail: this can indeed be seen for $N_{\rm in}=1$ and especially $N_{\rm in}=5$ in Fig.~\ref{fig-SDF}(a), where the $\cancel{\rm C}$QED model does not reproduce the CQED simulations. For the second set of parameter, in Fig.~\ref{fig-SDF}(b), a very good agreement is still observed, even at high powers, between the CQED and $\cancel{\rm C}$QED models.

Fig.~\ref{fig-SDF} also reveals an asymmetry in the Mollow triplet emission spectrum, induced by the non-zero detuning $\omega_{\rm c}-\omega_{\rm a}=0.5\kappa$ when using the second set of parameters in Tab.~\ref{table-of-parameters}. In such conditions, the lower energy emission line is more resonant with the cavity mode, and thus extracted more efficiently through the cavity, thus through the CC ports. As the total spectral density $S^{\rm (incoh)}_{\rm tot} =S^{\rm (incoh)}_{\rm AC}+S^{\rm (incoh)}_{\rm CC}$ must remain symmetrical, due to energy conservation, a reversed asymmetry is observed for the AC ports, where the higher-energy anti-Stokes line displays a higher amplitude in the AC ports than in the CC ports. It is an important feature of this $\cancel{\rm C}$QED approach that this Mollow triplet asymmetry, which is a well-known cavity-QED effect observed both numerically~\cite{Tian1992} and experimentally~\cite{Ulhaq2013}, can be quantitatively reproduced and analytically computed. We note that such an asymmetry can only be obtained in the presence of a detuning between one of the three characteristic frequencies, $\omega_{\rm a}, \omega_{\rm c}$ and $\omega_{\rm las}$. Also, it can only be significant if there is a significant fraction of the light emitted through the AC ports, so that $S_{\rm AC}^{\rm (incoh)}$ is not negligible compared to $S_{\rm CC}^{\rm (incoh)}$. In the case where $S_{\rm AC}^{\rm (incoh)}\ll S_{\rm CC}^{\rm (incoh)}$, i.e when $\gamma_{\rm a}\ll\Gamma$ and thus $C\gg 1$, $S_{\rm tot}^{\rm (incoh)}\approx S_{\rm CC}^{\rm (incoh)}$ and thus $S_{\rm CC}^{\rm (incoh)}$ remains symmetrical. For instance, when $C=40$ as in the first set of parameters of Table.~\ref{table-of-parameters}, we do indeed see that $S_{\rm AC}^{\rm (incoh)}\ll S_{\rm CC}^{\rm (incoh)}$, since the $S_{\rm AC}^{\rm (incoh)}$ spectrum had to by multiplied by 80, in Fig.~\ref{fig-SDF}(b), to be compared with $S_{\rm CC}^{\rm (incoh)}$. On the contrary, for the second set of parameters $C=1$, $S_{\rm CC}^{\rm (incoh)}$ and $S_{\rm AC}^{\rm (incoh)}$ are directly comparable, equally participating to the total incoherent spectrum $S_{\rm tot}^{\rm (incoh)}$.

\section{Computing second-order correlation functions under coherent CW excitation}\label{Chapter_7}
We now show how to compute second-order correlations within the self-consistent cavity elimination model. We start by computing the density matrix of the system conditioned to a photon detection, then link it with the second-order correlation function. Finally, we compare numerical predictions of the full CQED model with numerical and analytical predictions in the $\cancel{\rm C}$QED approach.

\subsection{Conditional density matrix after photon detection}

We start by modeling the effect of a photon detection event through an arbitrary output port, leading to a conditional density matrix for the effective atom. We first define Kraus operators~\cite{Breuer2002,Kraus1983} that allow describing the measurement of a photon through an atom-coupled output port $l$,  $K_{\rm AC}^{ (l)}$, or through a cavity-coupled output port $j$,  $K_{\rm CC}^{ (j)}$, and the Kraus operator $K_0$ describing the absence of photon detection:

\begin{align}
       & \hat{K}_{\rm AC}^{ (l)}=\hat{c}_{\rm out}^{ (l)}\nonumber\\
       & \hat{K}_{\rm CC}^{ (j)}=\hat{b}_{\rm out}^{ (j)}\\
        &\hat{K}_{\rm 0} = \left(\hat{\mathbb{I}}-\sum_j\hat{b}_{\rm out}^{ (j)\dagger}\hat{b}_{\rm out}^{ (j)}-\sum_l\hat{c}_{\rm out}^{ (l)\dagger}\hat{c}_{\rm out}^{ (l)}\right)^{1/2}\nonumber
\end{align}
where one can check the trace preservation criterion $\sum_{ l}\hat{K}_{\rm AC}^{ (l)\dagger}\hat{K}_{\rm AC}^{ (l)}+\sum_{ j}\hat{K}_{\rm CC}^{ (j)\dagger}\hat{K}_{\rm CC}^{ (j)}+\hat{K}_{\rm 0}^\dagger \hat{K}_{\rm 0}=\hat{\mathbb{I}}$. Each operator either describes a quantum jump of the system associated to the detection of a photon in one of the output port, or the absence of any quantum jump. An important feature of the self-consistent cavity elimination approach is that all these Kraus operators are expressed in terms of the two-level atom degrees of freedom. This formalism is thus adapted to the description of the back-action induced, on the atom state, by any photon detection event. Starting from any density matrix $\rho$, a photon detection in the AC port $l$ (resp. CC port $j$), leads to the conditional density matrix $\rho_{\text{|AC,}l}$ (resp. $\rho_{\textrm{|CC,}j}$) computed as:
\begin{align}\label{rho-tilde-def}
        \rho_{\textrm{|AC,}l} &= \frac{\hat{K}_{\rm AC}^{ (l)}\rho \hat{K}_{ AC}^{ (l)\dagger}}{\textrm{Tr}(\hat{K}_{\rm AC}^{ (l)}\rho \hat{K}_{\rm AC}^{ (l)\dagger})}\nonumber\\
        \rho_{\textrm{|CC,}j} &= \frac{\hat{K}_{\rm CC}^{ (j)}\rho \hat{K}_{ CC}^{ (j)\dagger}}{\textrm{Tr}(\hat{K}_{\rm CC}^{ (j)}\rho \hat{K}_{\rm CC}^{ (j)\dagger})}
\end{align}

In the general case, the Kraus operators $\hat{K}_{\rm CC}^{(j)} = \hat{b}_{\rm out}^{(j)}$ and $\hat{K}_{\rm AC}^{(l)} = \hat{c}_{\rm out}^{(l)}$ are defined in a large Hilbert space taking into account the input fields, themselves being connected to potential excitation sources providing non-trivial quantum states. As such, the back action induced by a single photon detection in an output port can have an effect on both the cavity-QED system and its excitation sources. Tracing over the other degrees of freedom, one can focus on the back action induced on the TLS population and coherence after a detection event. We thus define:
\begin{align}
       & \langle\hat{\sigma}^\dagger\hat{\sigma}\rangle_{\textrm{|AC,}l} = Tr\left(\rho_{\textrm{|AC,}l}\hat{\sigma}^\dagger\hat{\sigma}\right) = \frac{\langle \hat{K}^{(l)\dagger}_{\rm AC}\hat{\sigma}^\dagger\hat{\sigma}\hat{K}^{(l)}_{\rm AC}\rangle}{\langle \hat{K}^{(l)\dagger}_{\rm AC}\hat{K}^{(l)}_{\rm AC}\rangle} \nonumber\\
        &\langle\hat{\sigma}\rangle_{\textrm{|AC,}l} = Tr\left(\rho_{\textrm{|AC,}l}\hat{\sigma}\right) = \frac{\langle \hat{K}^{(l)\dagger}_{\rm AC}\hat{\sigma}\hat{K}^{(l)}_{\rm AC}\rangle}{\langle \hat{K}^{(l)\dagger}_{\rm AC}\hat{K}^{(l)}_{\rm AC}\rangle}\\
    & \langle\hat{\sigma}^\dagger\hat{\sigma}\rangle_{\textrm{|CC,}j} = Tr\left(\rho_{\textrm{|CC,}j}\hat{\sigma}^\dagger\hat{\sigma}\right) = \frac{\langle \hat{K}^{(j)\dagger}_{\rm CC}\hat{\sigma}^\dagger\hat{\sigma}\hat{K}^{(j)}_{\rm CC}\rangle}{\langle \hat{K}^{(j)\dagger}_{\rm CC}\hat{K}^{(j)}_{\rm CC}\rangle}\nonumber \\
    &\langle\hat{\sigma}\rangle_{\textrm{|CC,}j} = Tr\left(\rho_{\textrm{|CC,}j}\hat{\sigma}\right) = \frac{\langle \hat{K}^{(j)\dagger}_{\rm CC}\hat{\sigma}\hat{K}^{(j)}_{\rm CC}\rangle}{\langle \hat{K}^{(j)\dagger}_{\rm CC}\hat{K}^{(j)}_{\rm CC}\rangle}\nonumber
\end{align}
the conditional population and coherences, conditioned to a photon detection event in an AC port $l$ or a CC port $j$, respectively.

Noting that $\hat{K}^{(l)}_{\rm AC} = \hat{c}_{\rm out}^{(l)} = \hat{c}_{\rm in}^{(l)}+\sqrt{\gamma_l}\hat{\sigma}$, we find that the expressions of $\langle\hat{\sigma}^\dagger\hat{\sigma}\rangle_{\textrm{|AC,}l}$ and $\langle\hat{\sigma}\rangle_{\textrm{|AC,}l}$ simplify to:
\begin{align}\label{rho-tilde-expression-AC-general}
        &\langle\hat{\sigma}^\dagger\hat{\sigma}\rangle_{\textrm{|AC,}l} = \frac{\langle\hat{c}_{\rm in}^{(l)\dagger}\hat{\sigma}^\dagger\hat{\sigma}\hat{c}_{\rm in}^{(l)}\rangle}{\langle\hat{c}_{\rm out}^{(l)\dagger}\hat{c}_{\rm out}^{(l)}\rangle} \rule[-25pt]{0pt}{0pt} \nonumber\\
        &\langle\hat{\sigma}\rangle_{\textrm{|AC,}l} = \frac{\sqrt{\gamma_l^*}\langle\hat{\sigma}^\dagger\hat{\sigma}\hat{c}_{\rm in}^{(l)}\rangle+\langle\hat{c}_{\rm in}^{(l)\dagger}\hat{\sigma}\hat{c}_{\rm in}^{(l)}\rangle}{\langle\hat{c}_{\rm out}^{(l)\dagger}\hat{c}_{\rm out}^{(l)}\rangle}
\end{align}
which for a vacuum or coherent input gives: 
 \begin{align}\label{rho-tilde-expression-AC}
        &\langle\hat{\sigma}^\dagger\hat{\sigma}\rangle_{\textrm{|AC,}l} =\left|\langle\hat{c}_{\rm in}^{(l)}\rangle\right|^2 \frac{\langle\hat{\sigma}^\dagger\hat{\sigma}\rangle}{\langle\hat{c}_{\rm out}^{(l)\dagger}\hat{c}_{\rm out}^{(l)}\rangle}\nonumber\\
     &\langle\hat{\sigma}\rangle_{\textrm{|AC,}l} = \frac{\sqrt{\gamma_l^*}\langle\hat{\sigma}^\dagger\hat{\sigma}\rangle\langle\hat{c}_{\rm in}^{(l)}\rangle+\left|\langle\hat{c}_{\rm in}^{(l)}\rangle\right|^2\langle\hat{\sigma}\rangle}{\langle\hat{c}_{\rm out}^{(l)\dagger}\hat{c}_{\rm out}^{(l)}\rangle}
 \end{align}
As such, detecting a photon in an AC port that has no input field, $\langle\hat{c}_{\rm in}^{(l)}\rangle=0$, projects the atom in the ground state: $\langle\hat{\sigma}^\dagger\hat{\sigma}\rangle_{\textrm{|AC,}l}=0$.

An important feature of the self-consistent cavity elimination approach is that a similar analysis can be applied to compute the back action induced by a photon detection in a cavity-coupled output port. Indeed, using the fact that $\hat{K}_{\rm CC}^{(j)}=\hat{b}_{\rm out}^{(j)}=\hat{b}_{\rm out,c}^{(j)}-\sqrt{\Gamma_j}\hat{\sigma}'$ (see Eq.~\eqref{effective-input-output}) and $\hat{\sigma}' =\hat{\sigma}-\left(\hat{\sigma}_z-x\hat{\mathbb{I}}\right)\frac{\hat{\Omega}}{2\kappa'}$ (Eq.~\eqref{sigma-prime-Gamma-j-def}), one can derive:
\vspace{\fill}
\begin{widetext}
    \begin{equation}\label{rho-tilde-expression-CC-general}
     \begin{cases}
       & \langle \hat{\sigma}^\dagger\hat{\sigma}\rangle _{\textrm{|CC,}j} = \frac{1}{\langle \hat{b}_{\textrm{out}}^{(j),\dagger} \hat{b}_{\textrm{out}}^{(j)}\rangle }\langle \left(\hat{b}_{\textrm{out,c}}^{(j)}  + \sqrt{\Gamma_j}(1-x)\frac{\hat{\Omega}}{2\kappa'}\right)^\dagger \hat{\sigma}^\dagger\hat{\sigma}\left( \hat{b}_{\textrm{out,c}}^{(j)}  + \sqrt{\Gamma_j}(1-x)\frac{\hat{\Omega}}{2\kappa'}\right)\rangle\\[0.3cm]
      & \langle \hat{\sigma}\rangle _{\textrm{|CC,}j} = \frac{1}{\langle \hat{b}_{\textrm{out}}^{(j),\dagger} \hat{b}_{\textrm{out}}^{(j)}\rangle }\Bigg[-\sqrt{\Gamma_j^*}\langle\hat{\sigma}^\dagger\hat{\sigma}\left( \hat{b}_{\textrm{out,c}}^{(j)}  + \sqrt{\Gamma_j}(1-x)\frac{\hat{\Omega}}{2\kappa'}\right)\rangle
      +\langle\left( \hat{b}_{\textrm{out,c}}^{(j)}  - \sqrt{\Gamma_j}(1+x)\frac{\hat{\Omega}}{2\kappa'}\right)^\dagger\hat{\sigma}\left( \hat{b}_{\textrm{out,c}}^{(j)}  + \sqrt{\Gamma_j}(1-x)\frac{\hat{\Omega}}{2\kappa'}\right)\rangle\Bigg]
    \end{cases}
\end{equation}
\end{widetext}
In the case where only coherent or vacuum inputs are used in all input ports, the operators $\hat{b}_{\rm out,c}^{(j)}$ and $\hat{\Omega}$ are fully decorrelated from the TLS so the conditional population and coherence simplify to:

\begin{equation}
     \begin{cases}
       & \langle \hat{\sigma}^\dagger\hat{\sigma}\rangle _{\textrm{|CC,}j} = \frac{\langle \hat{\sigma}^\dagger\hat{\sigma}\rangle}{\langle \hat{b}_{\textrm{out}}^{\textrm{(j)},\dagger} \hat{b}_{\textrm{out}}^{\textrm{(j)}}\rangle }\left|\langle \hat{b}_{\textrm{out,c}}^{(j)}  + \sqrt{\Gamma_j}(1-x)\frac{\hat{\Omega}}{2\kappa'}\rangle\right|^2\\
      & \langle \hat{\sigma}\rangle _{\textrm{|CC,}j} = \frac{1}{\langle \hat{b}_{\textrm{out}}^{\textrm{(j)},\dagger} \hat{b}_{\textrm{out}}^{\textrm{(j)}}\rangle }\big[-\sqrt{\Gamma_j^*}\langle\hat{\sigma}^\dagger\hat{\sigma}\rangle+\langle\big( \hat{b}_{\textrm{out,c}}^{(j)}  \\
      &- \sqrt{\Gamma_j}(1+x)\frac{\hat{\Omega}}{2\kappa'}\big)^\dagger\rangle\langle\hat{\sigma}\rangle\big]\langle \hat{b}_{\textrm{out,c}}^{(j)}  + \sqrt{\Gamma_j}(1-x)\frac{\hat{\Omega}}{2\kappa'}\rangle
    \end{cases}\label{rho-tilde-expression-CC}
\end{equation}

Though these conditional quantities can be applied starting from any density matrix $\rho$, at any time, we will now focus on the case of a coherent CW excitation, in the stationary regime where the population and coherence before the back action are taken as $\langle\hat{\sigma}^\dagger\hat{\sigma}\rangle_{\rm st}$ and $\langle\hat{\sigma}\rangle_{\rm st}$, previously computed in Eqs.~\eqref{steady-states}.

\subsection{Second order coherences}

We now compute the second-order auto-correlation function for both the AC and CC ports, under a coherent CW excitation. This correlation function is defined for the AC port $l$ as:
\begin{align}\label{def-g-2-port-AC}
         g_{\textrm{AC,}l}^{\textrm{(2)}}(\tau)&=\frac{\langle\hat{c}^{(l)\dagger}_{\textrm{out}}\hat{c}^{(l)\dagger}_{\textrm{out}}(\tau) \hat{c}^{(l)}_{\textrm{out}}(\tau)\hat{c}^{(l)}_{\textrm{out}}\rangle_{\textrm{st}}}{\langle\hat{c}^{(l)\dagger}_{\textrm{out}}\hat{c}^{(l)}_{\textrm{out}}\rangle_{\rm st}^2}
\end{align}
wich can be connected to the density matrix conditioned to a photon detection in an atom-coupled port $l$ as:
\begin{align}
        &\langle\hat{c}^{(l)\dagger}_{\textrm{out}}\hat{c}^{(l)\dagger}_{\textrm{out}}(\tau) \hat{c}^{(l)}_{\textrm{out}}(\tau)\hat{c}^{(l)}_{\textrm{out}}\rangle=\textrm{Tr}(\rho_{st}\hat{c}^{(l)\dagger}_{\textrm{out}}\hat{c}^{(l)\dagger}_{\textrm{out}}(\tau) \hat{c}^{(l)}_{\textrm{out}}(\tau)\hat{c}^{(l)}_{\textrm{out}})\nonumber\\
        &=\langle\hat{c}^{(l)\dagger}_{\textrm{out}}\hat{c}^{(l)}_{\textrm{out}}\rangle\textrm{Tr}(\rho_{\textrm{|AC,}l}\hat{c}^{(l)\dagger}_{\textrm{out}}(\tau) \hat{c}^{(l)}_{\textrm{out}}(\tau))\nonumber\\
       &=\langle\hat{c}^{(l)\dagger}_{\textrm{out}} \hat{c}^{(l)}_{\textrm{out}}\rangle\left. \langle\hat{c}_{\textrm{out}}^{(l)\dagger}(\tau) \hat{c}^{(l)}_{\textrm{out}}(\tau)\rangle\right|_{\textrm{AC,}l}
\end{align}
With $\rho_{\rm st}$ the density matrix of the system in the stationary regime. This leads to a new expression of the second-order correlation function:
\begin{align}\label{g-2-def-back-action-AC}
        g_{\textrm{AC},l}^{\textrm{(2)}}(\tau) = \frac{\left. \langle\hat{c}_{\textrm{out}}^{(l)\dagger}(\tau) \hat{c}^{(l)}_{\textrm{out}}(\tau)\rangle\right|_{\textrm{AC,}l}}{\langle\hat{c}^{(l)\dagger}_{\textrm{out}}\hat{c}^{(l)}_{\textrm{out}}\rangle_{\rm st}} 
\end{align}
In this expression, the numerator $\left. \langle\hat{c}_{\textrm{out}}^{(l)\dagger}(\tau) \hat{c}^{(l)}_{\textrm{out}}(\tau)\rangle\right|_{\textrm{AC,}l}$ represents the conditional photon flux in the AC port $l$, at a delay $\tau$ after the conditioning detection event in the same port. It can be inferred by accounting for the system's evolution, starting from the conditional density matrix $\rho_{|\textrm{AC},l}$ deduced from Eq.~\eqref{rho-tilde-expression-AC-general}. Noting that the general expressions for the evolution of $\langle\hat{\sigma}^\dagger\hat{\sigma}\rangle(t)$ and $\langle\hat{\sigma}\rangle(t)$ have already been defined in Eq.~\eqref{Bloch-solution}, and that $\hat{c}_{\rm out}^{(l)}=\hat{c}_{\rm in}+\sqrt{\gamma_l}\hat{\sigma}$, one can derive:
\begin{align}\label{cdag-c-evolution}
       &\left. \langle\hat{c}_{\textrm{out}}^{(l)\dagger}(\tau) \hat{c}^{(l)}_{\textrm{out}}(\tau)\rangle\right|_{\textrm{AC,}l}=\alpha_\textrm{1}^{(l)}(\tau)\left(\langle\hat{\sigma}^\dagger\hat{\sigma}\rangle_{\textrm{|AC},l}-\langle\hat{\sigma}^\dagger\hat{\sigma}\rangle_{\rm st}\right) \nonumber\\
        &\quad+\alpha_2^{(l)}(\tau)\left(\langle\hat{\sigma}^\dagger\rangle_{\textrm{| AC},l}-\langle\hat{\sigma}^\dagger\rangle_{\textrm{st}}\right)
        +\alpha_3^{(l)}(\tau)\left(\langle\hat{\sigma}\rangle_{\textrm{| AC,}l}-\langle\hat{\sigma}\rangle_{\rm st}\right) \nonumber\\
       &\qquad+\langle\hat{c}_{\textrm{out}}^{(l)\dagger} \hat{c}_{\textrm{out}}^{(l)}\rangle_{\textrm{st}}
\end{align}
with:
\begin{align}\label{alpha-i-label}
        \alpha_1^{(l)}(\tau) & = 2\sqrt{\gamma_{l}}\textrm{Re}\left(\langle \hat{c}_{\rm in}^{(l)}\rangle_{\rm st}A_1(\tau)\right)+\gamma_{l}B_i(\tau)\\
        \alpha_2^{(l)}(\tau) & = \sqrt{\gamma_{l}}\left(\langle \hat{c}_{\rm in}^{(l)\dagger}\rangle_{\rm st}A_2(\tau)+A_3^*(\tau)\langle \hat{c}_{\rm in}^{(l)}\rangle_{\rm st}\right)+\gamma_{l}B_2(\tau)\nonumber \\
        \alpha_3^{(l)}(\tau) & = \alpha_2^{(l)*}(\tau) \nonumber
\end{align}
This expression holds in the general case, yet it is only in the resonant case that the $\alpha_i^{(l)}(\tau)$ can be expressed in a simple form by using Eqs.~\eqref{A-i-sol} and~\eqref{B-i-sol}. Still, in the general case, we know that at zero delay the only non-zero coefficients are $A_3(0)=B_1(0)=1$, thus $ \alpha_1^{(l)}(0) = \gamma_{l}$, $\alpha_2^{(l)}(0) = \sqrt{\gamma_{l}}\langle \hat{c}_{\rm in}^{(l)}\rangle_{\rm st}$ and $\alpha_3^{(l)}(0)  = \alpha_2^{(l)*}(0)$. This allows expressing the second-order correlation function at zero delay as:
\begin{align}
       &g^{\rm (2)}_{\textrm{AC},l}(0)=1+\frac{1}{\langle\hat{c}^{(l)\dagger}_{\textrm{out}}\hat{c}^{(l)}_{\textrm{out}}\rangle_{\rm st}}\Bigg[\gamma_{l}\left(\langle\hat{\sigma}^\dagger\hat{\sigma}\rangle_{\textrm{|AC},l}-\langle\hat{\sigma}^\dagger\hat{\sigma}\rangle_{\rm st}\right)\nonumber\\ 
       &+\sqrt{\gamma_{l}}\left(\langle \hat{c}_{\rm in}^{(l)}\rangle_{\rm st}\left(\langle\hat{\sigma}^\dagger\rangle_{\textrm{| AC},l}-\langle\hat{\sigma}^\dagger\rangle_{\textrm{st}}\right)+\rm c.c\right)\Bigg]
\end{align}
We note that computing $\frac{\langle\hat{c}^{(l)\dagger}_{\textrm{out}}\hat{c}^{(l)\dagger}_{\textrm{out}} \hat{c}^{(l)}_{\textrm{out}}\hat{c}^{(l)}_{\textrm{out}}\rangle_{\textrm{st}}}{\langle\hat{c}^{(l)\dagger}_{\textrm{out}}\hat{c}^{(l)}_{\textrm{out}}\rangle_{\rm st}^2}$ yields the same analytical expression. The zero-delay second-order correlation can exhibit either bunching ($g_{\textrm{AC,}l}^{\textrm{(2)}}(0)>1$) or anti-bunching ($g_{\textrm{AC,}l}^{\textrm{(2)}}(0)<1$), depending on the value of $\langle\hat{\sigma}^\dagger\hat{\sigma}\rangle_{\textrm{| AC,}l}-\langle\hat{\sigma}^\dagger\hat{\sigma}\rangle_{\rm st}$ and $\langle\hat{\sigma}^\dagger\rangle_{\textrm{| AC,}l}-\langle\hat{\sigma}^\dagger\rangle_{\textrm{st}}$, that is, depending strictly on the back-action induced by the detection of a first photon.

We now perform a similar analysis for a CC port $j$. Using Eq.~\eqref{rho-tilde-def} we write:
\begin{align}\label{g-2-def-back-action-CC}
        g_{\textrm{CC,}j}^{\textrm{(2)}}(\tau) = \frac{\left. \langle\hat{b}_{\textrm{out}}^{(j)\dagger}(\tau) \hat{b}^{(j)}_{\textrm{out}}(\tau)\rangle\right|_{\textrm{CC,}j}}{\langle\hat{b}_{\textrm{out}}^{ (j)\dagger} \hat{b}_{\textrm{out}}^{ (j)}\rangle_{\textrm{st}}} 
\end{align}
To explicit this quantity, we use the effective $\cancel{\rm C}$QED input-output relations given by Eq.~\eqref{effective-input-output} and compute $ \langle\hat{b}_{\textrm{out}}^{(j)\dagger}(\tau) \hat{b}^{(j)}_{\textrm{out}}(\tau)\rangle$, then, we infer its evolution from the expressions~\eqref{sigma_prime_evolution}. Taking into account the back-action induced by a first photon detection in the CC port $j$, Eq.~\eqref{rho-tilde-expression-CC}, then yields:
\begin{widetext}
\begin{align}\label{bdag-b-evolution}
       \left. \langle\hat{b}_{\textrm{out}}^{(j)\dagger}(\tau) \hat{b}^{(j)}_{\textrm{out}}(\tau)\rangle\right|_{\textrm{CC,}l}&=\beta_1^{(j)}(\tau)(\langle\hat{\sigma}^\dagger\hat{\sigma}\rangle_{\textrm{| CC,}j}-\langle\hat{\sigma}^\dagger\hat{\sigma}\rangle_{\rm st})
        +\beta_2^{(j)}(\tau)(\langle\hat{\sigma}^\dagger\rangle_{\textrm{| CC,}j}-\langle\hat{\sigma}^\dagger\rangle_{\rm st})
        +\beta_3^{(j)}(\tau)(\langle\hat{\sigma}\rangle_{\textrm{| CC,}j}-\langle\hat{\sigma}\rangle_{\rm st})\nonumber\\
      &\quad +\langle\hat{b}_{\textrm{out}}^{(j)\dagger} \hat{b}_{\textrm{out}}^{(j)}\rangle_{\textrm{st}}
\end{align}
\end{widetext}
with:
\begin{align}\label{beta-i-def}
         \beta_1^{(j)}(\tau)& =-2Re\left(\sqrt{\Gamma_j}\langle\hat{b}_{\textrm{out,c}}^{(j)\dagger}\rangle C_1(\tau)\right)+\left|\Gamma_{j}\right|D_{1}(\tau)\nonumber\\
         \beta_2^{(j)}(\tau)& =-\sqrt{\Gamma_j}\langle\hat{b}_{\textrm{out,c}}^{(j)\dagger}\rangle C_2(\tau)-\sqrt{\Gamma_j^*}\langle\hat{b}_{\textrm{out,c}}^{(j)}\rangle C_3^*(\tau)\nonumber\\
        &\quad+\left|\Gamma_{j}\right|D_{2}(\tau)\nonumber\\
         \beta_3^{(j)}(\tau)&= \beta_2^{(j)*}(\tau)\raisetag{-0.4cm}
\end{align}
where the $C_i(\tau)$ and $D_i(\tau)$ coefficients are expressed as a function of $A_i(\tau)$ and $B_i(\tau)$ using Eqs.~\eqref{C-i-expr},~\eqref{D-i-expr}. In the general case, we thus know that at zero delay the only non-zero coefficients are: 

\begin{align}
         \beta_1^{(j)}(0)& = 2Re\left(\sqrt{\Gamma_j}\langle\hat{b}_{\textrm{out,c}}^{(j)\dagger}\rangle\frac{\Omega}{\kappa'} \right)+\left|\Gamma_{j}\right|\left(1-\left|\frac{\Omega}{\kappa'}\right|^2\textrm{Re}(x)\right)\nonumber\\
            \beta_2^{(j)}(0)& =-\sqrt{\Gamma_j^*}\langle\hat{b}_{\textrm{out,c}}^{(j)}\rangle+\left|\Gamma_{j}\right|\frac{\Omega(1+x)}{2\kappa'}\\
         \beta_3^{(j)}(0)&= \beta_2^{(j)*}(0)\nonumber
\end{align}

This gives access to an analytical formula for the second order correlation function at zero delay for the light scattered in the cavity-coupled port $j$:
\begin{widetext}
    \begin{align}
         g_{\textrm{CC,}j}^{\textrm{(2)}}(0) &= 1+\frac{1}{\langle\hat{b}_{\textrm{out}}^{ (j)\dagger} \hat{b}_{\textrm{out}}^{ (j)}\rangle_{\textrm{st}}}\Bigg\{\left[2\textrm{Re}\left(\sqrt{\Gamma_j}\langle\hat{b}_{\textrm{out,c}}^{(j)\dagger}\rangle\frac{\Omega}{\kappa'} \right)+\left|\Gamma_{j}\right|\left(1-\left|\frac{\Omega}{\kappa'}\right|^2\textrm{Re}(x)\right)\right]\left(\langle\hat{\sigma}^\dagger\hat{\sigma}\rangle_{\textrm{| CC,}j}-\langle\hat{\sigma}^\dagger\hat{\sigma}\rangle_{\rm st}\right)\nonumber\\
         &\quad+\left[\left(-\sqrt{\Gamma_j^*}\langle\hat{b}_{\textrm{out,c}}^{(j)}\rangle+\left|\Gamma_{j}\right|\frac{\Omega(1+x)}{2\kappa'}\right)\left(\langle\hat{\sigma}^\dagger\rangle_{\textrm{| CC,}j}-\langle\hat{\sigma}^\dagger\rangle_{\rm st}\right)+\rm c.c\right]\Bigg\}
\end{align} 
\end{widetext}

The second order auto-correlation function can display a strong bunching depending on the interference between the empty cavity output field described by $\langle\hat{b}_{\textrm{out,c}}^{(j)}\rangle$ and the atom-induced optical response described by the term $-\sqrt{\Gamma_j}\hat{\sigma}'$. This potentially strong bunching effect will be illustrated in the next subsection.

\subsection{Practical case: resonant excitation from a single cavity port}\label{Sec:practical-case-g2}

We now study a practical case considering the same conditions as in Sec.~\ref{Sec:practical-case-intensity} and Sec.~\ref{Sec:practical-case-SDF}. We consider a single, CW coherent drive incident on the first CC port, resonant with the effective atom frequency: $\omega_{\rm las}=\omega_{\rm a'}$. In such regime, the time-dependent coefficients $A_i$ and $B_i$ have been analytically derived, from which the other coefficients $C_i,D_i$ (see Eq.~\eqref{C-i-expr},~\eqref{D-i-expr}) and $\alpha_i,\beta_i$ (see Eqs.~\eqref{alpha-i-label} and~\eqref{beta-i-def}) are analytically deduced, hence also the second-order auto-correlation functions. In the following, we focus on two auto-correlation functions which display a qualitatively different behavior: the one associated to the first CC output port (reflected light), $g_{\rm CC,1}^{(2)}(\tau)$, and the one associated to any atom-coupled output port, $g_{\textrm{AC,}l}^{(2)}(\tau)$.

\begin{figure*}[t]
    \centering
    \makebox[\textwidth][c]{\includegraphics[width=2\columnwidth]{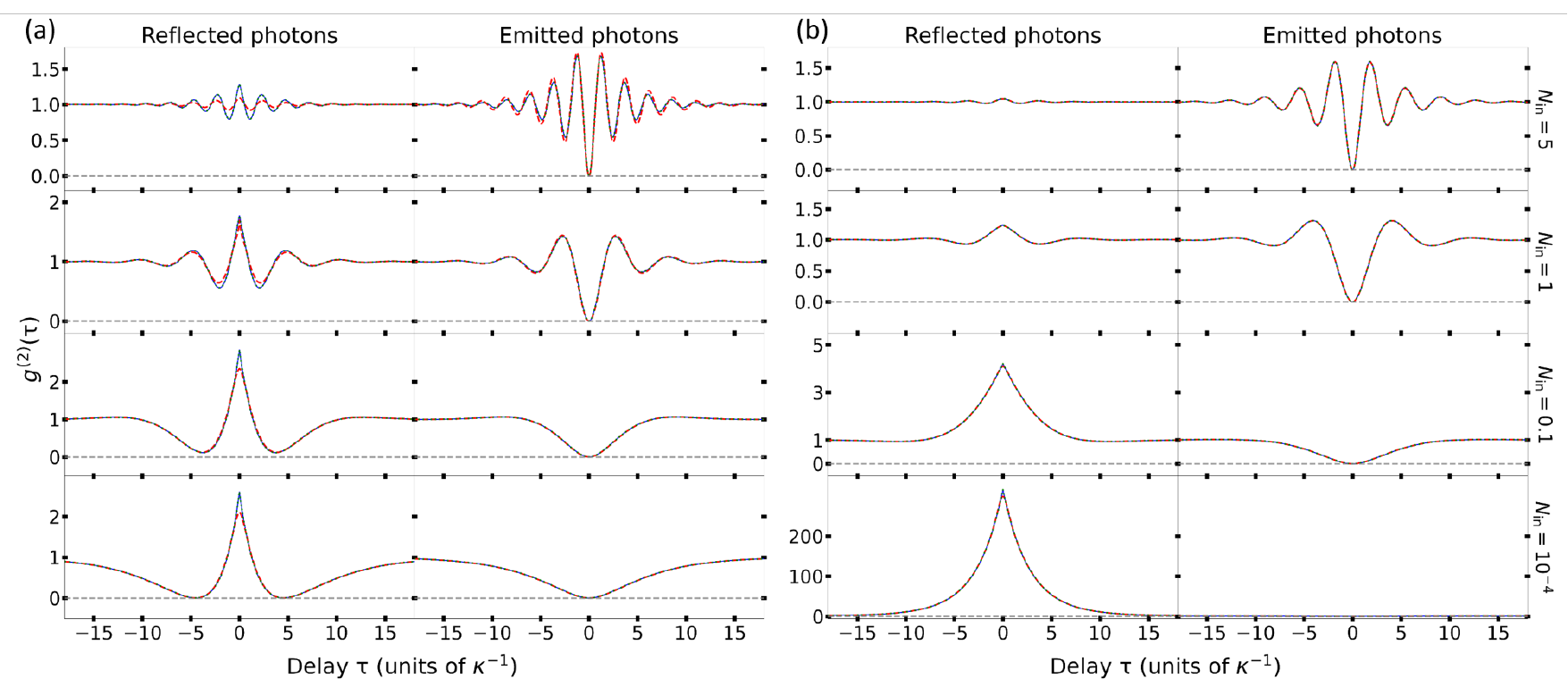}}
    \caption{Second-order auto-correlation functions $g^{(2)}_{\rm CC,1}$ (reflected photons, left columns) and $g^{(2)}_{\textrm{AC},l}$ (emitted photons, right columns). The full CQED simulations (dotted red) are compared with the analytical $\textrm{\cancel{C}QED}$ calculations (blue) and the numerical $\textrm{\cancel{C}QED}$ simulations (dotted green). Each row corresponds to a different incoming photon flux $\phi_{\rm in}$ with (from the top to the bottom row): $N_{\rm in}$=5; 1; 0.1 and $10^{-4}$ respectively. (a) First set of parameters in Table.~\ref{table-of-parameters}, with $C=40$ and $\omega_{\rm a}=\omega_{\rm c}$. (b) Second set of parameters in Table.~\ref{table-of-parameters}, with $C=1$ and $\omega_{\rm a}-\omega_{\rm c}=0.5\kappa$.}\label{fig-g-2}
\end{figure*}

As before, Fig.~\ref{fig-g-2} compares the analytical calculations and the numerical $\textrm{\cancel{C}}$QED approach with the simulations performed with the full CQED model. Fig.~\ref{fig-g-2}.(a) (resp. Fig.~\ref{fig-g-2}.(b)) displays the predictions obtained with the first set of parameters (resp. second) in Tab.~\ref{table-of-parameters} with increasing input power from the top to the bottom row. The left column (resp. right column) displays the auto-correlation function associated to the detection of two reflected (resp. emitted) photons, as a function of the delay $\tau$ between two detection events, i.e $g^{(2)}_{\rm CC,1}(\tau)$ (resp. $g^{(2)}_{\textrm{AC,}l}(\tau)$).

We first discuss the predictions in the right column, corresponding to $g^{(2)}_{\textrm{AC,} l}(\tau)$, which shows the characteristics of a second-order correlation function measured for a TLS~\cite{NgBoon2022,Kimble1976}. At zero delay, the two level system has just emitted a photon and is consequently in the ground state, leading to a zero probability of detecting a second photon and thus to a strong anti-bunching. At low power, $g_{\textrm{AC,}l}^{(2)}(\tau)$ monotonously increases to unity with a timescale governed by $\Gamma^{-1}$. When increasing the excitation power, Rabi oscillations occur between the excited and ground states, leading to an oscillating probability of emitting a second photon, and thus, to an oscillation of $g^{(2)}_{\textrm{AC,}l}(\tau)$ at a timescale governed by the inverse of the Rabi frequency $\Omega$. At long delays, the conditional density matrix $\rho_{|\textrm{AC,} l}$ evolves back to the stationary density matrix and $g_{\textrm{AC,}l}^{(2)}(\tau)$ gets back to unity.

For the reflected photons, on the contrary, $g^{(2)}_{\textrm{CC,}1}(\tau)$ exhibits a clear bunching effect at zero delay. Such bunching effects have been observed experimentally, e.g in the case of single atoms inside a cavity~\cite{Kubanek2008,Tomm2024}, and in waveguide-QED systems~\cite{LeJeannic2021}. They have also been theoretically investigated for CQED devices, e.g. in Ref.~\cite{Shi2011,SanchezMunoz2014}. This bunching effect arises from the interference between the reflected laser and the atom-resonance fluorescence escaping from the port 1, potentially decreasing the probability of reflecting one photon (and the average flux $\langle \hat{b}_{\rm out}^{(1)\dagger}\hat{b}_{\rm out}^{(1)}\rangle$), while the probability of reflecting two photons is relatively less decreased (compared to the decrease that would be obtained with a linear optical loss yielding the same reflected flux). In the full CQED framework, this interference is seen through the input-output relation $\hat{b}_{\rm out}^{(1)} = \hat{b}_{\rm in}^{(1)}+\sqrt{\kappa_1}\hat{a}$: the light directly reflected by the cavity mirror interferes with the light that entered the cavity and interacted with the atom. An interesting feature of the self-consistent cavity elimination approach is that this interference can be more directly interpreted using the effective input-output relation $\hat{b}_{\rm out}^{(j)}= \hat{b}_{\rm out,c}^{(j)}-\sqrt{\Gamma_j}\hat{\sigma}'$, where the empty cavity response directly interferes with the atom-induced response.

As can be seen in the left columns of both Fig.~\ref{fig-g-2}(a) and~\ref{fig-g-2}(b), oscillations also appear in the auto-correlation function $g_{|\textrm{CC,}1}^{(2)}(\tau)$ as the power is increased, which reflect the Rabi oscillations of the effective atom state in its evolution from the conditional density matrix $\rho_{|\rm CC,1}$ to the stationary regime. The contrast of the zero-delay bunching and of the corresponding oscillations, however, decreases with increasing power, exhibiting the saturation of the atom-induced optical response. Indeed, when the atom resonance fluorescence saturates, the reflected light is dominated by the empty-cavity response described by $\hat{b}_{\rm out,c}^{(1)}$, yielding a negligible interference effect and a mostly coherent reflected light, with $g_{\rm CC,1}^{(2)}(\tau)\approx 1$. We note that Fig.~\ref{fig-g-2}(b) displays a higher bunching at low input power than Fig.~\ref{fig-g-2}(a). This arises from the fact that the interference is highly destructive for the reflected field at low power, when $\omega_{\rm las}=\omega_{\rm a'}$, for the second set of parameters in Tab.~\ref{table-of-parameters} (as can be seen in the low fraction of reflected light at the atom-laser resonance in Fig.~\ref{fig-flux-weak-2}).

The $\cancel{\rm C}$QED analytical, exactly matching the $\cancel{\rm C}$QED numerical simulations, display a quantitatively good agreement with the full CQED model simulations. However, we note that this agreement is not perfect at short delays (i.e $t\approx\kappa$), even at low power. This is a consequence of the long-time approximation $t\gg\kappa$ required for the cavity to evolve from its initial state and couple to the atom (see Eqs.~\eqref{eq-a-with-time-approx} and~\eqref{Integrated_langevin_eq}). Additionally, the oscillations are not perfectly reproduced at high power for the first set of parameters, a regime where the atom dynamics is not fully described by the optical Bloch equations of the reduced model, as already illustrated in Figs.~\ref{fig-flux-weak-1} and~\ref{fig-SDF} at the highest excitation powers.

\section{Current limitations of the self-consistent cavity elimination approach}\label{Chapter_8}
In the previous sections, we have obtained analytical and numerical predictions in good agreement with numerical simulations of the full CQED model, in the weak coupling regime. It is only natural to test the limits of the approach at or above the threshold of the strong-coupling regime, as will be illustrated in subsection~\ref{Sec:plot-strong}. Before that, we discuss in subsection~\ref{Sec:strong-discussion} the issues that should be typically encountered when applying the self-consistent cavity–elimination approach to strongly-coupled systems. We show that the current limitations of the self-consistent cavity-elimination model are directly linked to the validity of the important approximation made in subsection~\ref{Sec:approximation}, i.e. the approximation of a slowly-evolving effective drive.

\subsection{Beyond the weak-coupling regime: expected limitations.}\label{Sec:strong-discussion}
The idea of treating a strongly-coupled CQED device with a reduced model, in a two-level Hilbert space, seems entirely in contradiction with the necessity to describe effects associated to the structure of the Jaynes-Cummings ladder~\cite{Fink2008,Shore1993,Larson2024}. As such, the self-consistent cavity elimination approach has initially been designed strictly for operation in the weak-coupling regime. This is visible in the chosen ansatz of Eq.~\eqref{Ansatz},  i.e. $\Dot{\hat{\sigma}}=-\Gamma'\hat{\sigma}+\hat{F}$. At first sight, indeed, this ansatz considers only a mono-exponential dynamics governed by the value of the effective damping rate $\Gamma’$. As regards the effective drive operator $\hat{F}$, it seems well adapted to play the role of a Langevin force~\cite{Gardiner2004}, associated to the interaction with the various ports, with the expectation that $\langle \hat{F}\rangle=0$ if all input ports are only driven by vacuum. 

This is actually the case considering Eq.~\eqref{F-expression}, i.e. the expression $\hat{F}= (\hat{\sigma}_z-x\hat{\mathbb{I}})\frac{\hat{\Omega}}{2}$ with $\hat{\Omega}$ an operator gathering all inputs. Indeed, $\hat{\Omega}$ verifies $\hat{\Omega}\ket{0}_E= 0$ and, correspondingly $\bra{0}_E\hat{\Omega}^\dagger = 0$, with $\ket{0}_E$ corresponding to the environment in the vacuum state for all input ports. Therefore, since $\dot{\hat{\sigma}}=-\Gamma'\hat{\sigma}+\hat{F}$ and $\dot{\hat{\sigma}}_z=-\Gamma(\hat{\sigma}_z+\hat{\mathbb{I}})+\hat{F}^\dagger\hat{\sigma}+\hat{\sigma}^\dagger\hat{F}$, this leads to $\langle\dot{\hat{\sigma}}\rangle =-\Gamma'\langle\hat{\sigma}\rangle$ and $\langle\dot{\hat{\sigma}}_z\rangle=-\Gamma(\langle\hat{\sigma}_z\rangle+1)$ in the case of a vacuum input in all ports. Such situation implies a strictly mono-exponential decay with radiative lifetime $\Gamma$, hence no vacuum Rabi oscillations.

This argument is only valid, however, through the fact that the input operators $\hat{b}_{\rm in}^{(j)}$ and $\hat{c}_{\rm in}^{(l)}$ appear, via the operator $\hat{\Omega}$, on the right-hand side in the expression $\hat{F} = (\hat{\sigma}_z-x\hat{\mathbb{I}})\frac{\hat{\Omega}}{2}$. Had they not, the equality $\hat{F}\ket{0}_E=0$ would not be fulfilled and the expected dynamics would naturally go beyond the monoexponential behavior, for a vacuum input in all ports. Since the expression $\hat{F} = (\hat{\sigma}_z-x\hat{\mathbb{I}})\frac{\hat{\Omega}}{2}$ directly follows from the approximation of a slowly-evolving effective drive, i.e $\hat{F}-\frac{\Dot{\hat{F}}}{\kappa'}+\frac{\Ddot{\hat{F}}}{\kappa^{'2}}\approx\hat{F}$, it is only natural that more complex dynamics could be described by releasing, even partially, this approximation. We remind, indeed, that the complete self-consistency equation~\eqref{Self-consistency-eq} allows exactly verifying the Heisenberg-Langevin equations~\eqref{Heisenberg-Langevin system} in the limit $\kappa t\gg 1$. Releasing this approximation thus constitutes a clear path for future research, beyond the scope of the present paper.

An encouraging feature, for the proposed cavity elimination approach, lies in the fact that the second-order self-consistency equation for $\Gamma’$, Eq.~\eqref{Gamma-prime-identification-eq}, can be taken as exact: it yields two solutions $\Gamma'_-$ and $\Gamma’_+$ which entirely describe the system’s optical response in the low-power limit, even far in the strong coupling regime (where $\Gamma’_-$ and $\Gamma’_+$ describe the lower and upper polariton branches). This has been shown in Sec.~\ref{Subsec:Semiclassical}, where we have verified that the computed intensities in all output ports lead to the exact same CQED and $\cancel{\rm C}$QED predictions, in the case of a continuous-wave excitation in the low-power limit, where the semiclassal approximation $\hat{\sigma}_z=-1$ applies. 

Such a result, which could initially come as a surprise, is actually perfectly compatible with the validity range of the slowly-evolving effective drive approximation, $\hat{F}-\frac{\Dot{\hat{F}}}{\kappa'}+\frac{\Ddot{\hat{F}}}{\kappa^{'2}}\approx\hat{F}$. Within this approximation, one finds that  $\hat{F} = (\hat{\sigma}_z-x\hat{\mathbb{I}})\frac{\hat{\Omega}}{2}$ and thus $\Dot{\hat{F}} = \Dot{\hat{\sigma}}_z\frac{\hat{\Omega}}{2}+(\hat{\sigma}_z-x\hat{\mathbb{I}})\frac{\Dot{\hat{\Omega}}}{2}$. In the specific case of a continuous-wave excitation, considering the same excitation frequency in all input ports, one can choose the reference frequency $\omega_{\rm ref}$ to match the excitation frequency and ensure that $\Dot{\hat{\Omega}} =0$ exactly. In addition, in the low-power limit where $\hat{\sigma}_z$ can be replaced by -1, $\Dot{\hat{\sigma}}_z$ can be replaced by $0$. The combination of both conditions ensures that $\Dot{\hat{F}}$ and all its higher-order derivatives are exactly zero. This allows exactly satisfying, independently from the value of $\kappa’$, the approximation $\hat{F}-\frac{\Dot{\hat{F}}}{\kappa'}+\frac{\Ddot{\hat{F}}}{\kappa^{'2}}\approx\hat{F}$ initially made. Conversely, in the case of pulsed or, more generally, polychromatic excitation, $\Dot{\hat{\Omega}} \ne 0$ and the approximation can not be considered exact. Furthermore, as soon as the excitation power is increased or for any other situation where the system is not initially in the ground state, the semi-classical equation $ \hat{\sigma}_z=-1$ breaks and the approximation $\hat{F}-\frac{\Dot{\hat{F}}}{\kappa'}+\frac{\Ddot{\hat{F}}}{\kappa^{'2}}\approx\hat{F}$ becomes inexact. 

These considerations also allow understanding why good results could still be obtained in the weak-coupling regime, with a CW excitation from a single port, when considering an increased excitation power (see sections~\ref{Chapter_5} to~\ref{Chapter_7}). In such a case where $\Dot{\hat{\Omega}}$ is exactly zero, the only approximation comes from the non-zero value of $\Dot{\hat{\sigma}}_z $. The slowly-evolving effective drive approximation, however, will remain valid if the typical values of $\frac{\Dot{\hat{F}}}{\kappa'}=\frac{\Dot{\hat{\sigma}}_z\hat{\Omega}}{2\kappa'}$ can be neglected in comparison with the typical values of $\hat{F} = (\hat{\sigma}_z-x\hat{\mathbb{I}})\frac{\hat{\Omega}}{2}$. 

In a sufficiently weak coupling regime, the natural decay of the operator $\hat{\sigma}_z$ (in absence of input) is only influenced by the slowest damping rate $\Gamma=\rm \frac{\textrm{Re}(\Gamma’_-)}{2}$, much smaller than $\kappa$. Additionaly, if the excitation power is moderate-enough that $\Omega=\langle\hat{\Omega}\rangle$ remains significantly smaller than $\kappa$, the Rabi oscillations induced on the operator $\hat{\sigma}_z$ will stay much slower than $\kappa$. Typical values of the operator $\frac{\Dot{\hat{\sigma}}_z}{\kappa'}$ will then always remain much smaller than one, even close to saturation where $\hat{\sigma}_z$ approaches zero, in a regime where $\Gamma <\Omega\ll\kappa$. This explains why a quantitative agreement could be obtained with the full CQED model in Fig.~\ref{fig-flux-weak-2}, even at saturation, discrepancies only appearing in Figs.~\ref{fig-SDF} and~\ref{fig-g-2}, at the largest excitation powers, where $\Omega$ becomes of the order of $\kappa$.

For a strongly-coupled CQED system, it appears that the exact same considerations will also apply if the atom is largely-detuned from the cavity, i.e. when $\left|\omega_{\rm c}-\omega_{\rm a}\right|$ is significantly larger than both $\kappa$ and $g$. In such a case, the damping rate $\Gamma=\rm \frac{Re(\Gamma’_-)}{2}$ is also much smaller than $\kappa$, and the approximation $\hat{F}-\frac{\Dot{\hat{F}}}{\kappa'}+\frac{\Ddot{\hat{F}}}{\kappa^{'2}}\approx\hat{F}$ is expected to lead to satisfying results even at high excitation powers. However, for a strongly-coupled system with a low atom-cavity detuning, larger inaccuracies are expected at intermediate or high power. In such a case, indeed, the natural decay of $\hat{\sigma}_z$ (in absence of inputs) is governed by the damping rates Re$(\Gamma’_-)$ and Re$(\Gamma’_+)$ which are both of the same order as $\kappa$. Also, in such strong-coupling regime with a low atom-cavity detuning, where the system’s damping is governed by $\kappa$, reaching intermediate values of $\hat{\sigma}_z$ (between -1 and 0) implies using values of $\Omega$ of the order of $\kappa$. The convergence of all these frequency scales to the damping rate $\kappa$ should thus lead to a complete breaking of the approximation $\hat{F}-\frac{\Dot{\hat{F}}}{\kappa'}+\frac{\Ddot{\hat{F}}}{\kappa^{'2}}\approx\hat{F}$.

To illustrate the limitations of this approximation, we provide in the next subsection analytical and numerical predictions equivalent to those displayed in Figs.~\ref{fig-flux-weak-2},~\ref{fig-SDF} and~\ref{fig-g-2}, yet with parameters corresponding to a CQED system at or beyond the strong-coupling threshold.

\subsection{Analytical and numerical predictions above the strong-coupling threshold.}\label{Sec:plot-strong}

\begin{figure}[t]
    \centering
    \includegraphics[width=\linewidth]{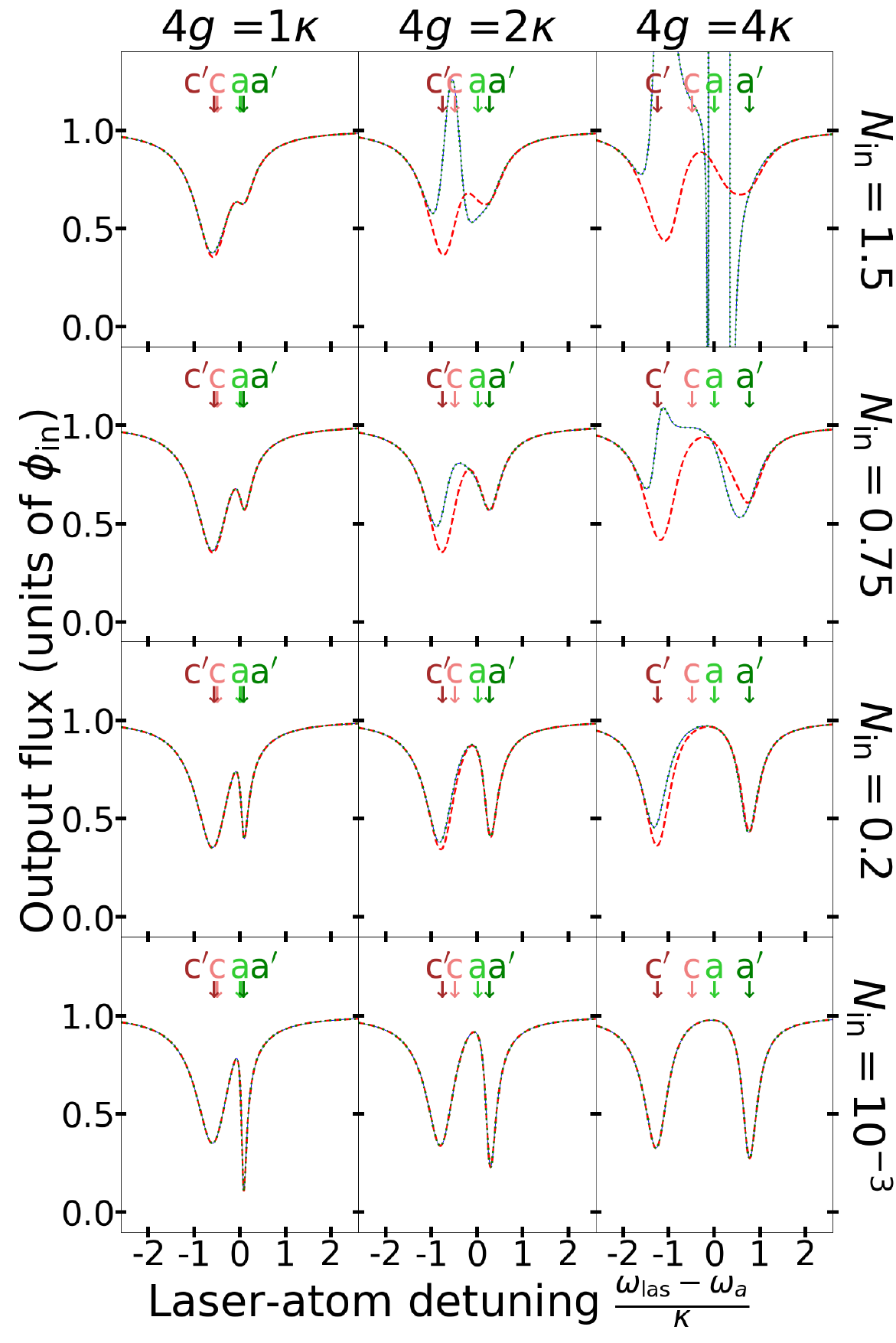}
    \caption{Reflected flux, in units of the input flux $\phi_{\rm in}$, with increasing atom-cavity coupling: $4g=\kappa$ (first column), $4g=2\kappa$ (second column) and $4g=4\kappa$ (third column). The full CQED simulations (dotted red) are compared with the analytical $\textrm{\cancel{C}QED}$ calculations (blue) and the numerical $\textrm{\cancel{C}QED}$ simulations (dotted green). Each row corresponds to a different incoming photon flux $\phi_{\rm in}$ with (from the top to the bottom row): $N_{\rm in}$=1.5; 0.75; 0.2 and $10^{-3}$ respectively. The second set of parameters in Table~\ref{table-of-parameters} has been used, except for the light-matter coupling, with an atom cavity detuning $\omega_{\rm a}-\omega_{\rm c}=0.5\kappa$.}\label{fig:strong_flux}
\end{figure}

\begin{figure}
    \centering
    \includegraphics[width=\linewidth]{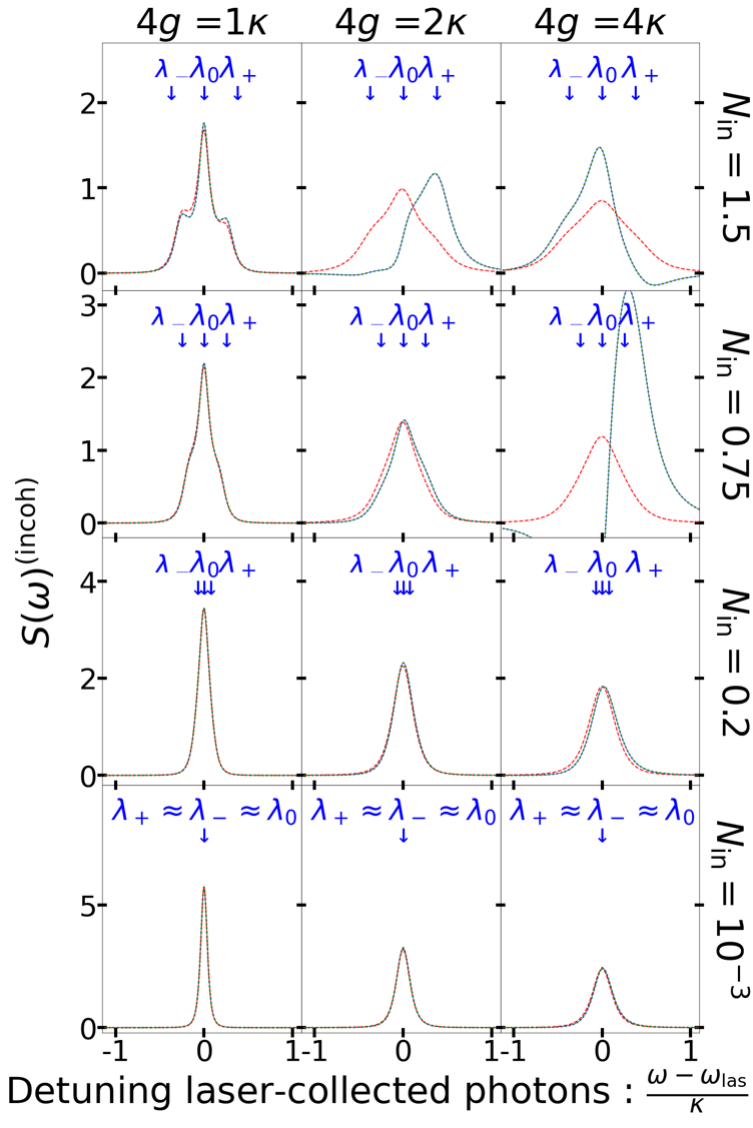}
    \caption{Spectral density of the reflected incoherent light, with increasing atom-cavity coupling: $4g=\kappa$ (first column), $4g=2\kappa$ (second column) and $4g=4\kappa$ (third column). The full CQED simulations (dotted red) are compared with the analytical $\textrm{\cancel{C}QED}$ calculations (blue) and the numerical $\textrm{\cancel{C}QED}$ simulations (dotted green). Each row corresponds to a different incoming photon flux $\phi_{\rm in}$ with (from the top to the bottom row): $N_{\rm in}$=1.5; 0.75; 0.2 and $10^{-3}$ respectively. The second set of parameters in Table~\ref{table-of-parameters} has been used, except for the light-matter coupling, with an atom cavity detuning $\omega_{\rm a}-\omega_{\rm c}=0.5\kappa$.}\label{fig:strong_sdf}
\end{figure}

\begin{figure}
    \centering
    \includegraphics[width=\linewidth]{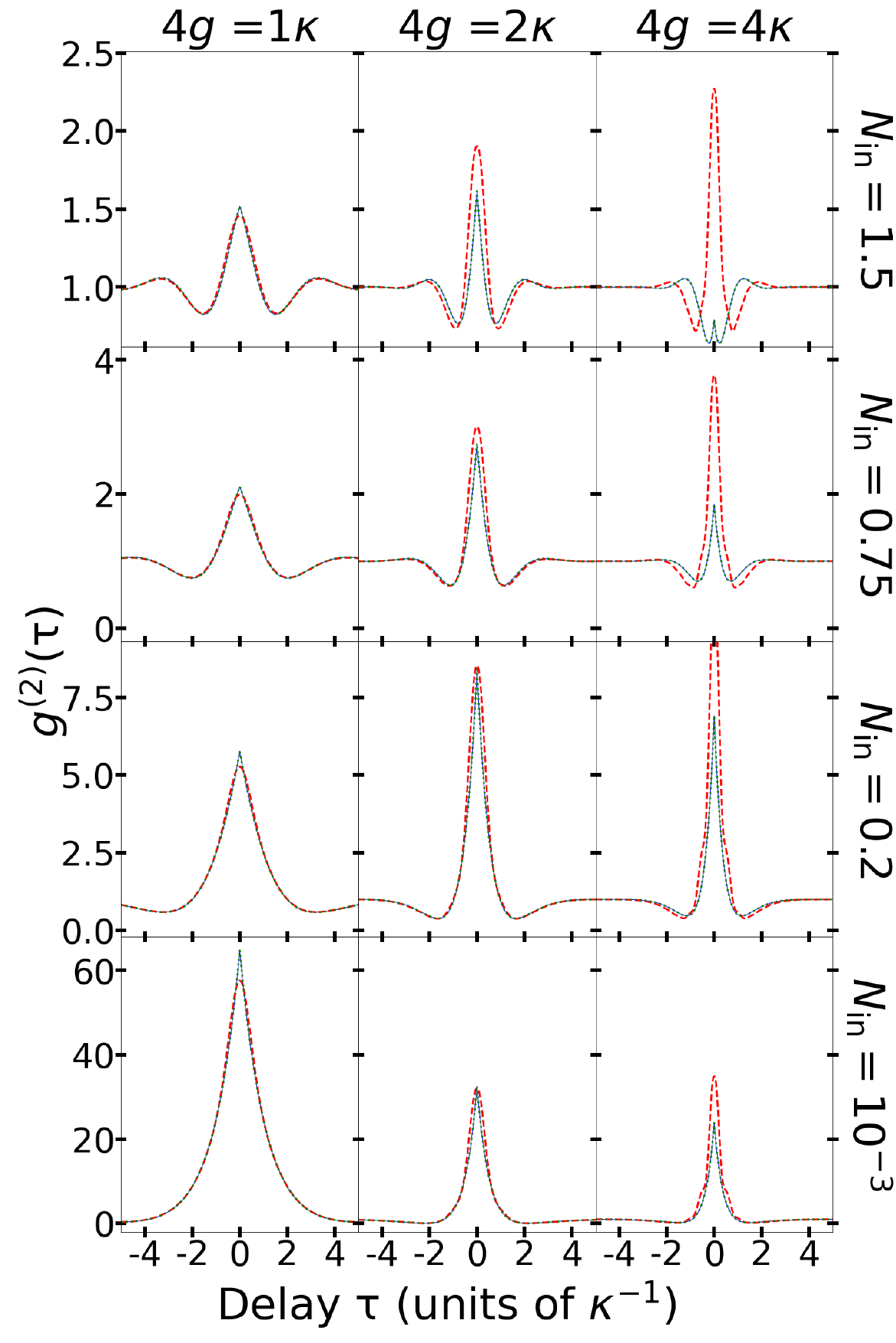}
    \caption{Second-order auto-correlation functions $g^{(2)}_{\rm CC,1}$, with increasing atom-cavity coupling: $4g=\kappa$ (first column), $4g=2\kappa$ (second column) and $4g=4\kappa$ (third column). The full CQED simulations (dotted red) are compared with the analytical $\textrm{\cancel{C}QED}$ calculations (blue) and the numerical $\textrm{\cancel{C}QED}$ simulations (dotted green). Each row corresponds to a different incoming photon flux $\phi_{\rm in}$ with (from the top to the bottom row): $N_{\rm in}$=1.5; 0.75; 0.2 and $10^{-3}$ respectively. The second set of parameters in Table~\ref{table-of-parameters} has been used, except for the light-matter coupling, with an atom cavity detuning $\omega_{\rm a}-\omega_{\rm c}=0.5\kappa$.}\label{fig:g_2_strong}
\end{figure}

In this section, we benchmark the $\cancel{\rm C}$QED model by assuming a single CW excitation from the first cavity-coupled port and comparing the analytical and numerical results of the self-consistent cavity elimination with the numerical simulations of the full CQED model. To do so, we respectively plot in Figs~\ref{fig:strong_flux},~\ref{fig:strong_sdf} and~\ref{fig:g_2_strong} the reflected (according to def.~\eqref{Name-of-ports}) photon flux, spectral density and second order auto-correlation functions. The parameters $\kappa_1 = 0.8\kappa$, $\gamma_{\rm a}=\kappa/32$ and $\omega_{\rm a}-\omega_{\rm c}=0.5\kappa$ were used, as in the second set of parameters of Tab.~\ref{table-of-parameters}. The atom-cavity coupling rate has been varied, taking the values (from the left column to the right one) $g=\frac{\kappa}{4},\frac{\kappa}{2},\kappa$, allowing to test the strong coupling threshold and beyond. In the three figures, each row corresponds to a different driving power: $N_{\rm in}=10^{-3},0.2,0.75,1.5$. 

Fig.~\ref{fig:strong_flux} shows the reflectivity of the CQED device depending on the driving frequency, computed analytically and numerically using the $\cancel{\rm C}$QED approach and numerically using the full CQED model. The reflectivity profile displays the two reflectivity dips expected from strong light-matter interaction~\cite{Fink2008}. These reflectivity dips are centered on the frequencies denoted as $\omega_{\rm a'}$ and $\omega_{\rm c'}$ such that $\omega_{\rm a'}-\omega_{\rm ref}=\rm Im(\Gamma'_-)$ and $\omega_{\rm c'}-\omega_{\rm ref}=\rm Im(\Gamma'_+)$ (their analytical expressions, which do not depend on $\omega_{\rm ref}$, are given in the Appendix~\ref{Subsec-Study-Gamma}). Due to the non-zero damping rates, $\omega_{\rm a'}$ (resp. $\omega_{\rm c'}$) do not match the frequencies $\omega_{\rm u}$ (resp. $\omega_{\rm l}$) of the upper (resp. lower) polariton eigenstates of the Jaynes Cummings Hamiltonian~\eqref{Full-Hamiltonian}. Still, in the present situation where $\omega_{\rm a}>\omega_{\rm c}$, the optical properties of the reflectivity dips are influenced by these respective polariton states, the upper one presenting a more "atom-like" nature and the lower one a more "cavity-like" nature. The reflectivity dip centered on $\omega_{\rm a'}$ is thus the one whose frequency is closer to $\omega_{\rm a}$ and with a narrower spectral width governed by Re($\Gamma'_-$), while the reflectivity dip centered on $\omega_{\rm c'}$ is the one whose frequency is closer to $\omega_{\rm c}$ and with a larger spectral width governed by Re($\Gamma'_+$).

The simulations made at the lowest power (bottom row) show an excellent agreement between the analytical and numerical predictions of the $\cancel{\rm C}$QED model and the simulations made using the full CQED model. This was expected from the fact that the $\cancel{\rm C}$QED approach is exact at low excitation power since it retrieves the predictions of the semi-classical approximation $\hat{\sigma}_z=-1$, as shown in Sec.~\ref{Subsec:Semiclassical}. In such a low excitation regime, where the atom mostly remains in the ground state, the system's optical response is coherent and governed by the complex amplitudes $\langle \hat{b}_{\rm out}^{(j)}\rangle$ and $\langle\hat{c}_{\rm out}^{(l)}\rangle$ given in Eqs.~\eqref{steady-states} and~\eqref{Sigma-prime-steady}. In contrary, as the excitation power increases, the atom leaves the ground state (i.e $\hat{\sigma}_z\ne-1$) inducing Rabi oscillations (i.e $\Dot{\hat{\sigma}}_z\ne0$). In this case, a stronger atom-cavity coupling leads to a smaller $\kappa$ and $\frac{\Dot{\hat{\sigma}}_z}{\kappa}$ is not negligible anymore, which explains the discrepancies between the CQED and $\cancel{\rm C}$QED models at high input power: $\cancel{\rm C}$QED predictions even become unphysical, with a reflectivity higher than one at large excitation powers for the simulations with the largest light-matter coupling. We finally note that there is no clear mark of the strong coupling threshold: instead, the $\cancel{\rm C}$QED predictions gradually lose accuracy, with respect to the CQED model, when $g$ is progressively increased.

Fig.~\ref{fig:strong_sdf} displays the incoherent spectral density computed for the reflected field. It is computed using the second set of parameters of Table~\ref{table-of-parameters} yet with a different light-matter coupling strength $g$ for each column, each row corresponding to a different excitation power. The excitation laser frequency is chosen to be the atom-like effective frequency, $\omega_{\rm las}=\omega_{a'}$, centered on the atom-like reflectivity dip (see Fig.~\ref{fig:strong_flux}). Such resonant excitation allows using the analytical formula of Eq.~\eqref{SDF-cavity-ports} to plot the incoherent part of the reflected spectral density. Still, the $\cancel{\rm C}$QED model is found to give predictions in agreement with the full CQED model at low input power in agreement with the fact that, in such regime, the typical values of $\Dot{\hat{\sigma}}_z$ are much lower than $\kappa$, and with the fact that $\Dot{\hat{\Omega}}=0$ for a monochromatic CW laser excitation at $\omega_{\rm las}=\omega_{\rm ref}$. However, when the power increases, the atom leaves the ground state ($\hat{\sigma}_z\ne -1$) and induces non-negligible Rabi oscillations, especially when the cavity damping rate $\kappa$ is comparable with $g,\Gamma,\Omega$. For a complete analysis of the spectral density behavior in the strong coupling regime, we refer to Ref.~\cite{DelValle2009}.

In Fig.~\ref{fig:g_2_strong}, we plot the second order auto-correlation function for the reflected field, $g_{\textrm{CC,}1}^{\rm (2)}(\tau)$ as a function of the delay $\tau$ between photon detection events. It is computed using the second set of parameters of Tab.~\ref{table-of-parameters}, yet with a different light-matter coupling strength $g$ for each column, and a different input power for each row. We note that the discrepancies at high input power are caused by the approximation $\hat{F}-\frac{\Dot{\hat{F}}}{\kappa'}+\frac{\Ddot{\hat{F}}}{\kappa^{'2}}\approx\hat{F}$ not being valid anymore, leading to unphysical predictions of the $\cancel{\rm C}$QED model. However, Fig.~\ref{fig:g_2_strong} shows an additional discrepancy between the $\cancel{\rm C}$QED approach and the CQED model even at low input power, in the strongest coupling regime. This discrepancy is a combination of two effects:
\begin{itemize}
    \item The system's dynamics becomes faster with increasing atom-cavity coupling, occuring at timescales where $\kappa t\approx1$, thereby violating the long time approximation made to obtain Eq.~\eqref{eq-a-with-time-approx} in Sec.~\ref{Chapter_3}. In such regime, the entire dynamics happen in timescales so short that the cavity did not forget its initial state. Such a regime is outside the scope of our current cavity elimination technique.
    \item Even if at low power the atom stays mainly in the ground state ($\langle \hat{\sigma}_z\rangle_{\rm st}\approx-1$), the detection of a reflected photon projects the atom's state according to Eqs.~\eqref{rho-tilde-expression-CC}. This projection then enables conditional values of $\langle\hat{\sigma}_z\rangle_{|\textrm{CC,}1}\ne -1$, which will evolve back to the stationary value $\langle\hat{\sigma}_z\rangle_{\rm st}$ with a dynamics governed by Re($\Gamma'_+$) and Re($\Gamma'_-$), both of the order of $\kappa$ when strongly-coupled system are considered. Interestingly, though, the values of the auto-correlation function retrieve an agreement, between the $\cancel{\rm C}$QED and CQED predictions at low power, for values of $\kappa t$ significantly above 1.
\end{itemize}

\section{Conclusion}\label{Chapter_9}

In this work, we have introduced a self-consistent cavity elimination technique directly adapted to the input-output description of CQED devices. In this framework, we use Heisenberg-Langevin equations to describe a two-level atom coupled to a single-mode cavity, interacting with their environment through both cavity-coupled and atom-coupled ports. An ansatz of the form $\Dot{\hat{\sigma}} = -\Gamma'\hat{\sigma}+\hat{F}$ is introduced to describe the dynamics of the atom lowering operator $\hat{\sigma}$, with $\Gamma’$ an effective complex damping rate and $\hat{F}$ an effective drive operator. This allows deriving a self-consistency equation fully describing the dynamics of the cavity-QED system at times exceeding the photon lifetime in the cavity. The two eigenvalues describing exactly the cavity-QED dynamics in the single-excitation regime are exactly retrieved in this self-consistent approach, the one corresponding to the slowest damping rate being the best suited to be used as the effective damping rate $\Gamma'$. Moreover, based on the approximation of a slowly-evolving effective drive, a simple expression of the effective drive operator can be derived as a function of the input fields and of the atom operators, hereby providing a reduced model for the effective atom dynamics. Under this approximation, an analytical expression is retrieved for the effective annihilation operator: this allows expressing all input-output relations, for the cavity-coupled ports, as a function of the atom operators only. In addition, the effective dynamics of the atom can be derived through optical Bloch equations for the atom operators, as a function of all the input fields.

Within this general framework, we first studied the low-excitation regime which allows applying a semi-classical treatment, where the population of the excited state is always considered negligible. In such a case, the optical response of the cavity-QED system can be described by scattering amplitudes which can be analytically derived. We find that, in this low-excitation limit and under continuous-wave excitation -- or excitation with long optical pulses compared to the effective atom lifetime -- the scattering amplitudes derived from the reduced model exactly match the ones predicted by directly applying the semi-classical treatment to the full CQED model. This exact correspondence is obtained independently from the strength of the light-matter coupling.

Another specific case of interest is the one where coherent excitation is applied. In such a case, the optical Bloch equations can entirely be described in the Hilbert space of the effective atom: they can then be used both for analytical calculations or fast numerical simulations, using an effective Hamiltonian and two decoherence operators associated to two decoherence rates $\gamma_{\rm M}$ and $\gamma_{\rm NM}$. We find that, while $\gamma_{\rm M}$ is always positive and can be used to describe Markovian processes, $\gamma_{\rm NM}$ is intrinsically negative -- though it tends towards zero in the low-excitation regime. This illustrates the intrinsic non-Markovianity which is expected to arise when considering the cavity as part of the atom’s environment, and when the cavity memory time cannot be neglected compared to the timescales governing the atom dynamics. 

Based on these features, we focused on the regime of coherent CW excitation. In such regime, analytical expressions could be computed for the output photon flux in each port, beyond the low-power excitation regime. We analytically solved the system’s dynamics for resonant excitation, allowing us to compute the first-order coherences and the spectral densities for the emitted/scattered fields through each output port. Finally, we used the effective input-output relations in the reduced model to describe the measurement back-action induced by a photon detection, in any output port, on the effective density matrix describing the atom. This, in turn, allowed expressing the second-order correlation functions for all the output fields.

All of these analytical calculations were checked to give an exact agreement with the numerical simulations performed within the reduced model, in a reduced Hilbert space of dimension 2. Most importantly, they were also shown to provide an excellent agreement with the numerical predictions obtained with the full CQED model, when the atom-cavity coupling is weak, in a broad range of excitation powers. We then extended the analysis to the strong coupling regime, showing a complete agreement between the cavity-eliminated model and the full CQED model, in the low-excitation regime, when computing the output flux and the spectral density of the output field in each port. The approximation of a slowly-evolving effective drive operator happens to be valid in this regime of low-power CW excitation, where the excited state is negligibly populated. When considering second-order correlations, however, the relevant physics is described within a transient regime immediately following the measurement back-action induced by a single photon detection. In the strong coupling regime, the self-consistent cavity elimination approach fails at describing this transient behavior whose dynamics occurs at the timescale of the photon lifetime in the cavity. 

In its present state, the self-consistent cavity elimination approach already provides a very practical tool both for the analytical understanding and for the efficient numerical simulation of various cavity-QED phenomena. The analytical expressions of the output operators, in particular, allow studying experimentally-relevant quantities of all sorts, including, but not limited to, one-time and two-times expectation values. It could also be used to analyze, not only the measurement back-action induced on the atom by the detection of a single photon, but the actual atom-photon entanglement before any photon detection occurs. 

In presence of coherent excitation, the reduced model is also simply implemented numerically, through a few modifications of the full CQED model. Essentially, one has to modify the CQED master equation by using the effective Hamiltonian and decoherence operators and rates introduced in Section~\ref{sub-sec-non-markov}, and to replace the annihilation operator $\hat{a}$ by the effective annihilation operator expressed in Section~\ref{Subsec:effective-input-output}, as a function of the atom operators $\hat{\sigma}$ and $\hat{\sigma}_z$. This allows performing all simulations in a Hilbert space of dimension 2 instead of 2(N+1) for the full CQED model, with N the total number of photons considered in the truncated cavity Hilbert space.

Based on this work, a number of questions and perspectives would need to be addressed by future research. An open question remains, for instance, regarding the physical interpretation of the ansatz, $\Dot{\hat{\sigma}}=-\Gamma’\hat{\sigma}+\hat{F}$, and of the approximation of a slowly-evolving effective drive, $\hat{F}-\frac{\Dot{\hat{F}}}{\kappa'}+...\approx\hat{F}$. Indeed, $\hat{F}$ is an operator which must take into account all the input field operators, together with effective atom operators. Its complex dynamics can thus be governed by multiple timescales, which makes it hard to define simple figures of merit governing the validity of the approximation. 

As discussed in section~\ref{Sec:strong-discussion}, it is only in the framework of the approximation $\hat{F}-\frac{\Dot{\hat{F}}}{\kappa'}+\frac{\Ddot{\hat{F}}}{\kappa^{'2}}-...\approx\hat{F}$, leading to the expression $\hat{F} = \left(\hat{\sigma}_z-x\hat{\mathbb{I}}\right)\frac{\hat{\Omega}}{2}$, that the effective drive operator $\hat{F}$ can be directly interpreted as describing an effective Langevin force applied to the two-level atom. This expression indeed leads, in presence of vacuum inputs in all ports, to the expectation value $\langle\hat{F}\rangle=0$ and to a mono-exponential decay governed by $\langle\Dot{\hat{\sigma}}\rangle=- \Gamma’ \langle\hat{\sigma}\rangle$ and $\langle\Dot{\hat{\sigma}}_z\rangle = - \Gamma \langle\hat{\sigma}_z+\hat{\mathbb{I}}\rangle$, with $\Gamma=2\textrm{Re}(\Gamma’_-)$. This corresponds to a purely Markovian evolution in the case of vacuum inputs in all ports, with no recoherence occurring, in agreement with the fact that the negative decoherence rate $\gamma_{\rm NM}$ in Eq.~\eqref{effective_decoherence_rates} equals zero when $\langle\hat{\Omega}\rangle=0$. As such, the expression $\hat{F} = \left(\hat{\sigma}_z-x\hat{\mathbb{I}}\right) \frac{\hat{\Omega}}{2}$ is found to describe a non-Markovian behavior, as discussed in Section~\ref{sub-sec-non-markov}, only due to the presence of driving fields. It fails at describing the fundamental non-Markovianity obtained with cavity-QED devices in the strong-coupling regime, where vacuum Rabi oscillations occur in absence of external drive, if the CQED system has been initially excited. 

Releasing the approximation of a slowly-evolving effective drive operator, i.e. considering the time derivatives of $\hat{F}$ within the self-consistency equation introduced in Section~\ref{Sec:approximation}, seems an important perspective for the present work. Though the full CQED dynamics may not be analytically solvable in the most general case, useful results may be obtained by finding new relevant approximations, less stringent than the present one. For instance, it may be found that the expression $\hat{F} = \left(\hat{\sigma}_z-x\hat{\mathbb{I}}\right)\frac{\hat{\Omega}}{2}$ is just the first term of a more complex expression, which can be used to extend the range of validity of the self-consistent cavity elimination approach to larger coupling strengths and/or larger excitation powers.

Another main perspective lies in the extension of the approach, beyond the two-level system, to more realistic atomic systems coupled to several cavity modes. Such an extension would be fundamental to describe useful atom-photon~\cite{Hamsen2018,Brekenfeld2020,Aoki2009,Bechler2018} or spin-photon~\cite{Tomm2024,Borregaard2019,Mehdi2024,Bhaskar2020,Atature2018} interfaces, especially when using them as photon receivers in quantum network configuration~\cite{Kimble2008,Reiserer2015}. It may be interesting, for instance, to develop reduced models allowing to describe a multi-level natural or artificial atom, in interaction with two cavity modes associated to orthogonal polarisation states~\cite{Giesz2016,Lodahl2015,Mehdi2024,medeiros2025}. In such cases, two cavity modes may need to be self-consistently eliminated, associated to the two polarization eigenmodes of the cavity. The necessity to address several non-degenerate optical transitions, e.g. in presence of Zeeman-split transitions induced by external magnetic fields, will constitute an additional challenge for the extension of the self-consistent cavity elimination approach to realistic systems.

Another important perspective is the use of cavity elimination in presence of non-trivial inputs, beyond the CW coherent excitation which has been mostly used here. First, the slowly-evolving effective drive approximation may certainly find some limitation in presence of short optical pulses compared to the photon lifetime in the cavity, where $\Dot{\hat{\Omega}}$ cannot be neglected. It is quite promising, though, that the present approach already takes into account the cavity-filtering effect through the CC input operators $\hat{b}_{\rm in}^{(j\rm ,filt)}$ i.e. a convolution product whose evolution cannot be faster than the cavity damping. This may allow the self-consistent cavity elimination approach to remain useful even in the short pulse regime, at least in situations where a weakly-coupled cavity-QED device is driven from cavity-coupled ports. 

Another side of this perspective lies in the use of the present cavity elimination approach to describe CQED systems driven by non-trivial quantum inputs, potentially arising from other quantum systems. Within the input-output formalism, this could directly be performed by modeling cascaded systems~\cite{Carmichael1993,Gardiner1993,Christiansen2025}, including the case where a general quantum input is modeled as the output from a virtual cavity~\cite{Kiilerich2019}. More generally, the interaction with quantum inputs may be modeled through quantum collision models~\cite{Ciccarello2017,Maffei2021}, particularly when considering their generalization to non-Markovian dynamics~\cite{Ciccarello2022,Diosi2012,Magnifico2025}. Similarly, it would be interesting to reformulate the self-consistent cavity elimination approach within the SLH framework~\cite{Combes2017}, considering the proposals for its non-Markovian extensions~\cite{Zhang2017,Gough2016,Xue2017} that are based on the preliminary works~\cite{Imamoglu1994,Jack1999}. Indeed, the SLH formalism is ideally adapted to describe cascaded quantum systems in quantum networks.

In the long run, self-consistent cavity elimination approaches could thus be used for the realistic description of quantum network architectures~\cite{Kimble2008,Christiansen2025}, where the non-markovianity of each node and its internal dynamics is taken into account, while preserving a low Hilbert space dimension. This would be very relevant, in particular, for the description of quantum communication and quantum computing protocols where CQED devices are used both as emitters and receivers of single photons~\cite{Cirac1997,Thomas2022,Bechler2018,Nguyen2019,Stas2022,Ritter2012,Pichler2017,Ferreira2024}.

\section{Acknowledgments}
We thank Adrià Medeiros and Petr Steindl for fruitful discussions and advice. This work was partially supported by the French National Research Agency (ANR) through project ANR-22-PETQ0013. It was conducted within the research program of the QDLight joint laboratory (C2N/Quandela).

\newpage
\appendix
\section{Multiport CQED input-output formalism}\label{Annex-Heisenberg-Langevin}
The Heisenberg-Langevin equations derived within the input-output formalism are well known in the case of devices interacting with their environment through a single port, e.g for single-atoms~\cite{Gardiner1993} or CQED systems~\cite{Gardiner1985}. In this section, we follow Ref.~\cite{Gardiner1985} and extend the input-output formalism to describe CQED devices with a general geometry, that is, with an arbitrary number of input-output ports both atom-coupled (AC) and cavity-coupled (CC). Following this derivation, we check the consistency of the input-output formalism under the derivation of a product and we continue by explicitly proving the input-output relations.

\subsection{General Heisenberg-Langevin equation for a multi-port CQED system}\label{Subsec-appendix-derivation-Heisenberg-Langevin}

We first derive the Heisenberg-Langevin equations for an arbitrary system operator $\hat{O}_{\rm sys}\in\mathbb{H}_{\rm cav}\otimes\mathbb{H}_{\rm atom}$. Following~\cite{Gardiner1985}, we start by deriving the differential equation governing the evolution of the environment annihilation operators describing the AC ports, $\hat{c}_{l}(\omega)$, and describing the CC ports, $\hat{b}_{j}(\omega)$. To this purpose, we use the Hamiltonian~\eqref{Full-Hamiltonian} in the main text to compute the Heisenberg-equation of motion of an arbitrary environment operator, $\hat{O}_{\rm env} \in\mathbb{H}_{\rm env}$:
\begin{widetext}
      \begin{align}
        &\Dot{\hat{O}}_{\rm env} =-i[\hat{O}_{\rm env},\hat{\mathcal{H}}_{\textrm{sys}}+\hat{\mathcal{H}}_{\textrm{int}}+\hat{\mathcal{H}}_{\textrm{B}}]\\
        &=\sum_l\int_{-\infty}^\infty d\omega\sqrt{\frac{\gamma_{l}}{2\pi}}\left([\hat{O}_{\rm env},\hat{c}_{l}^{\dagger}(\omega)]\hat{\sigma}-\hat{\sigma}^\dagger[\hat{O}_{\rm env},\hat{c}_{l}(\omega)]\right)+\sum_j\int_{-\infty}^\infty d\omega\sqrt{\frac{\kappa_{j}}{2\pi}}\left([\hat{O}_{\rm env},\hat{b}_{j}^{\dagger}(\omega)]\hat{a}-\hat{a}^\dagger[\hat{O}_{\rm env},\hat{b}_{j}(\omega)]\right)\nonumber\\
        &-i \sum_j \int_{-\infty}^\infty d\omega (\omega-\omega_{\rm ref})[\hat{O}_{\rm env},\hat{b}^\dagger_{\rm j}(\omega)\hat{b}_{\rm j}(\omega)]-i\sum_l \int_{-\infty}^\infty d\omega (\omega-\omega_{\rm ref})[\hat{O}_{\rm env},\hat{c}^\dagger_{\rm j}(\omega)\hat{c}_{\rm j}(\omega)]\nonumber
\end{align}  
\end{widetext}

This general equation is then simplified by taking $\hat{O}_{\rm env}=\hat{c}_{l}(\omega)$ or $\hat{O}_{\rm env}=\hat{b}_{j}(\omega)$:
\begin{align}
        \Dot{\hat{c}}_{l}(\omega) &= -i(\omega-\omega_{\rm ref})\hat{c}_{l}(\omega)+\sqrt{\frac{\gamma_{l}}{2\pi}}\hat{\sigma}\label{d-j-equation}\\
        \Dot{\hat{b}}_{j}(\omega) &= -i(\omega-\omega_{\rm ref})\hat{b}_{j}(\omega)+\sqrt{\frac{\kappa_{j}}{2\pi}}\hat{a}\label{d-j-equation-2}
\end{align}
These two equations describe the evolution of the environment annihilation operators $\hat{b}_{j}(\omega)$ and $\hat{c}_{l}(\omega)$, annihilating photons at a given frequency $\omega$. Integrating them yields:
\begin{align}
        \hat{c}_{l}(\omega) &= e^{-i(\omega-\omega_{\rm ref})(t-t_0)}\hat{c}_{l,0}(\omega)\nonumber\\
        &\qquad +\sqrt{\frac{\gamma_{l}}{2\pi}}\int_{t_0}^t e^{-i(\omega-\omega_{\rm ref})(t-t')}\hat{\sigma}(t')dt'\label{eq-A-4}\\
        \hat{b}_{j}(\omega) &= e^{-i(\omega-\omega_{\rm ref})(t-t_0)}\hat{b}_{j,0}(\omega)\nonumber\\ 
        &\qquad+\sqrt{\frac{\kappa_{j}}{2\pi}}\int_{t_0}^t e^{-i(\omega-\omega_{\rm ref})(t-t')}\hat{a}(t')dt'\label{eq-A-4-2}
\end{align}
Where $\hat{c}_{l,0}(\omega)$ (resp. $\hat{b}_{j,0}(\omega)$) is the operator $\hat{c}_{l}(\omega)$ (resp. $\hat{b}_{j}(\omega)$) evaluated at an initial time $t_0$, following the same canonical commutation relations. We note that the term proportionnal to $\int_{t_0}^t e^{-i(\omega-\omega_{\rm ref})(t-t')}\hat{\sigma}(t')dt'$ explicitly describes the effect of the atom emitting a photon through an AC port, leading to a change in the environment annihilation operator $\Dot{\hat{c}}_{l}(\omega)$. Similarly, the term proportionnal to $\int_{t_0}^t e^{-i(\omega-\omega_{\rm ref})(t-t')}\hat{a}(t')dt'$ describes the effect of the cavity leaking a photon through a CC port, leading to a change in the environment annihilation operator $\hat{b}_{j}(\omega)$.

Following~\cite{Gardiner1985}, we now compute the Heisenberg equation of motion of an arbitrary system operator $\hat{O}_{\rm sys}\in\mathbb{H}_{\rm atom}\otimes\mathbb{H}_{\rm cav}$:
\begin{widetext}
       \begin{align}\label{eq-A3}
        \Dot{\hat{O}}_{\rm sys} &=-i[\hat{O}_{\rm sys},\hat{\mathcal{H}}_{\textrm{sys}}+\hat{\mathcal{H}}_{\textrm{int}}+\hat{\mathcal{H}}_{\textrm{B}}]\\
        &=\sum_l\int_{-\infty}^\infty d\omega\sqrt{\frac{\gamma_{l}}{2\pi}}\left(\hat{c}_{l}^{\dagger}(\omega)[\hat{O}_{\rm sys},\hat{\sigma}]-[\hat{O}_{\rm sys},\hat{\sigma}^\dagger]\hat{c}_{l}(\omega)\right)+\sum_j\int_{-\infty}^\infty d\omega\sqrt{\frac{\kappa_{j}}{2\pi}}\left(\hat{b}_{j}^{\dagger}(\omega)[\hat{O}_{\rm sys},\hat{a}]-[\hat{O}_{\rm sys},\hat{a}^\dagger]\hat{b}_{j}(\omega)\right)-i[\hat{O}_{\rm sys},\hat{\mathcal{H}}_{\textrm{sys}}]\nonumber
\end{align} 
\end{widetext}

Where we used the Hamiltonian defined in Eq.~\eqref{H-int-def} of the Main text. We then insert the expressions~\eqref{eq-A-4} and~\eqref{eq-A-4-2} in Eq.~\eqref{eq-A3} and use the following definition:
\begin{equation}\label{eq-A-5}
    \int_{-\infty}^\infty d\omega e^{-i\omega(t-t')} = 2\pi\delta(t-t')
\end{equation}
This leads to the general Heisenberg-Langevin equation that describes the evolution of a system operator, for a cavity-QED system with an arbitrary number of AC and CC ports:
\begin{widetext}
      \begin{align}\label{General-langevin-eq}
      &  \Dot{\hat{O}}_{\rm sys} =-i[\hat{O}_{\rm sys},\hat{\mathcal{H}}_{\textrm{int}}]\nonumber\\
        &-\sum_l\left[[\hat{O}_{\rm sys},\hat{\sigma}^\dagger]\left(\frac{\gamma_{l}}{2}\hat{\sigma}+\sqrt{\gamma_{l}}\hat{c}^{(l)}_{\textrm{in}} \right)-\left(\frac{\gamma_{l}}{2}\hat{\sigma}+\sqrt{\gamma_{l}}\hat{c}^{(l)}_{\textrm{in}} \right)^\dagger[\hat{O}_{\rm sys},\hat{\sigma}]\right]
        -\sum_j \bigg[[\hat{O}_{\rm sys}, \hat{a}^\dagger] \left(\frac{\kappa_{j} }{2} \hat{a} + \sqrt{\kappa_{j}} \hat{b}_{\textrm{in}}^{(j)} \right) - \left(\frac{\kappa_{j}}{2} \hat{a} + \sqrt{\kappa_{j}} \hat{b}^{(j)}_{\textrm{in}}\right)^\dagger [\hat{O}_{\rm sys}, \hat{a}] \bigg]
\end{align}  
\end{widetext}

Where we defined the input operators $\hat{c}^{(l)}_{\textrm{in}},\hat{b}^{(j)}_{\textrm{in}}$, at a given time t, as:
\begin{align}
        \hat{c}^{(l)}_{\textrm{in}}  &= \frac{1}{\sqrt{2\pi}}\int_{-\infty}^\infty d\omega e^{-i(\omega-\omega_{\rm ref})(t-t_0)}\hat{c}_{l,0} (\omega)\label{input-def}\\
        \hat{b}^{(j)}_{\textrm{in}}  &= \frac{1}{\sqrt{2\pi}}\int_{-\infty}^\infty d\omega e^{-i(\omega-\omega_{\rm ref})(t-t_0)}\hat{b}_{j,0} (\omega)\label{input-def-2}
\end{align}
Eq.~\eqref{General-langevin-eq} allows retrieving exactly the Heisenberg-Langevin equations describing the evolution of $\hat{a}$, $\hat{\sigma}$ and $\hat{\sigma}_z$ shown in Eq.~\eqref{eq:a_dot},~\eqref{eq:sigma_dot} and~\eqref{eq:sigma_z_dot} of the Main Text, respectively. We also note that the input operators defined in Eqs.~\eqref{input-def} and~\eqref{input-def-2} can be treated as noise terms~\cite{Gardiner1985} originating either from an AC port ($\hat{c}^{(l)}_{\textrm{in}}$) or from a CC port ($\hat{b}^{(j)}_{\textrm{in}}$), if the input states are incoherent thermal states. Conversely, $\hat{c}^{(l)}_{\textrm{in}}$ and $\hat{b}^{(j)}_{\textrm{in}}$ can be taken as to represent classical driving, i.e. coherent inputs, from their respective ports $j$ and $l$. We emphasize, however, that the interest of such input-output framework lies in its ability to provide a global description for arbitrary input geometries and arbitrary input states.

\subsection{Calculus rule: derivation of a product of system operators}\label{Subsec-appendix-commutation-relation}

In this appendix, we check the consistency of the Heisenberg-Langevin framework by explicitly showing that, for any system operators $\hat{O}_1,\hat{O}_2\in\mathbb{H}_{\rm cav}\otimes\mathbb{H}_{\rm atom}$, the fundamental calculus rule:
\begin{align}\label{calculus_rule_appendix}
       \frac{d}{dt}\hat{O}_1\hat{O}_2 = \Dot{\hat{O}}_1\hat{O}_2+\hat{O}_1\Dot{\hat{O}}_2
\end{align}
is satisfied. First, we note that, in the input-output formalism, the derivative $\frac{d}{dt}\hat{O}_1\hat{O}_2$ should be computed using Eq.~\eqref{General-langevin-eq}. The first term of this equation being a commutator, it satisfies automatically~\eqref{calculus_rule_appendix}. Following Ref.~\cite{Gardiner1985}, integrating Eq.~\eqref{eq-A-4} and using Eq.~\eqref{eq-A-5} also yields:
\begin{align}
    \sqrt{\gamma_{l}}\int_{-\infty}^{\infty} d\omega\hat{c}_{l} (\omega) &= \sqrt{\gamma_{l}}\hat{c}^{(l)}_{\textrm{in}}  + \frac{\gamma_{l}}{2}\hat{\sigma}\label{commutation-rel}\\
    \sqrt{\kappa_{j}}\int_{-\infty}^{\infty} d\omega\hat{b}_{j} (\omega) &= \sqrt{\kappa_{j}}\hat{b}^{(j)}_{\textrm{in}}  + \frac{\kappa_{j}}{2}\hat{a}\label{commutation-rel-2}
\end{align}
As $\hat{c}_{l},\hat{b}_{j}$ are environment operators that commute with every system operator at a given time, we deduce that $\sqrt{\gamma_{l}}\hat{c}^{(l)}_{\textrm{in}}+\frac{\gamma_{l}}{2}\hat{\sigma}$ and $\sqrt{\kappa_{j}}\hat{b}^{(j)}_{\textrm{in}}  + \frac{\kappa_{j}}{2}\hat{a}$ also commute with every system operator $\hat{O}_{\rm sys}$. It is then straightforward to prove the calculus rule~\eqref{calculus_rule_appendix}, by using this property when calculating $\frac{d}{dt}\hat{O}_1\hat{O}_2$ with~\eqref{General-langevin-eq}. Then, by using the calculus rule~\eqref{calculus_rule_appendix}, one can prove that the identity operators remain constant in the input-output framework, i.e $\frac{d}{dt}\hat{\mathbb{I}}_{\rm atom} = \frac{d}{dt}\{\hat{\sigma},\hat{\sigma}^\dagger\} =0$ and $\frac{d}{dt}\hat{\mathbb{I}}_{\rm cavity} = \frac{d}{dt}[\hat{a},\hat{a}^\dagger] =0$, thereby preserving the fundamental commutation rules for both the fermionic and bosonic operators. Similarly, one can compute the derivative of $\hat{\sigma}_z=2\hat{\sigma}^\dagger\hat{\sigma}-\hat{\mathbb{I}}$ using the rule~\eqref{calculus_rule_appendix}, retrieving the one computed when using the general Heisenberg-Langevin equation~\eqref{General-langevin-eq}.

\subsection{Input-output relations}\label{Subsec-input-output-relations}
We now prove the input-output relations for both the AC and CC ports. These relations are central to the formalism since they are used to compute any observable quantity in a very practical way. Following~\cite{Gardiner1985}, we start by solving Eqs.~\eqref{d-j-equation} and~\eqref{d-j-equation-2} in a time-reversed way, yielding:
\begin{align}
        \hat{c}_{l}(\omega) &= e^{-i\omega(t-t_1)}\hat{c}_{l,1}(\omega)-\sqrt{\frac{\gamma_{l}}{2\pi}}\int_{t}^{t_1}e^{-i(\omega-\omega_{\rm ref})(t-t')}\hat{\sigma}(t')dt'\label{time-reversed}\\
        \hat{b}_{j}(\omega) &= e^{-i\omega(t-t_1)}\hat{b}_{j,1}(\omega)-\sqrt{\frac{\kappa_{j}}{2\pi}}\int_{t}^{t_1}e^{-i(\omega-\omega_{\rm ref})(t-t')}\hat{a}(t')dt'\label{time-reversed-2}
\end{align}
Where $\hat{c}_{l,1},\hat{b}_{j,1}$ are the operators $\hat{c}_{l},\hat{b}_{j}$ evaluated at a given final time $t_1>t$, i.e the final-state annihilation operators for a given photon frequency $\omega$. Integrating Eq.~\eqref{time-reversed} and~\eqref{time-reversed-2} then yields:
\begin{align}
        \sqrt{\gamma_{l}}\int_{-\infty}^\infty d\omega\hat{c}_{l} (\omega) &= \sqrt{\gamma_{l}}\hat{c}^{(l)}_{\textrm{out}}  -\frac{\gamma_{l}}{2}\hat{\sigma}\label{final-time-input-output}\\
        \sqrt{\kappa_{j}}\int_{-\infty}^\infty d\omega\hat{b}_{j} (\omega) &= \sqrt{\kappa_{j}}\hat{b}^{(j)}_{\textrm{out}}  -\frac{\kappa_{j}}{2}\hat{a}\label{final-time-input-output-2}
\end{align}
Where we defined the output operators $\hat{c}^{(l)}_{\textrm{out}},\hat{b}_{\textrm{out}}^{(j)}$ as:
\begin{align}
        \hat{c}^{(l)}_{\textrm{out}}  &= \frac{1}{\sqrt{2\pi}}\int_{-\infty}^\infty d\omega e^{-i(\omega-\omega_{\rm ref})(t-t_1)}\hat{c}_{l,1} (\omega)\\
        \hat{b}_{\textrm{out}}^{(j)}  &= \frac{1}{\sqrt{2\pi}}\int_{-\infty}^\infty d\omega e^{-i(\omega-\omega_{\rm ref})(t-t_1)}\hat{b}_{j,1} (\omega)\nonumber
\end{align}
Contrary to the input operators $\hat{b}_{\textrm{in}}^{(j)}$ and $\hat{c}^{(l)}_{\textrm{in}}$ which are constructed using the environment annihilation operators at an initial time $t_0$, the output operators are constructed using the environment annihilation operators at a given final time $t_1$.

Substracting Eq.~\eqref{commutation-rel} (resp.~\eqref{commutation-rel-2}) from Eq.~\eqref{final-time-input-output} (resp.~\eqref{final-time-input-output-2}) then leads to the input-output relation for an AC port $l$ (resp., CC port $j$):
\begin{align}
        \hat{c}^{(l)}_{\textrm{out}} &= \hat{c}^{(l)}_{\textrm{in}} + \sqrt{\gamma_{l}}\hat{\sigma}\\
        \hat{b}_{\textrm{out}}^{(j)} &= \hat{b}_{\textrm{in}}^{(j)}+\sqrt{\kappa_{j}}\hat{a}
\end{align}
We note that the input-output relations related to the AC ports are characteristic of waveguide-QED systems~\cite{Fan2010,Lalumiere2013}, depending solely on the inputs fields and the atomic lowering operator $\hat{\sigma}$, thereby providing a direct access to the atomic influence on the output fields. In contrast, the input-output relations related to the CC ports only reveal the cavity's effect on the output fields through the operator $\hat{a}$, preventing from directly describing the effect of the atom on these output fields. This distinction is vital for experiments where the primary system of interest is usually the atomic system through its interaction with the input fields~\cite{Kimble1998,Blatt2012,Senellart2017,Lodahl2015}, while the cavity's main interest lies in the Purcell enhancement and efficient extraction of photon emission through a given port.

\section{Cavity elimination}
In this appendix, we derive and analyse some key analytical points of the self-consistent cavity elimination. We first compute the integral $\int_0^\infty\hat{\sigma}(t-t')e^{-\kappa' t'}dt'$, thereby proving the main result which allows computing the self-consistency equation, and so, eliminating the cavity. We then proceed with the analysis of the complex damping rates $\Gamma'_{+}$ and $\Gamma'_-$. Finally, we give details on the derivation of the effective dynamics by solving the system of effective Bloch equations and providing its eigenvectors. 

\subsection{Calculation of $\int_0^\infty\hat{\sigma}(t-t')e^{-\kappa' t'}dt'$}\label{Subsec-appendix-integral-calc}
We first aim at re-expressing the integral $\int_0^\infty\hat{\sigma}(t-t')e^{-\kappa' t'}dt'$, which is the convolution product between the atom's lowering operator $\hat{\sigma}$ and an exponential decay function governed by the cavity damping rate $\kappa$.

We note that the non-Markovian behavior of the self-consistent cavity-elimination model, shown when computing the effective Lindblad-like equation, Eq.~\eqref{Lindblad-def} in the Main text, arises from these calculations, and particularly, from taking into account the past of the atom through the expression $\hat{\sigma}(t-t')$. To compute $\int_0^\infty\hat{\sigma}(t-t')e^{-\kappa' t'}dt'$, we first use Taylor's theorem to write $\hat{\sigma}(t-t')$ as:
\begin{equation}
    \hat{\sigma}(t-t') = \hat{\sigma}(t)+\sum_{n=1}^\infty \frac{(-t')^n}{n!}\frac{d^n}{dt^n}\hat{\sigma}(t)
\end{equation}
This decomposition then allows expanding $\int_0^\infty\hat{\sigma}(t-t')e^{-\kappa' t'}dt'$ as: 
\begin{align}
    &\int_0^\infty\hat{\sigma}(t-t')e^{-\kappa' t'}dt' =  \hat{\sigma}(t)\int_0^\infty e^{-\kappa' t'}dt'\nonumber\\ 
    &\qquad + \frac{d^n}{dt^n}\hat{\sigma}(t)\sum_{n=1}^\infty\int_0^\infty\frac{(-t')^n}{n!} e^{-\kappa' t'}dt'
\end{align}
And, using the general result:
\begin{equation}
    \int_0^\infty \frac{(-t')^n}{n!} e^{-\kappa' t'}dt' = -\frac{1}{(-\kappa')^{n+1}}
\end{equation}
we derive:
\begin{align}\label{firts-int-res}
        \int_0^\infty\hat{\sigma}(t-t')e^{-\kappa' t'}dt' &= \sum_{n=0}^\infty \frac{1}{(-\kappa')^{n+1}}\frac{d^n}{dt^n}\hat{\sigma}(t)\\
            &=-\frac{\hat{\sigma}}{\kappa'}+\frac{\Dot{\hat{\sigma}}}{\kappa'^2}-\frac{\Ddot{\hat{\sigma}}}{\kappa'^3}+...\nonumber
\end{align}
As such, the integral has been expressed as an infinite sum of derivatives which is not very practical to use. To go further, we insert the ansatz $\Dot{\hat{\sigma}}=-\Gamma'\hat{\sigma}+\hat{F}$ (see Main Text) in the calculations, and more particularly, we use it to express the $n^{th}$ derivative of $\hat{\sigma}$, for $n\geq 1$, as:
\begin{equation}
    \frac{d^n}{dt^n}\hat{\sigma}  = (-\Gamma')^n\hat{\sigma}+\sum_{k=0}^{n-1}(-\Gamma')^{n-1-k}\frac{d^k}{dt^k}\hat{F}
\end{equation}
Inserting this expression in Eq.~\eqref{firts-int-res} yields:
\begin{align}
       &\int_0^\infty\hat{\sigma}(t-t')e^{-\kappa' t'}dt'\nonumber\\ 
       &= \frac{1}{\kappa'}\sum_{n=0}^\infty \left(\frac{\Gamma'}{\kappa'} \right)^n\hat{\sigma}+\frac{1}{\kappa'}\sum_{n = 1}^\infty \frac{1}{(-\kappa')^n}\sum_{k=0}^{n-1}(-\Gamma')^{n-1-k}\frac{d^k}{dt^k}\hat{F} \label{eq-B-2}\\
        &= \frac{1}{\kappa'-\Gamma'}\hat{\sigma}+\frac{1}{\kappa'}\sum_{n = 1}^\infty \frac{1}{(-\kappa')^n}\sum_{k=0}^{n-1}(-\Gamma')^{n-1-k}\frac{d^k}{dt^k}\hat{F} \label{eq-B-3}
\end{align}
Where Eq.~\eqref{eq-B-3} only holds if the convergence condition:
\begin{equation}
    \left|\frac{\Gamma'}{\kappa'}\right|<1\label{Convergence-condition}
\end{equation}
is satisfied. To understand this condition, we use the expressions $\Gamma' =\frac{\Gamma}{2}+i(\omega_{\rm a'}-\omega_{\rm ref})$ and $\kappa' =\frac{\kappa}{2}+i(\omega_{\rm c}-\omega_{\rm ref})$ to rewrite~\eqref{Convergence-condition} as:
\begin{equation}
    \left|\frac{\Gamma}{2}\right|^2+\left|\omega_{\rm a'}-\omega_{\rm ref}\right|^2<\left|\frac{\kappa}{2}\right|^2+\left|\omega_{\rm c}-\omega_{\rm ref}\right|^2 \label{Convergence-condition-final}
\end{equation}
This new expression shows explicitly that the convergence condition~\eqref{Convergence-condition} depend on the reference frequency $\omega_{\rm ref}$. However, this reference frequency can be arbitrarily chosen. In addition, $\Gamma$ is always smaller than $\kappa$ (as discussed in more details in Sec.~\ref{Subsec-Study-Gamma}) so that this convergence condition can generally be satisfied for a suitable choice of the reference frequency $\omega_{\rm ref}$.
We then continue the calculations by using the rule:
\begin{equation}
    \sum_{n = 1}^\infty u_n \left( \sum_{k = 0}^{n-1} v_k \right) = \sum_{k=0}^\infty v_k \left( \sum_{n=k+1}^\infty u_n \right)
\end{equation}
to simplify the equation~\eqref{eq-B-3} as:
\begin{align}
       &\int_0^\infty\hat{\sigma}(t-t')e^{-\kappa' t'}dt'\nonumber\\
       &= \frac{1}{\kappa'-\Gamma'}\hat{\sigma}+\frac{1}{\kappa'}\sum_{k=0}^{\infty}\frac{d^k}{dt^k}\hat{F}\sum_{n=k+1}^\infty \frac{1}{(-\kappa')^n}(-\Gamma')^{n-1-k}\\
        &= \frac{1}{\kappa'-\Gamma'}\hat{\sigma}+\frac{1}{\kappa'}\sum_{k=0}^{\infty}\frac{1}{(-\kappa')^{k+1}}\frac{d^k}{dt^k}\hat{F}\sum_{n=0}^\infty \left(\frac{\Gamma'}{\kappa'} \right)^n\\
         &= \frac{1}{\kappa'-\Gamma'}\hat{\sigma}-\frac{1}{\kappa  '(\kappa'-\Gamma')}\sum_{k=0}^{\infty}\frac{1}{(-\kappa')^{k}}\frac{d^k}{dt^k}\hat{F}\\ 
         &= \frac{1}{\kappa'-\Gamma'}\hat{\sigma}-\frac{1}{\kappa  '(\kappa'-\Gamma')}\left(\hat{F}-\frac{\Dot{\hat{F}}}{\kappa^{'}}+\frac{\Ddot{\hat{F}}}{\kappa^{'2}}-...\right)\label{Result-integral}
\end{align}
Which is the result used in Eq.~\eqref{int-result} of the Main Text to derive the self-consistency equation, leading to the $\cancel{\rm C}$QED model. It is, as in Eq.~\eqref{firts-int-res}, an infinite sum of derivatives. However, Eq.~\eqref{Result-integral} is an improvement from Eq.~\eqref{firts-int-res} since computing $\hat{F}$ is equivalent to computing $\Dot{\hat{\sigma}}$, meaning that an order of differentation has been removed in the computational complexity.

\subsection{Study of $\Gamma'_+$ and $\Gamma'_-$}\label{Subsec-Study-Gamma}

Here we provide the complete formulas for $\Gamma'_+$ and $\Gamma'_-$ by expressing explicitly the complex square root $\sqrt{(\kappa'-\gamma')^2-4g^2}$. We then study their real and imaginary parts and, more specifically, their dependence on the light-matter coupling strength $g$ and on the atom-cavity detuning $\omega_{\rm a}-\omega_{\rm c}$. 

As shown in Eq.~\eqref{identification-eq} of the Main Text, the exact expressions for $\Gamma'_+$ and $\Gamma'_-$ can be deduced from the self-consistency equation by identifying the terms only related to the TLS behavior: $-\Gamma'\hat{\sigma} = -(\gamma'+\frac{g^2}{\kappa'-\Gamma'})\hat{\sigma}$. This identification leads to the expressions:
\begin{equation}
    \Gamma_\pm'=\frac{\kappa'+\gamma'\pm\sqrt{(\kappa'-\gamma')^2-4g^2}}{2}
\end{equation}
As $(\kappa'-\gamma')^2$ is a complex number, the expression of $\sqrt{(\kappa'-\gamma')^2-4g^2}$ is not trivial and we express it as:
\begin{equation}
    \sqrt{(\kappa'-\gamma')^2-4g^2} = f_1 + if_2
    \label{expression-srt}
\end{equation}
Where $f_1,f_2\in\mathbb{R}$ are two real coefficients defined as:
\begin{align}
    f_1 &= \sqrt{\frac{1}{2}(\sqrt{g_1^2 + g_2^2} + g_1)}\\
    f_2 &= \pm\sqrt{\frac{1}{2}(\sqrt{g_1^2 + g_2^2} - g_1)}\nonumber
\end{align}
With $g_1,g_2$ two real-valued constants defined as:
\begin{align}
        g_1 &= \frac{1}{4}(\kappa-\gamma_{\rm a})^2-(\omega_{\rm c}-\omega_{\rm a})^2-4g^2\\
        g_2 &= (\kappa-\gamma_{\rm a})(\omega_{\rm c}-\omega_{\rm a})\nonumber
\end{align}
We note that the sign of $f_2$ should be choosen to be the same than the one of $(\omega_{\rm c}-\omega_{\rm a})(\kappa-\gamma_{\rm a})$ thus depending on the device parameters. These two eigenvalues are exactly the same than the ones computed and analyzed in~\cite{Auffeves2008}, where it is shown that they correspond to the complex frequencies of the first manifold's dressed states of the atom-cavity system.

\begin{figure*}[t]
    \centering
    \makebox[\textwidth][c]{\includegraphics[width=\linewidth]{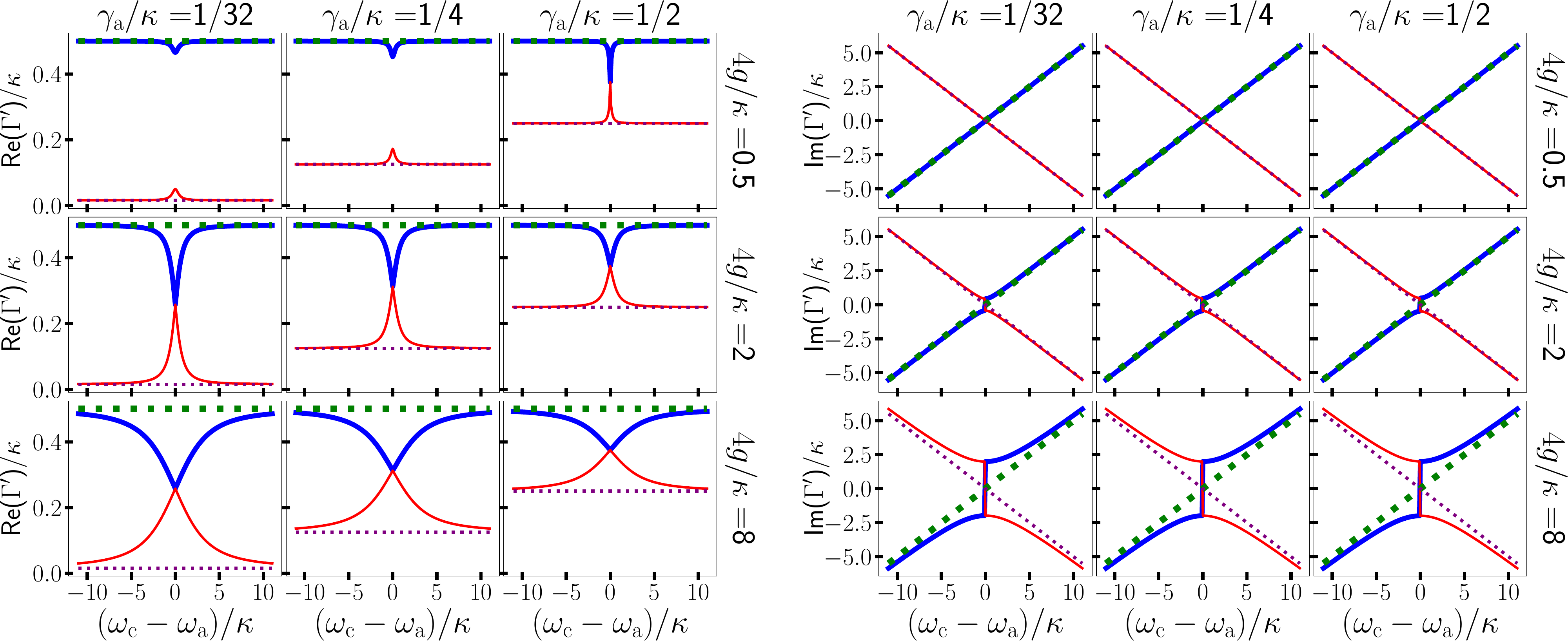}}
    \caption{Study of $\Gamma'_+$ and $\Gamma'_-$, depending on the atom-cavity detuning $\omega_{\rm a}-\omega_{\rm c}$. The left panel displays the real parts of $\Gamma'_\pm$ and the right one displays their imaginary parts. Each column represents different values of $\gamma_{\rm a}=\kappa/2$, $\kappa/4$, $\kappa/32$ and each row represent different values of atom-cavity coupling rate: $4g=0.5\kappa$, $2\kappa$, $8\kappa$. In both panels, the red thin curve represents $\Gamma'_-$ and the blue thick one represents $\Gamma'_+$. In the left panel, the thin purple dotted line represents Re($\gamma'$) and the thick green dotted one represents Re($\kappa'$). In the right panel, the thick green dotted line represents $\frac{\omega_{\rm c} - \omega_{\rm a}}{2}$ and the thin purple dotted one represents $-\frac{\omega_{\rm c} - \omega_{\rm a}}{2}$.}\label{fig:Gamma-prime-study}
\end{figure*}

In Fig.~\ref{fig:Gamma-prime-study}, we use Eq.~\eqref{expression-srt} to compare the real (left panel) and imaginary (right panel) parts of $\Gamma'_+$ and $\Gamma'_-$. Each column corresponds to different values of $\gamma_{\rm a}$ ($=\kappa/2$, $=\kappa/4$ and $=\kappa/32$), and each row represents a different atom-cavity coupling strength $4g/\kappa$ ($=0.5$, $=2$ and $=8$). In both panels, the thick blue lines represents $\Gamma'_+$, while the thin red ones represents $\Gamma'_-$.  In the left panel, the thin purple dotted line represents Re($\gamma'$) and the thick green dotted one represents Re($\kappa'$). In the right panel, the thick green dotted line represents $\frac{\omega_{\rm c} - \omega_{\rm a}}{2}$ and the thin purple dotted one represents $-\frac{\omega_{\rm c} - \omega_{\rm a}}{2}$.

The left panel of Fig.~\ref{fig:Gamma-prime-study} demonstrates that $\textrm{Re}(\Gamma'_+) \ge \textrm{Re}(\Gamma'_-)$, with equality only reached at the atom-cavity resonance (i.e. $\omega_{\rm a}=\omega_{\rm c}$) in the strong coupling regime (i.e. $4g > |\kappa-\gamma_{\rm a}|$). 
We thus use this fact and combine it with the long-time approximation made in Eq.~\eqref{eq-a-to-solve-final} of the Main Text ($t \gg 1/\kappa$) to justify the choice of $\Gamma'=\Gamma'_-$. Indeed, the scope of the self-consistent cavity elimination approximation is only to capture the dynamics of the CQED system at long time scales, i.e. the dynamics governed by the damping rate with the smallest real part. Furthermore, we note that when the atom and the cavity are largely detuned, the real parts of $\Gamma'_-$ and $\Gamma'_+$ tend towards coupling-independent values: Re$(\Gamma'_-)\rightarrow$Re($\gamma'$) and Re$(\Gamma'_+)\rightarrow$ Re($\kappa'$). This behavior highlights the analogy made between the cavity-like ($\Gamma'_+$) and atomic-like ($\Gamma'_-$) eigenvalues.
However, fixing $\omega_{\rm a}-\omega_{\rm c}=0.5\kappa$ and $\gamma_{\rm a}=\kappa/32$ (as done in the Main Text) leads to the following coupling-dependent values for the effective damping rates: $\frac{4g}{\kappa} = 0.5 \rightarrow \left(\textrm{Re}(\Gamma'_+) = 0.48 , \textrm{Re}(\Gamma'_-) = 0.03\right) $, $\frac{4g}{\kappa} = 2 \rightarrow \left(\textrm{Re}(\Gamma'_+) = 0.37 , \textrm{Re}(\Gamma'_-) = 0.14\right) $ and $\frac{4g}{\kappa} = 8\rightarrow \left(\textrm{Re}(\Gamma'_+) = 0.28, \textrm{Re}(\Gamma'_-) = 0.22\right) $. These numerical values shows  highly coupling-dependent damping rates and, most importantly, they show that Re($\Gamma'_+$)$\approx$Re($\Gamma'_-$) for the largest atom-cavity coupling rate: the second eigenvalue $\Gamma'_+$ is not negligible anymore to represent the complete dynamics of the CQED system.

Finally, we note that the first row of this plot (i.e. $4g/\kappa=0.5$) illustrates the importance of taking the bare atom's emission rate $\gamma_{\rm a}$ in the coupling strength definition. Indeed, taking $\gamma_{\rm a}=\frac{\kappa}{32}$ or $\gamma_{\rm a}=\frac{\kappa}{4}$ result in $\textrm{Re}(\Gamma'_+)\ne\textrm{Re}(\Gamma'_-)$ at resonance, as can be seen in the first two columns (top row) in Fig.~\ref{fig:Gamma-prime-study}. In contrast, the third column (i.e. $\gamma_{\rm a}=\kappa/2$) yields $4g/|\kappa-\gamma_{\rm a}|=1$, corresponding to the threshold of the strong coupling regime and leading to Re($\Gamma'_-$)=Re($\Gamma'_+$) when $\omega_{\rm a}=\omega_{\rm c}$.

The right panel of Fig.~\ref{fig:Gamma-prime-study} displays the imaginary parts of the two damping rates depending on the atom-cavity detuning. In the strong coupling regime, this plots show a discontinuity in the values of Im($\Gamma'_+$) and Im($\Gamma'_-$) at resonance. This effect is mathematically caused by the term $\sqrt{(\kappa'-\gamma')^2-4g^2}$ being purely complex in this regime, leading to a jump in the values of Im($\Gamma'_+$) and Im($\Gamma'_-$) when its sign flips along with the sign of the atom-cavity detuning. The stronger the coupling is, the larger the amplitude of $\sqrt{(\kappa'-\gamma')^2-4g^2}$ is, and, consequently, the larger the jump is. Finally, we note that when the coupling regime is weak, $\textrm{Im}(\Gamma'_\pm)$ evolves linearly as Im($\Gamma'_-$)$=\omega_{\rm a'}\approx\omega_{\rm a}$ and Im$(\Gamma'_+$)$=\omega_{\rm c'}\approx\omega_{\rm c}$. As the coupling increases, the atom-cavity interaction becomes significant and the effective frequencies $\omega_{\rm a'}$ and $\omega_{\rm c'}$ deviate qualitatively from the bare atom and cavity frequencies $\omega_{\rm a}$ and $\omega_{\rm c}$.

\subsection{Solution of the reduced effective dynamics}\label{Subsec-effective-dynamics-solution}

We now detail the resolution of the effective system of Bloch equations provided in Eq.~\eqref{Bloch-equation-final-system} of the Main Text, within the self-consistent cavity-elimination model. As done in the Main Text, we focus on the case of a single continuous-wave (CW) excitation resonant with the effective frequency ($\omega_{\rm laser}=\omega_{\rm a'}$), which allows solving the effective dynamics analytically. We note that the non-resonant case is also analytically solvable, however it yields particularly complex expressions -- we therefore rely on numerical solutions in this regime.

We start by writing the system of effective Bloch equations in a matrix form:
\begin{equation}
    \Dot{X} = AX+B\label{Bloch-equation-matrix}
\end{equation}
where we defined the matrices $X,$ $A$ and $B$ as:
\begin{align}
        X =\begin{pmatrix}
        \langle\hat{\sigma}_z\rangle\\
        \langle\hat{\sigma}\rangle\\
        \langle\hat{\sigma}^\dagger\rangle
    \end{pmatrix}\nonumber\\ 
    &A = \begin{pmatrix}
        -\Gamma & & -(1+x)^*\Omega^* & & -(1+x)\Omega\\
        \frac{\Omega}{2} & & -\frac{\Gamma}{2} & & 0\\
        \frac{\Omega^*}{2} & & 0 & & -\frac{\Gamma}{2}
    \end{pmatrix}\nonumber\\ 
    &
    B = \begin{pmatrix}
        -\Gamma \\
        -\frac{\Omega}{2}\\
        -\frac{\Omega^*}{2}
    \end{pmatrix}
\end{align}
To solve this system of linear equations, we diagonalize $A$ and write it as:
\begin{equation}
    A = PDP^{-1}
\end{equation}
where $D$ is a diagonal matrix with eigenvalues:
\begin{align}\label{eigenvalues}
            \lambda_0 &= -\frac{\Gamma}{2}\\
            \lambda_\pm &= \frac{\Gamma}{4}\left(-3\pm\sqrt{1-\frac{16|\Omega|^2}{\Gamma^2}Re\left(1+x\right)}\right)\nonumber
\end{align}
and $P$ is the change-of-basis matrix made of the eigenvectors of $A$, $v_0,v_+,v_-$, which are written as:
\begin{align}
    v_0 = \begin{pmatrix}
    0\\
        -\Omega(1+x)\\
        \Omega^*(1+x)^*
    \end{pmatrix}
\end{align}
and:
\begin{align}
        v_\pm =\begin{pmatrix}
        \frac{\Gamma}{4}\left(-1\pm\sqrt{1-\frac{16|\Omega|^2}{\Gamma^2}Re\left(1+x\right)}\right)\\
        \frac{\Omega}{2}\\
        \frac{\Omega^*}{2}
    \end{pmatrix}
\end{align}
These eigenvectors depend both on the inputs and on the atom-cavity coupling, through $\Omega$ and $x$, and are used to compute the solution of the system of effective Bloch equations. Using this decomposition, we write the solution of Eq.~\eqref{Bloch-equation-matrix} as:
\begin{equation}
    X(t)-X_{\rm st} = Pe^{Dt}P^{-1}(X(t)-X_{\rm st})
\end{equation}
with: 
\begin{equation}
    X_{\rm st} = \begin{pmatrix}
        \langle\hat{\sigma}_z\rangle_{\rm st}\\
        \langle\hat{\sigma}\rangle_{\rm st}\\
        \langle\hat{\sigma}^\dagger\rangle_{\rm st}
    \end{pmatrix}
\end{equation}
Simplifying this result yields the expressions provided in Eq.~\eqref{Bloch-solution} of the Main Text, i.e the evolution of $\langle\hat{\sigma}\rangle(t)$,$\langle\hat{\sigma}^\dagger\rangle(t)$ and $\langle\hat{\sigma}_z\rangle(t)$.

\subsection{EXPRESSIONS OF THE FIRST-ORDER CORRELATION FUNCTION COEFFICIENTS UNDER A SINGLE, COHERENT, CW EXCITATION.}\label{Appendix}
Finally, we give explicit formulas for the coefficients that appear in the first-order auto-correlation function of the AC ports~\eqref{g-1-atom-ports} (resp. the CC ports~\eqref{incohrent-g-1}), that is the expression of $\mathcal{N}_{0}$, $\mathcal{N}_{1}$ and $\mathcal{N}_{2}$ (resp. $\mathcal{N}_{0}'$, $\mathcal{N}_{1}'$ and $\mathcal{N}_{2}'$). To compute the coefficients $\mathcal{N}_{0}$, $\mathcal{N}_{1}$, $\mathcal{N}_{2}$, we first recall the general expression of the incoherent part of the first-order auto-correlation function, for any AC port~\eqref{g-1-atom-incoh-general}. Then, by assuming a single CW excitation with $\omega_{\rm las} = \omega_{\rm a'}$, we insert the explicit expressions of the coefficients $A_i$~\eqref{A-i-sol} in the general formula~\eqref{g-1-atom-incoh-general}, and, by factorizing around the exponentials, we express the effective first-order auto-correlation function in Eq.~\eqref{g-1-atom-ports} with the coefficients:
\begin{widetext}
    \begin{align}
        \mathcal{N}_{0} &= \frac{|\Omega|^4|1+x|^2(1+x)^*}{4\lambda_+\lambda_-(\lambda_+-\lambda_0)(\lambda_--\lambda_0)} \left[ 1 - \frac{x-x^*}{2\lambda_0^2|1+x|^2} \left( |\Omega|^2|1+x|^2 - \lambda_+\lambda_-(1+x) \right) \right] \nonumber\\
       \mathcal{N}_{1} &= \frac{|\Omega|^4|1+x|^2}{4\lambda_+\lambda_-(\lambda_--\lambda_+)} \left[ (1+x)\left(2-\frac{\lambda_++\lambda_-}{\lambda_0}\right) - \frac{|\Omega|^2|1+x|^2}{\lambda_+\lambda_-\lambda_0}(\lambda_0-\lambda_--\lambda_+) \right] \\
        \mathcal{N}_{2} &= -\frac{|\Omega|^4|1+x|^2}{4(\lambda_+\lambda_-)^2\lambda_0(\lambda_--\lambda_+)} \left( |\Omega|^2|1+x|^2 - \lambda_+\lambda_-(1+x) \right) \nonumber
    \end{align}
\end{widetext}    

Following the same procedure, we recall the general expression for the incoherent part of the first-order auto-correlation function for the CC ports~\eqref{general-g-1-expression}. Then, the $C_i$ coefficients are expressed in the case of a single CW excitation as~\eqref{C-i-expr}, leading to the expression~\eqref{incohrent-g-1} with the coefficients:
\begin{widetext}
    \begin{align}
        \mathcal{N}_{0}' &= \frac{|\Omega|^4|1+x|^2(1+x)^*}{4\lambda_+\lambda_-(\lambda_+-\lambda_0)(\lambda_--\lambda_0)} \left[ 1 - \frac{1}{2\lambda_0^2\kappa'} \left( \left(|\Omega|^2 - \frac{\lambda_+\lambda_-}{(1+x)^*}\right)(2\lambda_0 - \kappa'(x^* - x)) - |\Omega|^2\lambda_0(x^* - x) \right) \right] \nonumber\\
        \mathcal{N}_{1}' &= \frac{|\Omega|^4|1+x|^2(1+x)}{4\lambda_+\lambda_-(\lambda_+-\lambda_-)\kappa'^*} \left[ 1 + \frac{1}{\lambda_0\kappa'(1+x)} \left( |\Omega|^2(1+x)^* - \lambda_+\lambda_- - \left( \frac{|\Omega|^2|1+x|^2}{\lambda_+\lambda_-}(\kappa'+\lambda_0) - \kappa'(1+x) \right)(\lambda_0-\lambda_+-\lambda_-) \right) \right] \\
        \mathcal{N}_{2}' &= \frac{|\Omega|^4|1+x|^2}{4\lambda_0\lambda_+\lambda_-(\lambda_+-\lambda_-)|\kappa'|^2} \left( \kappa'(1+x) + 2\lambda_0 - \frac{|\Omega|^2|1+x|^2}{\lambda_+\lambda_-}(\kappa'+\lambda_0) \right)\nonumber
    \end{align}
\end{widetext}
We note that, under a coherent CW excitation at $\omega_{\rm las} = \omega_{\rm a'}$, these coefficients are very practical since they allow computing analytically the first-order auto-correlation function for both type of ports, as well as spectral densities.

\bibliography{references_final}

\end{document}